\documentclass[12pt]{article}
\usepackage{graphicx}
\usepackage{amsmath}
\usepackage{amssymb}
\usepackage{color}
\usepackage{amsfonts}
\usepackage{hyperref}

\usepackage[T1]{fontenc}
\usepackage[utf8x]{inputenc}
\usepackage{fancyhdr}
\usepackage{lipsum}
\usepackage{authblk}
\usepackage{mathrsfs}

\usepackage[backend=biber,style=numeric,sorting=none,doi=false,isbn=false,url=true,eprint=true,maxbibnames=99]{biblatex}
\renewbibmacro{in:}{}

\DeclareFieldFormat[article,inproceedings,incollection]{title}{\mkbibquote{#1}}

\DeclareFieldFormat{pages}{#1}

\usepackage{csquotes}
\usepackage[makeroom]{cancel}
\usepackage[normalem]{ulem}
\graphicspath{ {imagenes/} }

\usepackage[T1]{fontenc}

\numberwithin{equation}{section}

\definecolor{blue-violet}{rgb}{0.54, 0.17, 0.89}
\definecolor{PineGreen}{cmyk}{0.92, 0, 0.59, 0.25}
\definecolor{OliveGreen}{cmyk}{0.64, 0, 0.95, 0.40}
\definecolor{RawSienna}{cmyk}{0, 0.72, 1, 0.45}
\definecolor{Gray}{cmyk}{0, 0, 0, 0.50}
\definecolor{MidnightBlue}{cmyk}{0.98, 0.13, 0, 0.43}
\definecolor{Orange}{cmyk}{0, 0.61, 0.87, 0}
\definecolor{LimeGreen}{cmyk}{0.50, 0, 1, 0}
\definecolor{Green}{cmyk}{1, 0, 1, 0}
\definecolor{brightube}{rgb}{0.82, 0.62, 0.91}
\definecolor{JazzberryJam}{rgb}{0.65, 0.04, 0.37}

\newcommand{\diff}{\mathrm{d}}

\begin{document}

\title{\bf
Edge observables in Maxwell theory on null boundaries}

\author[1]{Du\v san \DJ or\dj evi\'c\thanks{dusan.djordjevic@ff.bg.ac.rs }}
\author[2,3]{Olivera Miskovic\thanks{olivera.miskovic@pucv.cl}}
\author[2]{Antonia Montecinos\thanks{antonia.montecinos@pucv.cl }}
\author[4]{Tatjana~Vuka\v{s}inac\thanks{tatjana.vukasinac@umich.mx}}
\affil[1]{\small \it Faculty of Physics, University of Belgrade, Studentski Trg 12-16, 11000 Belgrade, Serbia}
\affil[2]{\it Instituto de F\'isica, Pontificia Universidad Cat\'olica de Valpara\'iso,\newline   Avda.~Universidad 330, Curauma, Valpara\'iso, Chile}
\affil[3]{\it DISAT, Politecnico di Torino, Corso Duca degli Abruzzi, 24, 10129 Torino, Italy}
\affil[4]{\it Facultad de Ingenier\'ia Civil, Universidad Michoacana de San Nicol\'as de Hidalgo,\newline
Morelia, Michoac\'an 58000, M\'exico}


\maketitle

\begin{abstract}

We study edge observables in three-dimensional Maxwell theory on null
boundaries using a Hamiltonian formulation adapted to null foliations. We consider three backgrounds:  Minkowski spacetime, the BTZ black hole, and de Sitter space, where the relevant boundaries are, respectively, null infinity, the black hole horizon, and the cosmological horizon. In all three cases we find, besides the usual $\mathrm U(1)$ charge, a
second edge observable associated with a shift transformation intrinsic to the null boundary. The two charges are well defined quasilocally. They form a centrally extended algebra which, in Fourier modes, becomes two independent Abelian Kac--Moody algebras with opposite levels.  Thus, the same centrally extended edge structure appears at both asymptotic and finite-distance null boundaries. The phase space also admits charge-conjugation twisted configurations, in which the modes of the
algebra are half-integer. Finally, in Minkowski spacetime we find an exact
radiative solution in this twisted sector, with both charges nonvanishing.

\end{abstract}

\newpage

\tableofcontents

\section{Introduction}

Null hypersurfaces provide a natural setting for describing radiative
degrees of freedom. They are adapted to wave propagation and carry the
boundary data on which asymptotic symmetries act. At the same time, their geometry differs essentially from that of spacelike or timelike hypersurfaces: the induced metric is degenerate, and the canonical structure is sensitive to the choice of time foliation~\cite{Blagojevic:1993fp}.  Null foliations generally contain a characteristic zero mode ambiguity ~\cite{Dirac:1949cp,Maskawa:1975hx,Heinzl:1998kz,Alexandrov:2014rba}. This ambiguity may have physical
consequences, giving rise to additional soft charges or central extensions
of the boundary symmetry algebra ~\cite{Gonzalez:2023yrz,Simic:2023exz,Gonzalez:2024rho}. 

The purpose of this work is to investigate this structure in
three-dimensional Maxwell electrodynamics. This theory provides a
particularly simple model, and it already contains the essential ingredients
needed to study how null foliations give rise to boundary symmetries and
their associated soft degrees of freedom. In particular, Maxwell theory in three dimensions has a single local propagating degree of freedom.
Locally, the field strength is dual to a one-form, 
\begin{equation}
{}^{\ast }F\sim \diff\phi \,,
\label{scalar}
\end{equation}
so that the photon is locally dual to a scalar field \cite{Polyakov:1977}.

Three-dimensional electrodynamics also has distinctive infrared properties. The electrostatic potential of a point charge $q$ in flat space, analogous to the four-dimensional Coulomb potential, has the form
\begin{equation}
A_{t}(r)=\frac{q}{2\pi }\ln \left( \frac{r}{r_{0}}\right) .
\end{equation}
The corresponding electric field falls off as $1/r$, whereas the potential and the electrostatic energy grow logarithmically with distance. This logarithmic behaviour characterizes the nonradiative sector and must be
distinguished from the radiative behaviour of the propagating modes considered here.

In the radiative sector, source-free Maxwell theory in flat space,
expressed in terms of the scalar \eqref{scalar}, reduces to the wave equation. The asymptotic behaviour for harmonic modes of frequency $\omega>0$, in spherical coordinates,  is \begin{equation}
\phi(x)=\mathrm{Re}\left\{ \frac{ a^+(\varphi ,\omega )}{\sqrt{r}}\,\mathrm{e}^{-\mathrm{i}\omega (t-r)}+\frac{ a^-(\varphi ,\omega )}{\sqrt{r}}\,\mathrm{e}^{-\mathrm{i}\omega
(t+r)}\right\} +\mathcal{O}(r^{-3/2})\,,
\end{equation}
where the functions $a^\pm$ describe outgoing and incoming angular profiles. Introducing the null coordinates $u=t\pm r$, and restricting to one of the two branches,  we obtain the radiative falloff
\begin{equation}
\phi(x)=\frac{\phi_{(1)}(u,\varphi )}{\sqrt{r}}+\mathcal{O}(r^{-3/2})\,.
\label{expphi}
\end{equation}
It then follows from \eqref{scalar} the $u$-dependent falloff
$A_{u}=\mathcal{O}(r^{-1/2})$, $A_{r}=\mathcal{O}(r^{-3/2})$ and
$A_{\varphi}=\mathcal{O}(r^{1/2})$, in agreement with the radiative asymptotic conditions discussed in~\cite{Shimizu:2025hfl}. The half-integer radial behaviour reflects the
characteristic radiative falloff in three spacetime dimensions. We focus on this sector and exclude the logarithmic modes associated with static configurations.

These radiative boundary conditions define the asymptotic phase space on which the enlarged gauge symmetry acts. Just as the Poincar\'e symmetry of asymptotically flat gravity is enhanced
to the BMS symmetry at null infinity
\cite{Bondi:1962px,Sachs:1962wk,Strominger:2017zoo}, the global $\mathrm U(1)$
symmetry of electromagnetism is enlarged there to an
infinite-dimensional asymptotic symmetry generated by angle-dependent
large gauge transformations
\cite{He:2014cra,Kapec:2014zla,Kapec:2015ena}. A similar enhancement occurs in non-Abelian Yang--Mills theory
\cite{Strominger:2013lka,Mao:2017wvx}. Related infinite-dimensional electromagnetic symmetries have also been found
at spatial infinity within the standard Hamiltonian formulation, although their nontrivial realization requires twisted parity conditions and an
extension of the symplectic structure by surface degrees of freedom~\cite{Henneaux:2018gfi}. See also~\cite{Mukherjee:2023ihb} for a discussion of the relation
between Gauss's law and edge degrees of freedom in Maxwell
and $p$-form theories. In that construction, an additional charge appears together with the extra boundary degree of freedom. This raises the question of whether an analogous additional boundary observable can arise 
directly from the canonical structure of a null foliation.

To address this question, we apply the Hamiltonian formalism to radiative
systems, where it provides a systematic framework for identifying symmetries
and their charges. A null foliation adapted to wave propagation reveals a zero mode sector that is absent, or remains hidden, in the standard spacelike formulation. This sector suggests the existence of families of boundary configurations related by transformations acting only on the soft degrees of freedom, in analogy with the Goldstone interpretation of soft photon modes associated with asymptotic gauge symmetries~\cite{He:2014cra}. From this perspective, an additional shift symmetry intrinsic to the null boundary can arise, giving rise to a new edge observable, as
found for field theories on a Minkowski background~\cite{Gonzalez:2023yrz,Simic:2023exz,Gonzalez:2024rho}. The questions addressed
here are whether these additional edge observables survive on different
curved backgrounds and how the geometry and physical interpretation of the null boundary affect the resulting charge algebra.

\begin{figure}[h!]
    \centering
\includegraphics[width=0.85\textwidth]{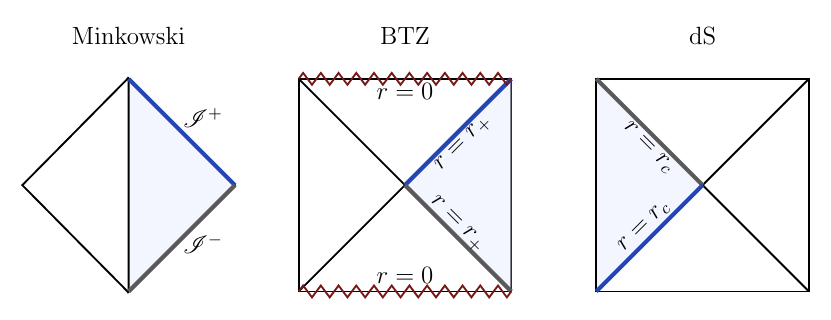}
    \caption{\it Penrose diagrams for the three spacetimes considered in this work: Minkowski space, the BTZ black hole, and de Sitter space. The shaded regions represent the coordinate patches used in the analysis, and the highlighted lines indicate the relevant null boundaries.}
    \label{Fig:three spacetimes}
\end{figure}

We consider three simple three-dimensional spacetimes that contain null hypersurfaces playing the role of effective boundaries:
Minkowski spacetime, the BTZ black hole, and de Sitter (dS) spacetime. In
Minkowski spacetime, the relevant boundary is future or past null infinity;
in the BTZ geometry, it is the black hole event horizon; and in the dS case, it is the cosmological horizon. The corresponding Penrose diagrams are shown in Fig.~\ref{Fig:three spacetimes}.

Despite these different interpretations, the same pair of edge observables
arises in all three cases, and their centrally extended algebra has a common
structure. These examples therefore show that the additional soft sector is
an intrinsic feature of the canonical formulation on null hypersurfaces. This property is generic to a broader class of systems in which zero modes give rise to additional asymptotic symmetries~\cite{Dordevic:2026dx}.

The zero mode sector also has a global aspect. Since charge conjugation is
a symmetry of pure Maxwell theory, the phase space admits configurations
twisted by charge conjugation around a defect or a black hole, depending on the background, in which the modes of the edge algebra are half-integer. In Minkowski spacetime we construct an exact radiative solution in this sector, with both charges nonvanishing.

\section{Edge modes in Minkowski spacetime} \label{Ch:Minkowski}

We are interested in the radiative sector of electromagnetic waves in flat
space. Therefore, the relevant asymptotic boundaries are future and past
null infinity, $\mathscr I^{\pm }$, obtained in the simultaneous limits $%
t\rightarrow \pm \infty $ and $r\rightarrow \infty $, while keeping $t\pm r$
constant. Outgoing radiation reaching $\mathscr I^{+}$, as well as incoming
radiation from $\mathscr I^{-}$, is naturally described in Bondi--Sachs
coordinates~\cite{Bondi:1962px,Sachs:1962wk,Bondi:1960jsa}, 
\begin{equation}
\diff s^{2}=g_{\mu \nu }\,\diff x^{\mu }\diff x^{\nu }=-\diff 
u^{2}-2\epsilon \,\diff u\,\diff r+r^{2}\diff \varphi ^{2}\,,
\label{metricBS}
\end{equation}
where $x^{\mu }=(u,r,\varphi )$ and
\begin{equation}
u=t-\epsilon r\in \mathbb{R}\,,\qquad \epsilon =\pm 1\,. 
\label{u}
\end{equation}
The choice $\epsilon =+1$ corresponds to retarded time and is adapted to
outgoing radiation near $\mathscr I^{+}$, while $\epsilon =-1$ corresponds
to advanced time and is adapted to incoming radiation near $\mathscr I^{-}$.
In both cases, the hypersurfaces $u=\mathrm{const}$ are null, and the asymptotic boundary is located at $r \to \infty$. The metric, its inverse, and the volume factor are 
\begin{equation}
g_{\mu \nu }=\left( 
\begin{array}{ccc}
-1 & -\epsilon  & 0 \\ 
-\epsilon  & 0 & 0 \\ 
0 & 0 & r^{2}%
\end{array}%
\right) ,\qquad g^{\mu \nu }=\left( 
\begin{array}{ccc}
0 & -\epsilon  & 0 \\ 
-\epsilon  & 1 & 0 \\ 
0 & 0 & \frac{1}{r^{2}}
\end{array}
\right) ,\qquad \sqrt{g}=r\,,  \label{metric}
\end{equation}
where $g=|\det g_{\mu \nu }|$.

This metric describes two independent coordinate patches with the same canonical structure for the Maxwell field. Their asymptotic data are related by matching conditions near spatial infinity $i^{0}$, with the relevant corners 
$\mathscr{I}_{-}^{+}$ defined by
$u|_{\epsilon=+1}\rightarrow-\infty$, and $\mathscr{I}_{+}^{-}$ defined by
$u|_{\epsilon=-1}\rightarrow+\infty$. Then antipodal matching identifies opposite
points on the circle, $\varphi\rightarrow\varphi+\pi$ (mod $2\pi$), so that any
asymptotic observable $f (u,\varphi)$ satisfies~\cite{Strominger:2017zoo,Kapec:2015ena,Campiglia:2015qka,Campiglia:2017dpg,Campiglia:2018see}  
\begin{equation}  
f(u,\varphi)\big|_{\mathscr{I}_{-}^{+}}
=
f(u,\varphi+\pi)\big|_{\mathscr{I}_{+}^{-}}\,.
\label{antipodal}
\end{equation}
This
leads to conservation laws compatible with Lorentz invariance in three~\cite{Prohazka:2017equ}, four~\cite{Strominger:2017zoo}, and higher dimensions 
\cite{Esmaeili:2019hom}.

In the following, we study the Hamiltonian evolution of the Maxwell field on this null foliation in a single coordinate patch, $\epsilon =1$ or $\epsilon =-1$. The second patch is described analogously, and the two sets of asymptotic data are related by the matching condition \eqref{antipodal}.

\subparagraph{Canonical momenta and Hamiltonian.}
Maxwell theory is described by the action 
\begin{equation}\label{lagrangiano}
    I_{\mathrm{EM}}
    =\int\limits \diff ^3x\,\mathcal{L}_{\mathrm{EM}}
    =-\frac{1}{4}\int\limits \diff ^3x\,\sqrt{g}\,F_{\mu\nu}F^{\mu\nu},
\end{equation}
where $g_{\mu\nu}(x)$ is the metric of the given background manifold and $A_\mu(x)$ is the electromagnetic gauge field. The corresponding field strength is $F_{\mu\nu}=\partial_\mu A_\nu-\partial_\nu A_\mu$.

The Hamiltonian analysis is performed on null hypersurfaces $\Xi $ of
constant $u$, parametrized by $x^{i}=(r,\varphi)$. The velocities $\dot{A}_{
{\mu }}=\partial _{{u}}A_{\mu }$ are defined with respect to the evolution vector field $\xi =\partial _{u}$~\cite{Nagarajan:1985xn}, which is transverse to the null hypersurfaces and timelike, $\xi \cdot \xi =g_{uu} =-1$.\footnote{This evolution
carries the fields from one null hypersurface $u=\mathrm{const}$ to another. This differs from the isolated horizon framework, where the canonical foliation is spacelike and the horizon is treated as an internal null boundary~\cite{Ashtekar:1999,Ashtekar:2000,Corichi:2025}. There, the evolution vector is
timelike in the exterior and becomes tangent and null at the horizon, so
that the evolution preserves the internal boundary.}

From the action \eqref{lagrangiano}, evaluated on the Minkowski background 
\eqref{metric}, the canonical momentum conjugate to the field $A_{\mu }$ is 
\begin{equation}
\pi ^{\mu }=\frac{\partial \mathcal{L}_{\mathrm{EM}}}{\partial \dot{A}_{\mu }
}=-\sqrt{g}\,F^{u\mu }\,.
\label{pi}
\end{equation}
Evaluating the components explicitly, the effect of the null foliation arises from $g^{uu}=0$. As a result, only the relation $\pi ^{r}$ can be inverted to obtain the corresponding velocity, 
\begin{equation}
\pi ^{r}=r ( \dot{A}_{r}-\partial _{r}A_{u} ) \,,
\label{pi.r}
\end{equation}
whereas the remaining two momenta give rise to primary constraints, 
\begin{equation}
\pi ^{u}\approx 0\,,\qquad \chi \equiv \pi ^{\varphi }-\frac{\epsilon }{r}\,F_{r\varphi }\approx 0\,.  \label{primary}
\end{equation}
The constraint $\pi ^{u}\approx 0$ is the standard primary constraint of
Maxwell theory, whereas $\chi \approx 0$ is characteristic of the null foliation because the kinetic term of a second order Lagrangian becomes linear in the relevant velocity in the null coordinates~\cite{Alexandrov:2014rba,Steinhardt:1979it,Majumdar:2022fut}.

The canonical momenta \eqref{pi} are naturally tensor densities. For a tensor density on $\Xi$ of weight $\Delta $, the covariant derivative $\nabla_{i}$ contains the additional term $-\Delta \,\Gamma _{ij}^{k}$ compared to the standard covariant derivative acting on tensors. In particular, since $\pi ^{i}$ is a vector density of weight $\Delta =1$, one obtains $\nabla _{i}\pi ^{i}=\partial _{i}\pi ^{i}$.
Then, the total Hamiltonian is 
\begin{equation}
H=\int\limits_{\Xi }\diff ^{2}x\,\left[\frac{1}{2r}\,(\pi ^{r})^{2}+\frac{r}{2}\,(\pi
^{\varphi })^{2}-A_{u}\,\nabla _{i}\pi ^{i}+\nabla _{i}(A_{u}\pi
^{i})+w_{u}\pi ^{u}+w\chi \right]\,,  \label{Ht}
\end{equation}
where the multipliers $w_{u}(x)$ and $w(x)$ are arbitrary spacetime functions. 
The canonical Poisson brackets are 
\begin{equation}
\{A_{\mu }(x),\pi ^{\nu }(x')\}_{u=u'}=\delta _{\mu }^{\nu
}\,\delta ^{(2)}(x-x')\,.
\end{equation}
From now on, in the brackets, we use the notation $f\equiv f(x)$ and $f'\equiv f(x')$, with $u=u'$. The two-dimensional Dirac delta $\delta ^{(2)}(x-x')=\delta (r-r')\delta (\varphi -\varphi
')$ is defined on the surface $\Xi $ with respect to the coordinate measure $\diff ^{2}x$, rather than the invariant measure. 

To obtain the evolution of a phase space function $f(x) = f(A(x),\pi(x),w_u(x),w(x))$
using the Poisson brackets, $\dot{f}(x)\approx \{ f(x),H\}$, the Hamiltonian must be differentiable. For now, we postpone this discussion and temporarily neglect all the boundary terms.
The bulk equations of motion are then 
\begin{equation}
\begin{array}[b]{llll}
\dot{A}_{u} & =w_{u}\,,\medskip  & \dot{\pi}^{u} & =\nabla _{i}\pi ^{i}\,,
\\ 
\dot{A}_{r} & =\dfrac{1}{r}\,\pi ^{r}+\partial _{r}A_{u}\,,\medskip  & 
\dot{\pi}^{r} & =\dfrac{\epsilon }{r}\,\partial _{\varphi }w\,, \\ 
\dot{A}_{\varphi } & =r\,\pi ^{\varphi }+\partial _{\varphi
}A_{u}+w\,,\quad  & \dot{\pi}^{\varphi } & =-\epsilon 
\,\partial _{r}\left( \dfrac{w}{r}\right) \,.
\end{array}
\label{dot}
\end{equation}
They are used to determine the boundary conditions for the
canonical variables.

\subparagraph{Radiative boundary conditions.}

We focus on the radiative sector of the Maxwell theory. Therefore, any nontrivial charge arising on the boundary has to be interpreted as a soft charge associated with the asymptotic gauge degrees of freedom.

A set of radiative boundary conditions was proposed by Shimizu and Sugishita in~\cite{Shimizu:2025hfl}. However, it does not allow for large gauge 
symmetries and leads to vanishing soft charges.
We therefore look for a weaker asymptotic behaviour of the
electromagnetic field in the radiative sector. We keep the behaviour $A_{\varphi }\sim \sqrt{r}$ and $\pi
^{r}\sim 1$/$\sqrt{r}$ proposed by Shimizu and Sugishita for the variables
that carry boundary degrees of freedom and, using Hamilton equations, we obtain 
\begin{equation}
\begin{array}[b]{llll}
A_{u} & =\sqrt{r}A_{(0)u}+\mathcal{O}(r^{-1/2})\,,\medskip  & \pi ^{u} & =%
\dfrac{\pi _{(2)}^{u}}{r\sqrt{r}}+\mathcal{O}(r^{-5/2})\,, \\ 
A_{r} & =\dfrac{A_{(1)r}}{\sqrt{r}}+\mathcal{O}(r^{-3/2})\,,\medskip  & \pi
^{r} & =\dfrac{\pi _{(1)}^{r}}{\sqrt{r}}+\mathcal{O}(r^{-3/2})\,, \\ 
A_{\varphi } & =\sqrt{r}A_{(0)\varphi }+\mathcal{O}(r^{-1/2})\,,\qquad
\medskip  & \pi ^{\varphi } & =\dfrac{\pi _{(2)}^{\varphi }}{r\sqrt{r}}+\mathcal{O}(r^{-5/2})\,, \\ 
w & =\sqrt{r}\,w_{(0)}+\mathcal{O}(r^{-1/2})\,, & w_{u} & =\sqrt{r}%
\,w_{(0)u}+\mathcal{O}(r^{-1/2})\,,
\end{array}
\label{rad}
\end{equation}
where $f_{(n)}(u,\varphi)$ denotes the coefficient multiplying the power $1/r^{n-1/2}$ in the expansion of an arbitrary function
$f(x)$ when $r\to \infty$. These relaxed boundary conditions are compatible with those of Ref.~\cite{Shimizu:2025hfl}  when $A_{(0)u}=0$. 

The leading order of Hamilton equations \eqref{dot} that satisfy the boundary conditions \eqref{rad} is given by
\begin{equation}
\begin{array}[b]{llll}
\dot{A}_{(0)u} & =w_{(0)u}\,,\medskip  & \dot{\pi}_{(2)}^{u} & =-\dfrac{1}{2}\,\pi _{(1)}^{r}+\partial _{\varphi }\pi _{(2)}^{\varphi }\,, \\ 
\dot{A}_{(1)r} & =\dfrac{1}{2}\,A_{(0)u}\,,\medskip  & \dot{\pi}_{(1)}^{r} & 
=\epsilon \,\partial _{\varphi }w_{(0)}\,, \\ 
\dot{A}_{(0)\varphi } & =\partial _{\varphi }A_{(0)u}+w_{(0)}\,,\quad  & 
\dot{\pi}_{(2)}^{\varphi } & =\dfrac{\epsilon }{2}\,w_{(0)}\,.
\end{array}
\label{dot(0)}
\end{equation}

On the other hand, null hypersurfaces do not form Cauchy surfaces for the full theory, since only massless excitations reach null infinity. Consequently, surface integrals $\oint_{\mathbb{S}^{1}}\diff \varphi \,f(u,\varphi )$
evaluated at fixed  $u$ at
null infinity can depend on the choice of the cut $u$. This ambiguity
can be removed by completing the cut with the corresponding segment of null
infinity up to its endpoint, such that the regularized boundary integral is defined as 
\begin{eqnarray}
\oint\limits_{\partial \Xi }\diff \varphi \,f &\equiv &\oint\limits_{\mathbb{S}^{1}}\diff \varphi \,f(u,\varphi )+\int\limits_{u}^{-\epsilon \infty
} \diff u'\oint\limits_{\mathbb{S}^{1}}\diff \varphi \,\partial' _{u}\,f(u',\varphi )  
=\oint\limits_{\mathbb{S}^{1}}\diff \varphi \,f(-\epsilon \infty ,\varphi )\,.\label{bndy-integral}
\end{eqnarray} The resulting functional is cut-independent, evaluated at the endpoint $u\rightarrow -\epsilon \infty $. Throughout the manuscript, we will suppress writing the limit in the argument of integrands. Further discussion can be found in~\cite{Donnelly:2016auv,Geiller:2017xad,Geiller:2017whh,
Hosseinzadeh:2018dkh,Speranza:2017gxd}.

With the falloff \eqref{rad} and the above prescription at hand, we can examine the differentiability of the Hamiltonian.

\subparagraph{Differentiability of the Hamiltonian.}
Let the smeared constraints be defined as 
\begin{equation}
\Pi [w_{u}]=\int\limits_{\Xi }\diff ^{2}x\,w_{u}\pi ^{u}\,,\qquad
S[w]=\int\limits_{\Xi }\diff ^{2}x\,w\chi \,,  \label{v.suav}
\end{equation}%
in the total Hamiltonian, which becomes 
\begin{equation}
H=H_{0}+\Pi [w_{u}]+S[w]\,.
\end{equation}%
This Hamiltonian is differentiable if the functional derivatives $
\delta H / \delta A_{\mu }$ and $\delta H / \delta \pi ^{\mu }$ are well-defined, that is, if no boundary terms remain in their derivation.
Denoting differentiable bulk terms by $\mathscr{D}$, the variation of the
total Hamiltonian gives 
\begin{equation}
\delta H=\mathscr{D}+(\delta H_{0})_{\partial \Xi }+(\delta S[w])_{\partial
\Xi }\,.
\end{equation}
Because $\varphi \in \mathbb{S}^{1}$ is periodic, the only boundary lies at $
r=\mathrm{const}\rightarrow \infty $, that is, $\mathbb{S}_{\infty }^{1}$. We then find 
\begin{eqnarray}
(\delta H_{0})_{\partial \Xi } &=&\int\limits_{\Xi }\diff ^{2}x\,\partial _{r}(\pi ^{r}\delta A_{u})=\oint\limits_{\mathbb{S}_{\infty}^{1}}\diff \varphi \,\pi ^{r}\delta A_{u}\,,  \notag \\
(\delta S[w])_{\partial \Xi } &=&-\int\limits_{\Xi }
\diff ^{2}x\,\partial _{r}\left( \frac{\epsilon}{r}w\,\delta A_{\varphi
}\right) =-
\oint\limits_{\mathbb{S}^{1}_\infty}\diff \varphi \,\frac{\epsilon}{r}\,w\,\delta A_{\varphi }\,.
\end{eqnarray}
Using the radiative boundary conditions \eqref{rad}, the integrands behave as
\begin{eqnarray}
\pi ^{r}\delta A_{u} &=&\pi _{(1)}^{r}\delta A_{(0)u}+\mathcal{O}(r^{-1})\,,
\notag \\
\frac{\epsilon }{r}\,w\,\delta A_{\varphi } &=&\epsilon
\,w_{(0)}\delta A_{(0)\varphi }+\mathcal{O}(r^{-1})\,.
\end{eqnarray}%
Therefore, if $A_{u}$ is fixed at the boundary, 
\begin{equation}
\delta A_{u}\big|_{\partial \Xi }=0\qquad \Rightarrow \qquad \delta A_{(0)u}=0\,,
\label{Au fijo}
\end{equation}
we obtain
\begin{eqnarray}
(\delta H_{0})_{\partial \Xi } = 0\,, \qquad
(\delta S[w])_{\partial \Xi } =-\epsilon\oint\limits_{\mathbb{S}_{\infty}^{1}}\diff \varphi \,w_{(0)}\delta A_{(0)\varphi }\,.
\label{diffH}
\end{eqnarray}
Thus, the canonical Hamiltonian $H_{0}$ is differentiable under the boundary
condition \eqref{Au fijo}, but the total Hamiltonian $H$ is not, because the
primary constraint $S[w]$ contributes  a nonvanishing boundary term. One possibility to deal with this is to impose conditions on $w_{(0)}$ or $A_{(0)\varphi }$ such that the boundary term vanishes, but it would break asymptotic symmetries. Another
possibility, which we will adopt here, is to add the boundary term 
\begin{equation}\label{Q}
Q_{S}[w]=\oint\limits_{\mathbb{S}^1_\infty}
\diff \varphi \,\frac{\epsilon}{r}\,w A_{\varphi }\,,
\end{equation}
where the multiplier $w(x)$ is field-independent. The improved Hamiltonian is then 
\begin{equation}
H_{Q}=H+Q_{S}[w]\,.
\end{equation}
With this improvement, Hamilton equations \eqref{dot} remain valid when the evolution is computed with the improved Hamiltonian, 
$f (x)\approx \{ f(x),H_Q\}$.

\subparagraph{Consistency of constraints and phase space.} 

Time evolution must preserve the constraint surface. This implies  that the primary constraint $\pi ^{u}\approx 0$ leads to the secondary constraint 
\begin{equation}
\dot{\pi}^{u}=0\quad \Rightarrow \quad \psi =\nabla _{i}\pi ^{i}\approx 0\,,
\label{secondary}
\end{equation}
while the consistency condition $\dot{\psi}=0$ is identically satisfied.

On the other hand, the consistency of $\chi \approx 0$ leads to a linear
differential equation for the multiplier $w$, 
\begin{equation}
\dot{\chi}=\left\{ \chi ,H_{0}\right\} -\frac{2\epsilon }{\sqrt{r}}
\,\partial _{r}\left( \frac{w}{\sqrt{r}}\right) =0\,.
\end{equation}%
Its general solution can be written as 
\begin{equation}
w(x)=\sqrt{r}\,W(u,\varphi )+\bar{w}(x)\,,  \label{W}
\end{equation}%
where the first term is the undetermined part of the multiplier, up to its radial
dependence. The second term, $\bar{w}(x)$, is the determined part and is given by 
\begin{eqnarray}
\bar{w}(x)
=\frac{\sqrt{r}}{2}\int\limits^{r}\frac{\diff r'}{\sqrt{r'}}\left[ -\partial _{r}\left( r\pi^{\varphi
}\right) +\frac{1}{r}\,\partial _{\varphi }\pi^{r}\right]_{r=r'} \,.
\label{bar w}
\end{eqnarray}
Using the asymptotic conditions \eqref{rad}, one finds $\bar{w}(x)=\mathcal{O}(r^{-1/2})$,
which is consistent with its expected behaviour.

When studying the differentiability of the Hamiltonian, we treated the multiplier $w(x)$ as field-independent in passing from \eqref{diffH} to
\eqref{Q}. This is in agreement with the decomposition \eqref{W}. Indeed, the field-dependent part $\bar w$ falls off sufficiently rapidly and therefore does not contribute to the
boundary term at $\partial\Xi$, that is, $Q_S[\bar w]=0$. The only contribution comes from the
field-independent part $\sqrt r\,W$. Hence, the improved Hamiltonian includes only the boundary term  $Q_S[\sqrt r\,W]$.

So far, we have discussed only the asymptotic region, whose conditions do not imply a regular extension of the fields to the origin. It is natural to ask whether these asymptotic conditions admit a regular
continuation to the origin. We address this question without necessarily imposing conditions at $r=0$, as we discuss later in detail. We analyze, in Minkowski space, the simplest sector, in which the homogeneous term $\sqrt{r}\,W$ remains the leading contribution to $w$. Let us denote the quantities independent of $r$ with a tilde. Then near the origin, $w\sim 
\sqrt{r}\,\tilde{w}$,  and Hamilton's equations allow the falloffs 
\begin{equation}
\begin{array}{ll}
A_{u}\sim \dfrac{\tilde{A}_{u}}{\sqrt{r}}\,,\medskip  & \pi ^{u}\sim \dfrac{%
\tilde{\pi}^{u}}{r\sqrt{r}}\,, \\ 
A_{r}\sim r\sqrt{r}\,\tilde{A}_{r}\,,\medskip  & \pi ^{r}\sim \dfrac{\tilde{%
\pi}^{r}}{\sqrt{r}}\,, \\[6pt]
A_{\varphi }\sim \sqrt{r}\,\tilde{A}_{\varphi }\,,\qquad  & \pi ^{\varphi
}\sim \dfrac{\tilde{\pi}^{\varphi }}{r\sqrt{r}}\,,%
\end{array}
\label{r=0Mink}
\end{equation}%
with $w_{u}\sim \tilde{w}_{u}/\sqrt{r}$, and no logarithmic subleading
terms. Substituting the momenta into the radial integral that
defines $\bar{w}$ in \eqref{bar w}, we find 
\begin{equation}
w\sim -\frac{2\partial _{\varphi }\tilde{\pi}^{r}+\tilde{\pi}^{\varphi }}{4\sqrt{r}}+\sqrt{r}\,W\,.
\end{equation}
The behaviour $w\sim \sqrt{r}\,W$ requires the first term to vanish, and the
equation of motion for $A_{r}$ gives the second condition, 
\begin{equation}
\tilde{\pi}^{\varphi }=-2\partial _{\varphi }\tilde{\pi}^{r}\,,\qquad \tilde{%
A}_{u}=2\tilde{\pi}^{r}\,.  \label{cancell}
\end{equation}%
At leading order, Hamilton's equations then become 
\begin{equation}
\begin{array}{ll}
\dot{\tilde{A}}_{\varphi }=W\,,\medskip  & \dot{\tilde{\pi}}^{u}=-\dfrac{1}{2%
}\,\tilde{\pi}^{r}+\partial _{\varphi }\tilde{\pi}^{\varphi }\,, \\[6pt]
\dot{\tilde{\pi}}^{r}=\epsilon \,\partial _{\varphi }W\,,\medskip \qquad  & 
\dot{\tilde{\pi}}^{\varphi }=\dfrac{\epsilon }{2}\,W\,, \\[6pt]
\dot{\tilde{A}}_{u}=\tilde{w}_{u}\,, & \tilde{w}_{u}=2\epsilon \,\partial
_{\varphi }W\,.%
\end{array}
\label{tilde.dot}
\end{equation}

The powers in \eqref{r=0Mink} describe one evolving sector, not the most
general near-origin behaviour. Since the equations are linear, stationary
data that solve the stationary equations and respect the global conditions
may be superposed, such as residual gauge terms $A_{i}=-\partial _{i}\Lambda
(r,\varphi )$, holonomies and Coulomb-type fields. They do not enter the
time derivatives, and we do not consider them here.

Requiring \eqref{cancell} to be preserved in time gives 
\begin{equation}
\left( \partial _{\varphi }^{2}+\frac{1}{4}\right) W=0\,,\qquad \tilde{w}%
_{u}=2\epsilon \,\partial _{\varphi }W\,,
\end{equation}
so that, in this sector, $W$ is constrained to
\begin{eqnarray}
W &=&W_{1/2}(u)\,\mathrm{e}^{\frac{\mathrm{i}}{2}\,\varphi }+W_{-1/2}(u)\,%
\mathrm{e}^{-\frac{\mathrm{i}}{2}\,\varphi }\,,  \notag \\
\tilde{w}_{u} &=&\mathrm{i}\epsilon \left( W_{1/2}(u)\,\mathrm{e}^{\frac{%
\mathrm{i}}{2}\,\varphi }-W_{-1/2}(u)\,\mathrm{e}^{-\frac{\mathrm{i}}{2}%
\,\varphi }\right) \,,  \label{excess1}
\end{eqnarray}%
with $W_{-1/2}=W_{1/2}^*$ for real $W$. In this sector the
constraints reduce to 
\begin{equation}
\tilde{\pi}^{u}\approx 0\,,\qquad -\frac{1}{2}\,\tilde{\pi}^{r}+\partial
_{\varphi }\tilde{\pi}^{\varphi }\approx 0\,,\qquad \tilde{\pi}^{\varphi }-%
\frac{\epsilon }{2}\,\tilde{A}_{\varphi }\approx 0\,,
\end{equation}%
and they are preserved by \eqref{tilde.dot}. Combining Gauss's law with %
\eqref{cancell} yields 
\begin{equation}
\left( \partial _{\varphi }^{2}+\frac{1}{4}\right) \tilde{\pi}^{r}=0\quad
\Rightarrow \quad \tilde{\pi}^{r}=\tilde{\pi}_{1/2}^{r}(u)\,\mathrm{e}^{%
\frac{\mathrm{i}}{2}\,\varphi }+\tilde{\pi}_{-1/2}^{r}(u)\,\mathrm{e}^{-%
\frac{\mathrm{i}}{2}\,\varphi }\,.  \label{excess2}
\end{equation}%
The remaining coefficients follow from $\tilde{\pi}^{\varphi }=-2\partial
_{\varphi }\tilde{\pi}^{r}$, $\tilde{A}_{\varphi }=2\epsilon \tilde{\pi}%
^{\varphi }$ and $\dot{\tilde{A}}_{\varphi }=W$, consistently with %
\eqref{tilde.dot}.

Every configuration of this sector with $\tilde{\pi}^{r}\neq 0$ is therefore
singular at $r=0$. The electromagnetic invariant behaves as 
\begin{equation}
F_{\mu \nu }F^{\mu \nu }=-\frac{2}{r^{3}}\,\left[ (\tilde{\pi}^{r})^{2}+(%
\tilde{\pi}^{\varphi })^{2}\right] +\mathcal{O}(r^{-2})\,.
\end{equation}%
Conversely, regularity at the origin requires $\tilde{\pi}^{r}=0$, and then %
\eqref{tilde.dot} and \eqref{excess1} imply $W=0$. On Minkowski space with
regular fields, this sector is therefore absent.

The angular dependence $\mathrm{e}^{\pm \frac{\mathrm{i}}{2}\varphi }$ in 
\eqref{excess1} and \eqref{excess2} shows that the fields are not
single-valued under the standard identification $\varphi \simeq \varphi +2\pi $, but they change sign. This is consistent, because charge conjugation $C:A_{\mu }\rightarrow -A_{\mu }$ is a symmetry of pure Maxwell theory. The
theory therefore admits $C$-twisted configurations, $\Phi^s(\varphi +2\pi
)=-\Phi^s(\varphi )$, with the twist around the origin~\cite{Shao:2025qvf}, where $\Phi^s \in \{ A_\mu, \pi^\mu, w, w_u, \ldots   \}$ are canonical fields, multipliers and later transformation parameters. All $C$-invariant
quantities, such as $F_{\mu \nu }F^{\mu \nu }$ and the energy density,
remain single-valued. The background metric is regular everywhere, including
at $r=0$. The topological defect lies in the space of field configurations,
not in the geometry.\footnote{%
There is a physically inequivalent completion with the defect in the geometry: flat space with a conical excess at $r=0$ (total angle $4\pi $),
on which all fields are $4\pi $-periodic. We do not consider it here, in order to keep a regular Minkowski background.}

The above analysis characterises one sector of the twisted configurations and shows that it is incompatible with regular fields at the origin. Imposing \eqref{r=0Mink} restricts $W$ to the two modes \eqref{excess1}, whereas other configurations with $W\neq0$ could behave
differently near the origin. We therefore do not impose conditions at $r=0$ in the definition of the phase space. This is natural rather than a limitation, as we explain below.\medskip

We restrict the canonical description to the punctured region $r>0$. The canonical variables and the observables built from them are then quasilocal: they refer to the region outside the origin, and the charges
discussed later are surface integrals over circles $\mathbb{S}_{r}^{1}$
enclosing it, evaluated as $r\rightarrow \infty $. For regular configurations this is no restriction because the Bondi chart covers only $r>0$, and regular fields extend continuously to the origin. For the twisted sector the puncture is necessary, because a field that changes sign along every loop
around the origin cannot be continued to $r=0$, where these loops become
contractible. In addition, the energy of the system in the region $r>r_{0}$ diverges as $r_{0}\rightarrow 0$ (as $1/r_{0}$ for the exact solution of Section~\ref{Ch:Exact}), like the self-energy of a point charge in four dimensions.\medskip

Explicitly, the phase space $\Gamma _{\infty }$ consists of canonical fields 
$(A_{\mu },\pi ^{\mu })$ on $\Xi \setminus \{0\}$ such that:\medskip 

(\textit{i})~they satisfy the asymptotic conditions \eqref{rad} as $r\rightarrow \infty $; 

(\textit{ii})~they do not have boundary conditions imposed as $r\to0$ beyond finiteness of the symplectic form; regular configurations and the sector \eqref{r=0Mink} are particular cases;

(\textit{iii})~all
fields, multipliers and parameters obey $\Phi^s (\varphi +2\pi )=\sigma\,\Phi^s (\varphi )$ with the same $\sigma=\pm
1 $. The zero mode sector, with $W\neq 0$ and the behaviour \eqref{r=0Mink},
has $\sigma=-1$. The case $\sigma=+1$ with regular fields is ordinary Minkowski
electrodynamics. Admissible variations preserve (\textit{i})--(\textit{iii}). Since $\sigma$ is discrete, two sectors are not connected by a variation.\medskip

The additional zero mode sector therefore requires the punctured radial
domain -- it exists in $\Gamma _{\infty }$, but not in the phase space of
regular fields on the full Minkowski space.    On the phase space of regular fields, $Q_S$ is still a well-defined
   function, but the flow it generates, with the fixed radial profile
   $\sqrt{r}\,\eta_{(0)}$, does not preserve regularity at the origin.
This is a physical feature of the model. The puncture is the worldline of a $C$-twist defect~\cite{Shao:2025qvf}, the endpoint of a
topological cut whose position is unobservable~\cite{Gaiotto:2014kfa}. The holonomy
around it is a discrete topological number, which defines a superselection
sector distinct from ordinary electrodynamics and is detected at null
infinity by the antiperiodicity of $W$. As for a point charge or an Aharonov--Bohm flux tube, the defect worldline is excised from the domain,
while its effect is measured by the fields at a distance. In the exact solution of Section~\ref{Ch:Exact}, the defect emits ($\epsilon=+1$) or absorbs ($\epsilon=-1$) the radiation described by $W$.

\subparagraph{Symmetry generators.} 

Within this restricted phase space, we now turn to the canonical generators associated with the constraints. The full set of constraints is $\{ \pi ^{u}, \psi,\chi\}$, and not all of them generate gauge symmetries.  For instance, $\pi ^{u}$ and $\psi $  are first class. The associated smeared generator is 
\begin{equation}
G[\varepsilon ]=\int\limits_\Xi \diff ^{2}x\,\left( \varepsilon \psi -\dot{\varepsilon}\pi ^{u}\right) \,,
\label{G.U(1)}
\end{equation}
where $\varepsilon (x)$ is a local parameter and the generator is constructed using Castellani's method~\cite{Castellani:1981us}.

The constraint $\chi$ requires a more careful analysis. Its Poisson bracket
with itself is nontrivial,
\begin{equation}
\{\chi,\chi'\} = \Omega(x,x') \equiv \frac{\epsilon}{r^{2}}\,\delta^{(2)} - \frac{2\epsilon}{r}\,\partial_{r}\delta^{(2)}
\,.  \label{Omega}
\end{equation}
The integral operator $\Omega(x,x')$ has zero modes:
\begin{equation}
\int \diff ^{2}x'\, \Omega(x,x')v(x') = -\frac{
2\epsilon}{\sqrt{r}}\, \partial_{r}\left(\frac{v(x)}{\sqrt{r}}\right)
=0 \quad \Rightarrow \quad v(x)=\sqrt{r}\,V(u,\varphi)\,.  \label{0}
\end{equation}
This shows that $\chi$ contains both first and second class sectors.
It may nevertheless be treated as entirely second class when its zero modes are excluded by the boundary conditions~\cite{Nagarajan:1985xn,Majumdar:2022fut,Goldberg:1991pb}.
In our approach, however, we do not exclude these modes. We use them to construct the first class generator by smearing $\chi$ with a particular choice of a parameter, 
\begin{equation}
S[\eta] = \int\limits_\Xi  \diff ^{2}x\,\eta(x)\chi(x)\,, \qquad \eta(x)=\sqrt{r}\,\eta_{(0)}(\varphi)\,.
\label{S}
\end{equation}
The radial dependence of $\eta $ in \eqref{S} is not truncated but fixed. This has several consequences. First, there are no independent subleading parameters. Then, the transformation cannot be localised near the boundary, since a radial cutoff $a(r)$ takes the smeared generator out of the zero mode sector, $\int 
\mathrm{d}^{2}x'\,\Omega (x,x')\,a(x')\,\eta(x') =-\frac{2\epsilon 
}{r}\, \eta\,\partial _{r}a$. Finally, the angular function $\eta _{(0)}(\varphi )$ remains arbitrary, so the family of parameters is infinite-dimensional.
The function $\eta_{(0)}$ is taken to be independent of $u$ since its $u$-dependent modes correspond to degenerate directions of the presymplectic form and are quotiented out of the covariant phase space~\cite{Lee:1990nz}. Note that the generator $S[\eta]$ has the same functional form as the
smeared constraint $S[w]$ defined in Eq.~\eqref{v.suav}. However, while
$S[w]$ contains both the first and second class sectors of $\chi$, because $w(x)$ is a generic multiplier, $S[\eta]$ selects only the first class sector.  Indeed, with this choice, $S[\eta]$ is first class by construction, because $\eta$ is a zero mode of $\Omega$~\cite{Gonzalez:2024rho}. Explicitly, 
\begin{equation}
\{S[\eta_{1}],S[\eta_{2}]\} = \int\limits_\Xi  \diff ^{2}x \int\limits_\Xi  \diff ^{2}x'\, \eta_{1}(x)\Omega(x,x')\eta_{2}(x') =0\,.
\end{equation}

The zero modes of $\Omega$ have a simple geometric meaning. The condition $%
\partial_r(\eta/\sqrt{r})=0$ states that $\eta/\sqrt{r}$ is constant along
each null generator of $\Xi$, so the intrinsic parameter $\eta_{(0)}(\varphi)
$ is a function on the space of generators, and $\sqrt{r}$ is the radiative
weight fixed by the characteristic constraint. Any other radial profile
would include the second class part of $\chi$. This is the light-front
zero mode problem~\cite{Maskawa:1975hx,Heinzl:1998kz}: the characteristic
constraint determines the fields along each generator only up to one
function per generator, which is not fixed by the evolution on $\Xi$ and has
to be supplied at an end of the generators. The undetermined part $\sqrt{r}
\,W$ of the multiplier in \eqref{W} is a zero mode of the same kind. The
canonical description on a null hypersurface is therefore naturally
quasilocal, as for phase spaces defined on finite segments of null
hypersurfaces~\cite{Chandrasekaran:2018aop}: the zero mode data are specified at the null boundary where the charges are evaluated, and no
condition is imposed at the other end of the generators, since that would over-determine them. This is why the phase space $\Gamma_\infty$ was defined without conditions
at $r=0$. In
particular, as found above, requiring regularity at the origin removes the
zero mode sector.
\medskip

The constraints are therefore classified as 
\begin{equation*}
\begin{array}{ll}
\textit{First class constraints:} & \pi^{u}\,,\;\psi\,,\;\chi|_{\text{zero
modes of }\Omega}\,, \\
\textit{Second class constraints:} & \chi|_{\text{nonzero modes of }\Omega}\,.
\end{array}
\end{equation*}


\subparagraph{Quasilocal charges.}


The boundary charges are obtained, as usual, by improving the first class constraints with surface terms~\cite{Regge:1974zd}. On the punctured domain $\Xi\setminus\{0\}$, the radial integration by parts in the variation of the smeared constraints has two ends, at $\mathbb{S}^1_{r\to\infty}$ and at $\mathbb{S}^1_{r\to0}$. The improved generators therefore carry two surface terms, $Q_{r\to\infty}-Q_{r\to0}$, each cancelling the boundary contribution of the variation at its own end. For the electric generator, the parameter $\varepsilon$ can be chosen to vanish near the origin, so that $Q_{r\to0}[\varepsilon]=0$. For the magnetic generator, the parameter $\eta=
\sqrt{r}\,\eta_{(0)}$ cannot always be globally non-vanishing. Its inner term vanishes for regular configurations, but in the twisted sector it is finite and does not need to vanish. The two surface terms are local functionals at disjoint ends of the radial domain, so they are defined independently and their Poisson brackets do not mix. The algebra of the improved generators therefore splits into an
asymptotic part and a defect part, each with its own possible central
extension.

In this work we study only the asymptotic part, which belongs to the phase
space $\Gamma _{\infty }$: the quasilocal charges $Q[\varepsilon ]\equiv
Q_{r\rightarrow \infty }[\varepsilon ]$ and $Q_{S}[\eta ]\equiv
Q_{S,r\rightarrow \infty }[\eta ]$, which are functionals of the edge
variables on the asymptotic null boundary. In the expressions below we
therefore analyze only the boundary terms at $\mathbb{S}_{\infty }^{1}$. We
do not analyze the inner contribution, which is associated with the defect
worldline rather than with a null boundary, and whose description would
require a model of the defect.

The generator \eqref{G.U(1)} is not differentiable, since it contains a boundary term in its variation,\footnote{  Integration by parts in $\varphi$ produces terms $\oint \mathrm d\varphi\,
\partial_\varphi(\varepsilon\,\delta\pi^\varphi)$. In the twisted sector
   both factors are antiperiodic, so the integrand is periodic and the term
   vanishes. Integrals of antiperiodic total derivatives would depend on the
   position $\varphi_0$ of the cut and are not observables -- they do not
   arise for the physical charges.}
\begin{equation}
\delta G[\varepsilon ] = -\int\limits_{\Xi }\diff ^{2}x\,\left( \partial
_{i}\varepsilon \,\delta \pi ^{i}-\dot{\varepsilon}\,\delta \pi ^{u}\right)
+\oint\limits_{\mathbb{S}_{\infty }^{1}}\diff \varphi \,\varepsilon
\,\delta \pi ^{r}\,.
\label{varG}
\end{equation}
The improved generator is obtained by adding a boundary contribution, the charge $Q$~\cite{Regge:1974zd}, 
\begin{equation}
G_{Q}[\varepsilon ]=G[\varepsilon ]+Q[\varepsilon ]\,.  \label{G}
\end{equation}%
If the parameter $\varepsilon$ is field-independent, $\delta \varepsilon=0$,
the charge is determined by 
\begin{equation}
\delta Q[\varepsilon ]=-\oint\limits_{\mathbb{S}_{\infty }^{1}}\diff %
\varphi \,\varepsilon \,\delta \pi ^{r}\quad \Rightarrow \quad Q[\varepsilon
]=-\oint\limits_{\mathbb{S}_{\infty }^{1}}\diff \varphi \,\varepsilon
\pi ^{r}.  \label{Qel}
\end{equation}
The resulting quantity is a quasilocal observable defined by the asymptotic data supported on $\mathbb{S}^1_\infty$. We refer to $Q[\varepsilon ]$ as the electric charge, since
it measures the electric flux $\pi ^{r}$ through $\mathbb{S}_{\infty }^{1}$.

In order for $Q[\varepsilon ]$ to remain finite and nonvanishing, the symmetry parameter has to behave asymptotically as 
\begin{equation}
\varepsilon (x)=\sqrt{r}\,\varepsilon _{(0)}(\varphi )+\mathcal{O}%
(r^{-1/2})\,.  \label{E asint}
\end{equation}
The leading term $\varepsilon_{(0)}$ has to be independent of $u$ to ensure that the boundary condition \eqref{Au fijo} is invariant under the corresponding symmetry transformations. 

The global charge corresponds to $\varepsilon _{(0)}=1$. Using the leading
Gauss constraint \eqref{secondary} in the asymptotic expansion \eqref{rad} to solve the leading momentum coefficient $\pi _{(1)}^{r}$,
\begin{equation}
\psi _{(2)}=-\frac{1}{2}\,\pi _{(1)}^{r}+\partial _{\varphi }\pi
_{(2)}^{\varphi }\approx 0\,,
\end{equation}
on the constraint surface we obtain
\begin{equation}
Q_{0}=Q[\varepsilon _{(0)}=1]=-\oint \diff \varphi \,\pi
_{(1)}^{r}=-2\oint \diff \varphi \,\partial _{\varphi }\pi
_{(2)}^{\varphi }=0\,.
\label{Q0=0}
\end{equation}
The last equality follows from angular periodicity. Thus, the global electric charge vanishes on the chosen radiative phase space.

The second soft charge requires a slightly different treatment. The
corresponding symmetry parameter has the exact radial dependence given by \eqref{S}, with no subleading terms, $\eta _{(n)}\equiv 0$ for $n\geq 1$, when $r \to \infty$. This means that the Dirac--Bergmann procedure has to be applied with care, as it is usually formulated for local symmetry parameters.  The consequences of this fixed radial profile are discussed after~\eqref{S},
   and its effect on the Hamilton equations after~\eqref{lead-var}.

The smeared generator $S[\eta ]$ is not differentiable. Its variation is 
\begin{equation}
\delta S[\eta ]=\int\limits_{\Xi }\diff ^{2}x\left[ \eta \,\delta \pi
^{\varphi }+\epsilon\,\partial _{r}\left( \frac{\eta }{r}
\right) \delta A_{\varphi }-\frac{\epsilon }{r}\partial _{\varphi }{\eta }\,\delta A_{r}\right] -\oint\limits_{\mathbb{S}^1_\infty} \diff \varphi \,\frac{\epsilon }{r}\,\eta \delta A_{\varphi }\,.
\label{varS}
\end{equation}
Thus, it also has to be improved by a boundary term, which defines the
new quasilocal charge, 
\begin{equation}
\delta Q_{S}[\eta ]=\oint\limits_{\mathbb{S}^1_\infty} \diff \varphi \,\frac{\epsilon }{r}\,\eta \delta A_{\varphi } \quad \Rightarrow \quad
Q_{S}[\eta ]=\oint\limits_{\mathbb{S}^1_\infty} \diff \varphi \,\frac{\epsilon }{r}\,\eta A_{\varphi }\,,
\label{Qshift'}
\end{equation}
where $\eta $ is assumed to be field-independent. The improved generator is therefore 
\begin{equation}
G_{S}[\eta ]=S[\eta ]+Q_{S}[\eta ]\,,  \label{G_S}
\end{equation}
and is differentiable. Taking into account the asymptotic form of $\eta$ given by \eqref{S} and the behaviour of $A_{\varphi }$ in the radiative sector \eqref{rad}, the soft charge is finite at the boundary, 
\begin{equation}
Q_{S}[\eta ]=\epsilon\oint\limits_{\mathbb{S}_{\infty }^{1}}
\diff \varphi \,\eta _{(0)}A_{(0)\varphi }\,.  \label{Qshift}
\end{equation}
We refer to $Q_{S}[\eta ]$ as the shift charge or the magnetic charge. Indeed,
in the untwisted sector, for $\eta _{(0)}=\mathrm{const}$, $Q_{S}$ is the leading coefficient of the holonomy integral at infinity, which by Stokes' theorem is the magnetic flux through the
region enclosed by $\mathbb{S}^1_\infty$.

Note that, including the inner charge at $r \to 0$,  the two magnetic terms may coincide on shell, as for the exact solution of Section~\ref{Ch:Exact}. Then $Q_{S}$ measures from infinity the data carried by the defect, just as the flux of a Coulomb field at infinity measures the charge of its
source.

The boundary term \eqref{Q} that improves the Hamiltonian has the same functional form as the above soft charge evaluated on the zero mode part of the multiplier. This is not a coincidence because the total Hamiltonian contains the constraint $\chi$, and its differentiability
requires the same boundary improvement as the generator $G_S[\eta]$.
There is no analogous contribution from the electric charge
$Q[\varepsilon]$ in the improved total Hamiltonian, because the constraint
$\psi$ appears with coefficient $A_u$, and the imposed Dirichlet
condition on $A_u$ cancels the corresponding boundary variation. The
charge $Q[\varepsilon]$ is therefore needed for the differentiability of
the gauge generator, but not of the total Hamiltonian. In the extended
Hamiltonian formalism, where $\psi$ is accompanied by an independent
multiplier, the corresponding boundary improvement would also be required, and
the improved extended Hamiltonian would then contain both $Q$ and $Q_S$.

It is worth emphasizing that the resulting nonvanishing functional derivatives can be obtained from the bulk terms in \eqref{varG} and \eqref{varS},
\begin{equation}
\begin{array}[b]{ll}
\dfrac{\delta G_{Q}[\varepsilon ]}{\delta \pi ^{\mu }}=-\partial _{\mu
}\varepsilon \,,\medskip \qquad  & \dfrac{\delta G_{S}[\eta ]}{\delta
A_{\varphi }}=\epsilon\,\partial _{r}\left( \dfrac{\eta }{r}
\right) , \\ 
\dfrac{\delta G_{S}[\eta ]}{\delta \pi ^{\varphi }}=\eta \,, & \dfrac{\delta
G_{S}[\eta ]}{\delta A_{r}}=-\dfrac{\epsilon }{r}\,\partial _{\varphi
}\eta \,.
\end{array}
\label{varGq}
\end{equation}
All other functional derivatives vanish.


\subparagraph{Symmetry transformations.}


The symmetry transformations generated by the two improved generators are 
\begin{equation}
\delta _{\varepsilon }=\{f,G_{Q}[\varepsilon ]\}\,,\qquad \delta _{\eta
}f=\{f,G_{S}[\eta ]\}\,,  \label{local}
\end{equation}
for any phase space function $f(x)\equiv f(x,A(x),\pi (x),w_{u}(x),w(x))$ on $\Gamma_\infty$.
To compute the transformations and the Poisson brackets of the
generators, one uses the functional derivatives \eqref{varGq}. 
In particular, the transformations generated by $G_{Q}[\varepsilon ]$ are
the usual $\mathrm U(1)$ gauge transformations, 
\begin{equation}
\delta _{\varepsilon }A_{\mu }=-\partial _{\mu }\varepsilon \,.
\label{trU(1)}
\end{equation}
Since $\dot{\varepsilon}_{(0)}=0$ in \eqref{E asint}, one has $\delta_{\varepsilon }A_{(0)u}=0$, consistently with the Dirichlet condition \eqref{Au fijo}. Invariance of Hamilton equations \eqref{dot} also implies 
\begin{equation}
\delta _{\varepsilon }w_{u}=-\ddot{\varepsilon}\,,\qquad \delta
_{\varepsilon }w=0\,.
\end{equation}
The transformations \eqref{trU(1)} preserve the asymptotic conditions \eqref{rad}. 

The generator \eqref{G_S}, associated with the asymptotic charge $Q_S$, produces nontrivial asymptotic transformations 
\begin{equation}
\delta _{\eta }A_{\varphi }=\eta \,,\qquad \delta _{\eta }\pi ^{r}=\dfrac{\epsilon }{r}\,\partial _{\varphi }\eta \,,\qquad \delta _{\eta }\pi
^{\varphi }=-\epsilon \,\partial _{r}\left( \dfrac{\eta }{r}\right) .
\label{trShift}
\end{equation} 
We supplement these transformations with a shift of the pure-gauge variable $%
A_u$, 
\begin{equation}
\delta_\eta A_u=\frac{2\epsilon}{r}\,\partial_\varphi\eta\,,  \label{dAu}
\end{equation}
generated by the first class constraint $\pi^u$, i.e., by $\Pi[\mu]$ given by \eqref{v.suav} with $\mu=\frac{2\epsilon}{r}\, \partial_\varphi\eta$. This shift has no boundary term and commutes with $G_Q$ and $G_S$, so it does not change the charges or their algebra. It is subleading at null infinity, so the Dirichlet condition \eqref{Au fijo} is preserved. The multipliers transform as
\begin{equation}
\delta _{\eta }w_{u}=0\,,\qquad \delta_\eta W=0\,,\qquad \delta _{\eta }w=\delta _{\eta }\bar{w}=-\frac{\epsilon }{2\sqrt{r}}\left( \partial
_{\varphi }^{2}+\frac{1}{4}\right) \eta _{(0)} \,,
\label{trShift2}
\end{equation}
where $\delta _{\eta }\bar{w}$ follows from \eqref{bar w}.
The boundary conditions are preserved by \eqref{trShift}--\eqref{trShift2}. The
corresponding nontrivial transformation laws for the asymptotic coefficients of the canonical fields and the multipliers are 
\begin{equation}\label{eta0}
\delta _{\eta }A_{(0)\varphi }=\eta _{(0)}\,,\qquad \delta _{\eta }\pi
_{(1)}^{r}=\epsilon \,\partial _{\varphi }\eta _{(0)}\,,\qquad \delta _{\eta
}\pi _{(2)}^{\varphi }=\frac{\epsilon }{2}\,\eta _{(0)}\,,
\end{equation}
while $A_{(0)u}$ and $w_{(0)}=W$ are invariant. 
Under these transformations, the Hamilton equations \eqref{dot} for $A_{u}$, $A_{r}$ and $\pi ^{u}$ are preserved exactly, while those for $A_{\varphi }$, $\pi ^{r}$ and $\pi ^{\varphi }$ change by 
\begin{equation}
\frac{3\epsilon }{2\sqrt{r}}\,\left( \partial _{\varphi }^{2}+\frac 14
\right) \eta _{(0)}\,,\quad -\frac{1}{2r\sqrt{r}}\,\partial _{\varphi
}\left( \partial _{\varphi }^{2}+\frac 14\right) \eta _{(0)}\,,\quad -
\frac{3}{4r^{2}\sqrt{r}}\left( \partial _{\varphi }^{2}+\frac{1}{4}\right)
\eta _{(0)}\,,
\label{lead-var}
\end{equation}
respectively. All non-vanishing terms are proportional to $(\partial _{\varphi }^{2}+
\frac{1}{4})\eta _{(0)}$ and are suppressed by $1/r$ relative to the leading
order of the corresponding equation. Hence the leading Hamilton equations \eqref{dot(0)} are invariant for arbitrary $\eta _{(0)}$, whereas the full
equations are invariant only for $\eta _{(0)}\propto \mathrm{e}^{\pm \frac{
\mathrm{i}}{2}\varphi }$. Such parameters are antiperiodic and exist only in
the twisted sector, where the shift maps solutions to solutions (see Section~\ref{Ch:Exact}). The violations reflect the fixed radial profile of $\eta $ (see the discussion after \eqref{S}). This is unlike $\varepsilon (x)$, which can be chosen at every order in $1/r$, because $\eta $ has no subleading parameters that could compensate them.

We conclude that $G_Q[\varepsilon ]$ generates large $\mathrm U(1)$ gauge transformations that preserve the asymptotic conditions and are improper in the canonical sense~\cite{Benguria:1976in}. On the other hand, $G_S[\eta]$ generates an asymptotic transformation that preserves the asymptotic conditions, but does not arise as the residual action of a bulk gauge transformation -- it is a symmetry of the intrinsic boundary
phase space on $\mathscr I^{\pm }$, rather than the $r\rightarrow \infty $ limit of a bulk gauge symmetry.

Finally, a consistent canonical realization of these symmetries also requires that they preserve the symplectic structure of the phase space. A direct evaluation shows that the canonical symplectic form, 
\begin{equation}
\mathit{\Omega }=\int\limits_{\Xi }\diff ^{2}x\,\delta \pi ^{\mu }\wedge \delta
A_{\mu }\,,
\label{s.form}
\end{equation}
remains invariant under the transformations \eqref{trU(1)}, \eqref{trShift} and \eqref{dAu} when the corresponding parameters are field-independent. Although the falloffs \eqref{rad} make the symplectic potential $\Theta $
logarithmically divergent at large $r$, its divergent contribution can be
written as an exact variation in field space. It therefore drops out of $\mathit{\Omega}=\delta \Theta $, since $\delta ^{2}=0$~\cite{Iyer:1994ys}, leaving a finite and well-defined symplectic form. For more details on the symplectic structure of the phase space and symplectic derivation of the charges, see Appendix \ref{Appendix}.


\subparagraph{Algebra and central charge.}


The Poisson brackets among the generators can be computed using their functional derivatives \eqref{varGq}. We find 
\begin{equation}
\{G_{Q}[\varepsilon _{1}],G_{Q}[\varepsilon _{2}]\}=0\,,\qquad \{G_{S}[\eta
_{1}],G_{S}[\eta _{2}]\}=0\,.
\end{equation}%
The first bracket vanishes because the $\mathrm U(1)$ gauge symmetry is Abelian,
whereas the second vanishes because the shift symmetry is Abelian. The mixed
Poisson bracket is nonvanishing,
\begin{equation}
\{G_{Q}[\varepsilon ],G_{S}[\eta ]\}=C[\varepsilon ,\eta ]\,,
\end{equation}
because of the field-independent central term 
\begin{equation}
C[\varepsilon ,\eta ]=-
\oint\limits_{\mathbb{S}^1_\infty} \diff \varphi \,\frac{\epsilon }{r}\,\varepsilon \,\partial _{\varphi }\eta =-
\epsilon \oint\limits_{\mathbb{S}^1_\infty} \diff \varphi \,\varepsilon _{(0)}\,\partial
_{\varphi }\eta _{(0)}\,,  \label{C}
\end{equation}
which defines a central extension of the algebra. It does not generate an additional transformation of the phase space variables, but  it affects the canonical algebra of the boundary observables. Such central extensions commonly arise in the Hamiltonian treatment of asymptotic symmetries~\cite{Brown:1986nw}.

The physical phase space is obtained by imposing the constraints and identifying configurations related by proper gauge transformations. In the canonical formulation, this reduction can be implemented by fixing the freedom associated with proper transformations, with Dirac brackets $\{ \;,\; \}^*$ replacing the Poisson brackets and describing the resulting algebra of observables. Improper boundary transformations with nontrivial charge variations remain physical.

On the constraint surface, the improved generators reduce to their boundary
charges, that is, $G_{Q}=Q$ and $G_{S}=Q_{S}$. For differentiable generators whose transformations preserve this surface and are well defined modulo proper gauge transformations, their Poisson algebra induces the following physical charge algebra:
\begin{equation}
\{Q[\varepsilon _{1}],Q[\varepsilon _{2}]\}^*=0\,,\quad \{Q_{S}[\eta
_{1}],Q_{S}[\eta _{2}]\}^*=0\,,\quad \{Q[\varepsilon ],Q_{S}[\eta
]\}^*=C[\varepsilon ,\eta ]\,.
\label{physQQ}
\end{equation}

In our case, constructing the Dirac brackets explicitly is technically involved because the characteristic constraint $\chi(x)$ contains both first and second class sectors. An explicit construction would require gauge conditions for $\pi^u$ and $\psi$ compatible with the Dirichlet data $A_{(0)u}$, together with the inverse of $\Omega$, that is, $\Omega^{-1}(x,x')=-\frac{\epsilon}{4}\sqrt{rr'}\, \mathrm{sgn}(r-r')\, \delta(\varphi-\varphi')$. This inverse is defined only up to zero mode
terms $\sqrt{r r'}\,h(\varphi,\varphi')$, and fixing them amounts to
specifying the zero mode data at an end of the null generators, which is
the light-front zero mode problem~\cite{Maskawa:1975hx,Heinzl:1998kz}. The symplectic formulation avoids these steps. The space of solutions satisfying the boundary conditions already satisfies all constraints, but still contains proper gauge redundancy. After identifying solutions related by proper gauge transformations, it gives the covariant description of the reduced physical phase space, provided the same initial and boundary data are used in both descriptions.

The charge algebra can therefore be computed using the presymplectic form on the space of solutions, without choosing explicit gauge-fixing conditions. For Hamiltonian vector fields $X_a$, we have
\begin{equation}
i_{X_a}\mathit{\Omega} =-\delta Q_a\,,\qquad \{Q_a,Q_b\}^*=i_{X_a}i_{X_b}\mathit{\Omega} \,.
\end{equation}
Proper gauge directions are null directions of $\mathit{\Omega}$ and do not affect
these contractions. In  Appendix \ref{Appendix}, we reproduce the charges and their algebra using the symplectic formulation. We obtain the charge variations from single contractions of the presymplectic form and the brackets, including the mixed central term \eqref{C}, from double contractions. We also show that the logarithmically divergent part of the symplectic potential is an exact variation in phase space and therefore does not contribute to the finite presymplectic form.

\subparagraph{Algebra in modes.}

All boundary fields and symmetry parameters are periodic ($\sigma=+1$) or
antiperiodic ($\sigma=-1$)  functions of $\varphi$ and admit a Fourier expansion in modes 
\begin{equation}
\Phi^s_n=\oint\limits_{\mathbb{S}^{1}}\frac{\diff \varphi}{2\pi}\, \Phi^s(\varphi)
\mathrm{e}^{-\mathrm{i}n\varphi}\,,\qquad n \in  \begin{cases}
\mathbb{Z}\,, & \sigma =+1\,, \\ 
\mathbb{Z+}\frac{1}{2}\,, & \sigma =-1\,.
\end{cases}
\end{equation}
Orthogonality $\oint \diff \varphi \,\mathrm{e}^{\mathrm{i}(n+m)\varphi }=2\pi
\delta _{n+m,0}$ holds because $n+m\in \mathbb{Z}$. They satisfy the antipodal matching condition \eqref{antipodal} that distinguishes even and odd modes,  
\begin{equation}
\Phi^s_n\big|_{\mathscr I_{-}^{+}}= \,\mathrm{e}^{\mathrm{i}\,\pi n}\Phi^s_n\big|_{\mathscr I_{+}^{-}}\,,
\end{equation}
that is, $\mathrm{e}^{\mathrm{i}n\pi}=(-1)^{n}$ for periodic configurations and $\mathrm{e}^{\mathrm{i}n\pi}=\mathrm{i}\,(-1)^{n-1/2}$ for antiperiodic
ones. In both sectors, the antipodal map is the rotation
$\varphi\to\varphi+\pi$. For antiperiodic configurations, the rotation $\varphi\to\varphi-\pi$ would give the opposite sign.

In particular, the
Fourier components of $A_{(0)\varphi}$, $\pi^r_{(1)}$, $
\varepsilon_{(0)}$, and $\eta_{(0)}$ will be denoted by $A_n$, $\pi_n$, $\varepsilon_n$, and $\eta_n$, respectively. The charges then
take the form 
\begin{eqnarray}
Q[\varepsilon] &=& \sum_{n }\varepsilon_{-n}\,Q_n\,, \qquad
Q_n=-2\pi\,\pi_n\,,  \notag \\
Q_S[\eta] &=& \sum_{n}\eta_{-n}\,S_n\,, \qquad S_n=2\pi\epsilon\,A_n\,,
\label{QFour}
\end{eqnarray}
where $Q_n$ is related to electric flux, whereas $S_n$ corresponds to the shift symmetry intrinsic to the null boundary. The Fourier mode algebra becomes
\begin{equation}
\{S_n,Q_m\}^{*} = \frac{\mathrm{i}\epsilon}{2}\,m\kappa\,\delta_{n+m,0}\,, \qquad
\kappa=4\pi\,,
\end{equation}
while all other brackets vanish. Equivalently, introducing the combinations that diagonalize it,
\begin{equation}
Q_n^\pm=Q_n\mp S_n\,,
\label{Qpm}
\end{equation}
we obtain two independent Abelian Kac--Moody algebras with opposite levels, 
\begin{eqnarray}
\{Q_n^\pm,Q_m^\pm\}^{*} &=& \pm \mathrm{i}\epsilon n\kappa\,\delta_{n+m,0}\,, 
\notag \\
\{Q_n^+,Q_m^-\}^{*} &=& 0\,.  \label{KM Abelian}
\end{eqnarray}
This algebra is an infinite-dimensional extension of $
\mathfrak{u}(1)\times\mathfrak{u}(1)$. 
In the twisted sector the modes are half-integer, and \eqref{KM Abelian}
becomes the twisted Abelian Kac--Moody algebra associated with the
automorphism $C$, as in the twisted sector of the $\mathbb{Z}_2$ orbifold
of a free boson~\cite{Dixon:1986qv}. It has the same level $\kappa$, but no zero
modes, so the global $\mathfrak{u}(1)\times\mathfrak{u}(1)$ part is absent. In the untwisted sector, the zero modes commute, 
\begin{equation}
\{Q_0^\pm,Q_0^\pm\}^{*}=0\,, \qquad \{Q_0^+,Q_0^-\}^{*}=0\,.
\end{equation}
On the chosen radiative phase space, however, \eqref{Q0=0} implies $Q_{0}^{+}+Q_{0}^{-}=2Q_{0}=0$, so only one independent global charge remains.
Thus, even in an Abelian gauge theory, the null boundary supports an infinite set of edge observables. The central extension further shows that these observables act nontrivially on the reduced phase space.

In the untwisted sector, the $n=0$ mode of \eqref{Qshift} has a natural relation to the
three-dimensional topological $\mathrm U(1)_{\mathrm{top}}$ symmetry. Its current is proportional to $^*F$ and is identically conserved by the local Bianchi identity $\diff F=0$~\cite{tong_gaugetheory}. The corresponding topological charge is the magnetic charge, given by the
magnetic flux $\int_{\Xi}F$, which can be written in terms of the Abelian holonomy integral $\oint_{\mathbb{S}^1_\infty}A$. Under the Maxwell--scalar duality \eqref{scalar}, this symmetry is mapped to the global shift symmetry of the dual massless scalar, $\phi \rightarrow \phi +c$~\cite{Turner:2019wnh}. In the twisted sector, on the other hand, the magnetic $\mathrm U(1)_{\mathrm{top}}$ is broken to $\mathbb Z_2$ and $S_0$ does not exist.

With our radiative boundary conditions, however, $A|_{\mathbb{S}_{\infty}^{1}}\sim \sqrt{r}$, so that the bulk holonomy and the corresponding magnetic flux diverge at null infinity. The canonical charge $Q_{S}$ instead depends on the finite leading boundary field $A_{(0)} =A_{(0)u}\diff u+A_{(0)\varphi }\diff\varphi$. In particular, its zero Fourier mode is $S_{0}=\epsilon \oint_{\mathbb{S}_{\infty }^{1}}A_{(0)}$. Thus, $\epsilon S_{0}$ is the holonomy integral of the leading asymptotic boundary connection $A_{(0)}$ or, equivalently, the leading finite coefficient of the three-dimensional magnetic charge. The zero mode $W$ sources the boundary magnetic flux, with the magnetic source located at $r=0$.
The higher Fourier modes provide an angle-dependent extension of this boundary zero mode and, together with the electric charges, form the centrally extended Kac--Moody algebra found above. Related Kac--Moody edge algebras have also been found for three-dimensional Maxwell theory with a finite boundary~\cite{Maggiore:2019wie}. A similar enhancement of magnetic
charges occurs at null infinity in four-dimensional electromagnetism~\cite{Gonzalez:2023yrz}.

In Ref.~\cite{Freidel:2018fsk}, an electric--magnetic central charge \eqref{C} was obtained within the covariant phase space formalism by extending the four-dimensional Maxwell phase space. Here, its charge--Goldstone interpretation follows from Maxwell--scalar duality. Using the asymptotic expansion of the scalar \eqref{expphi}, we introduce 
\begin{equation}
q=-2A_{(1)r}\,,\qquad p=\pi _{(1)}^{r}=\partial _{\varphi }\phi
_{(1)}\,,\qquad A_{(0)\varphi }=\epsilon \phi _{(1)}-\partial _{\varphi }q\,.
\end{equation}
For $\alpha =\varepsilon _{(0)}$, the electric transformations give $\delta
_{\alpha } q=\alpha $ and $\delta _{\alpha }\phi _{(1)}=0$, identifying $q$ as an electric Goldstone-like boundary coordinate because it transforms
inhomogeneously under the boundary symmetry. On the constraint surface, the
symplectic form contains the finite boundary contribution 
\begin{equation}
\Omega _{\partial }^{\infty }=-\oint \diff \varphi \,\delta \phi
_{(1)}\wedge \delta A_{(0)\varphi } =\oint \diff \varphi \,\delta
p\wedge \delta q\,,
\end{equation}
which pairs $q$ with $-p$ for nonzero angular modes, consistently with 
\eqref{Q0=0} (in the twisted sector, all modes are nonzero). The dependence of $Q_{S}$ on $\partial _{\varphi }q$ then
gives $\delta _{\alpha }Q_{S}[\eta ]=-C[\alpha ,\eta ]$, reproducing the
mixed central extension.  Here, the same structure arises at null infinity from the null constraint kernel of three-dimensional Maxwell theory, using boundary coefficients of the original Maxwell fields. The circle cocycle has the same form as the contour contributions around singularities in Ref.~\cite{Freidel:2018fsk}, but can be nonzero for smooth periodic parameters on the
boundary circle.

\subparagraph{Radiative properties of the system.}

To describe the radiative sector, we expand the radial-angular component of the field strength using the boundary conditions \eqref{rad}, 
\begin{equation}
F_{r\varphi }=\partial _{r}A_{\varphi }-\partial _{\varphi }A_{r}=\frac{1}{2
\sqrt{r}}\,\left( A_{(0)\varphi }-2\partial _{\varphi }A_{(1)r}\right) +
\mathcal{O}(r^{-3/2})\,.
\end{equation}
We therefore introduce the gauge-invariant amplitude
\begin{equation}
\mathcal{C}=A_{(0)\varphi }-2\partial _{\varphi }A_{(1)r}\,,  \label{shear}
\end{equation}
which we call the electromagnetic shear, by analogy with the gravitational
shear coefficient~\cite{Bosma:2023sxn}. The leading Hamilton equations then
give the electromagnetic news, 
\begin{equation}
\mathcal{N}=\dot{\mathcal{C}}=w_{(0)}=W\,.  \label{news}
\end{equation}%
Thus, the undetermined boundary value of the Hamiltonian multiplier
coincides on shell with the radiative data. Its role as a prescribed
boundary source is compatible with this interpretation.

Both $\mathcal{C}$ and $\mathcal{N}$ are invariant under electromagnetic
gauge transformations because the variations of $A_{(0)\varphi }$ and $%
2\partial _{\varphi }A_{(1)r}$ cancel in \eqref{shear}. The leading
constraints give 
\begin{equation}
\pi _{(2)}^{\varphi }=\frac{\epsilon }{2}\,\mathcal{C}\,,\qquad \mathcal{Q}%
\equiv \pi _{(1)}^{r}=\epsilon \,\partial _{\varphi }\mathcal{C}\,,
\label{q}
\end{equation}%
where $\mathcal{Q}$ is the electric charge aspect. In terms of the dual
scalar, the shear, news and charge aspect are 
\begin{equation}
\mathcal{C}=\epsilon \,\phi _{(1)},\qquad \mathcal{N}=\epsilon \,\dot{\phi}%
_{(1)}=W\,,\qquad \mathcal{Q}=\partial _{\varphi }\phi _{(1)}\,.
\label{scalar-news}
\end{equation}

The leading field strengths therefore take the form 
\begin{eqnarray}
F_{u\varphi } &=&\sqrt{r}\,\mathcal{N}+\mathcal{O}(r^{-1/2})\,,  \notag \\
F_{r\varphi } &=&\frac{\mathcal{C}}{2\sqrt{r}}+\mathcal{O}(r^{-3/2})\,, 
\notag \\
F_{ur} &=&\frac{\mathcal{Q}}{r^{3/2}}+\mathcal{O}(r^{-5/2})\,.
\label{F.physical}
\end{eqnarray}
In this phase space, the radial electric field does not contain independent $%
r^{-1}$ Coulombic contribution, and $\mathcal{Q}$ is determined by the
radiative amplitude. Its evolution, $\mathcal{\dot{Q}}=\epsilon \,\partial
_{\varphi }\mathcal{N}$, is the local electric-charge balance underlying the
integrated laws discussed below. The energy flux, by contrast, is quadratic
in the news.

Between two cuts, the integrated news gives $\Delta \mathcal{C}%
=\int_{u_{i}}^{u_{f}}\diff u\,\mathcal{N}$. This change controls the
leading electromagnetic kick memory~\cite{Bosma:2023sxn}. The corresponding
four-dimensional effect is discussed in~\cite{Pasterski:2015zua}.

The residual shift acts as 
\begin{equation}
\delta_{\eta}\mathcal{C}=\eta_{(0)}(\varphi)\,, \qquad \delta_{\eta}\mathcal{N}=0\,.
\end{equation}
It therefore changes the time-independent part of the radiative amplitude while preserving the news. This is analogous to the inhomogeneous part of
the transformation of gravitational shear under supertranslations.

\subparagraph{Matching and balance laws.}

For radiative systems at null infinity, it is useful to distinguish the endpoint charges at spatial infinity from their balance-law evolution along each null boundary. This is why, in this paragraph, $Q[\varepsilon ](u)$ and $Q_{S}[\eta ](u)$ denote the charges evaluated on a generic cut $u$ of a single null boundary, while their endpoint
values entering the matching conditions are obtained by taking $u\rightarrow
-\epsilon \infty $ according to the prescription \eqref{bndy-integral}.

Antipodal matching identifies the corresponding boundary data and symmetry parameters, and gives the conservation law 
\begin{equation}
Q[\varepsilon ]\big|_{\mathscr
I_{-}^{+}}= Q[\varepsilon ]
\big|_{\mathscr I_{+}^{-}}\,. \label{Qmatching}
\end{equation}
The endpoint charges can equivalently be expressed in terms of
the integrated soft and hard fluxes through $\mathscr I^{+}$ and
$\mathscr I^{-}$, respectively, together with the charge contributions
at $\mathscr I_{+}^{+}$ and $\mathscr I_{-}^{-}$. When these latter
contributions vanish, only the flux integrals remain~\cite{Strominger:2017zoo}.

The matching condition \eqref{Qmatching} relates charges on different null
boundaries at spatial infinity, whereas the balance law governs their evolution within a single patch (with fixed $\epsilon$). Using the functional derivatives in \eqref{varGq} and the Hamilton equations \eqref{dot}, we obtain
\begin{equation}
\{G_{Q}[\varepsilon ],H_{Q}\}=-\int\limits_{\Xi }\diff ^{2}x\,\dot{%
\varepsilon}\,\psi -\epsilon \oint\limits_{S_{\infty }^{1}}\diff \varphi
\,\frac{\varepsilon }{r}\,\partial _{\varphi }w\,.
\end{equation}
Only the zero mode contribution $w=\sqrt{r}\,W+\cdots $
survives at the boundary, and therefore
\begin{equation}
\{G_{Q}[\varepsilon ],H_{Q}\}\approx  C[\varepsilon ,W]\,.
\end{equation}
Here, $C[\varepsilon ,W]$ is the same central term as in \eqref{C}, with the shift symmetry parameter replaced by the zero mode multiplier $W$. On the reduced phase space, this gives the balance equation 
\begin{equation}
\dot{Q}[\varepsilon ]=\{Q[\varepsilon ],H_{Q}\}^{* }=C[\varepsilon ,W]\,.
\label{Qbalance}
\end{equation}
These balance laws are integrated forms of the leading Maxwell
equations. Their Hamiltonian derivation relates the evolution of
the boundary charges to the centrally extended edge algebra. Here,
$W$ acts as a boundary source for the evolution of the electric edge modes and coincides on shell with the electromagnetic news
$\mathcal{N}$.  Equation \eqref{Qbalance} therefore expresses their response to radiative data. Analogous expressions appear in Chern--Simons descriptions of AdS$_3$ boundary dynamics with chemical potentials, where the bulk equations
reproduce Ward identities of the deformed boundary theory~\cite{deBoer:2013gz,Ferlaino:2013vga}. For discussions of asymptotic charges and fluxes
in radiative systems, see~\cite{Sachs:1962wk,Strominger:2017zoo,Flanagan:2015pxa,Ciambelli:2023bmn}. Related examples in three-dimensional AdS gravity
were discussed in~\cite{Alessio:2020ioh}.

The result  \eqref{Qbalance} is particularly transparent in Fourier modes. The boundary part
of the improved Hamiltonian is 
\begin{equation}
H_Q\big|_{\partial\Xi} = Q_S[W] = \sum_{n}W_{-n}S_n\,,
\end{equation}
where $W_n(u)$ are the Fourier modes of the zero mode multiplier. The centrally extended algebra then gives 
\begin{equation}
\dot Q_n = -\frac{\mathrm{i}\epsilon}{2}\,n\kappa\,W_n\,.
\end{equation}
Hence, in the untwisted sector, the global electric charge remains zero, consistently with \eqref{Q0=0}, whereas the higher electric modes are driven by the corresponding angular modes of the boundary multiplier. A given mode is conserved when its associated source mode vanishes.

As regards the soft charge $Q_S[\eta]$, its endpoint values satisfy a matching condition analogous to \eqref{Qmatching}. The observable $Q_S[\eta]$, however, generates an asymptotic symmetry rather than a local symmetry of the Hamiltonian. As for the electric charge, $\{Q_S[\eta],H_Q\}^*$ does not vanish in general, and $Q_S$ is not conserved under the Hamiltonian evolution. Indeed, a computation analogous to the one above gives  the balance equation
\begin{equation}
    \dot{Q}_{S}[\eta ] =\epsilon\oint\limits_{\mathbb{S}^1_\infty}\diff \varphi\,\eta_{(0)}\left( \partial_\varphi A_{(0)u} + W \right)\,.
    \label{dotQs}
\end{equation}
In particular, the evolution of the $n=0$ sector obtained from \eqref{dotQs} is consistent with the
topological  $\mathrm U(1)_{\mathrm{top}}$ charge  previously discussed. When $\sigma=+1$, $\eta
_{(0)}=\mathrm{const}$, we get
for the $n=0$ mode 
\begin{equation}
\dot{S}_{0}=\epsilon\oint\limits_{\mathbb{S}_{\infty }^{1}}\diff\varphi \,W\,.
\end{equation}
The same expression follows from the local Bianchi identity $\diff F=0$, whereas the charge can change due to flux through the boundary. Indeed, at leading order, the Hamilton equation gives $F_{u\varphi }=\dot{A}_{\varphi }-\partial _{\varphi}A_{u}\sim \sqrt{r}W$, showing that $W$ controls this asymptotic magnetic flux. Thus, the zero mode is conserved when the corresponding boundary flux vanishes.

\subparagraph{Charges as edge observables.}

The balance laws clarify in what
sense $Q$ and $Q_S$ are physical. Each is a finite, differentiable
functional on $\Gamma_\infty$, invariant under proper gauge
transformations, and hence a well-defined function on the reduced phase
space. Its Hamiltonian flow preserves the symplectic form \eqref{s.form},
the constraints and the boundary conditions. Whether it also commutes with
the Hamiltonian is a separate, dynamical question: both charges obey
balance laws, \eqref{Qbalance} and \eqref{dotQs}, and are conserved only
when the corresponding flux vanishes. Angular momentum behaves in the same
way. It generates rotations and is a useful observable whether or not the
Hamiltonian is rotationally invariant, its evolution being governed by the
torque. The same holds for charges at radiative null
boundaries~\cite{Wald:1999wa,Flanagan:2015pxa}. Furthermore, the algebra
\eqref{physQQ} is a kinematical property of the Poisson structure,
independent of the dynamics. Its central term shows that $Q$ and $Q_S$ are
canonically conjugate, forming a Heisenberg pair for each nonzero mode,
like position and momentum. Physically, $Q_S$ measures the magnetic data
at the boundary, and its change between two cuts is controlled by the
integrated news, that is, by the electromagnetic memory.

\subparagraph{Endpoint boundary conditions.}

The endpoint prescription \eqref{bndy-integral} and the matching conditions 
\eqref{Qmatching} at $\mathscr I_{-}^{+}$ and $\mathscr I_{+}^{-}$ require
specifying the behaviour as $u\rightarrow u_{\infty }=-\epsilon \infty $.
From \eqref{dotQs}, a sufficient condition for convergence of the completion integral is that $\partial _{\varphi }A_{(0)u}+w_{(0)}$ be integrable in $u$
near the endpoint. Since $w_{(0)}=W$, a sufficient choice is 
\begin{equation}
A_{(0)u}=\mathcal{O}(|u|^{-1-\delta })\,,\qquad w_{(0)}=W=\mathcal{O}%
(|u|^{-1-\delta })\,,\qquad \delta >0\,,
\end{equation}%
with the same estimates for angular derivatives. Integrating the leading
Hamilton equations~\eqref{dot(0)} then gives 
\begin{eqnarray}
A_{(1)r} &=&A_{(1)r}^{\infty }(\varphi )+\mathcal{O}(|u|^{-\delta })\,, 
\notag \\
A_{(0)\varphi } &=&A_{(0)\varphi }^{\infty }(\varphi )+\mathcal{O}%
(|u|^{-\delta })\,.
\end{eqnarray}%
The leading momentum coefficients also approach finite limits through the
constraints. The other multiplier satisfies $w_{(0)u}=\mathcal{O}(|u|^{-2-\delta })$.

These conditions allow $A_{(0)u}$ and $W$ to be nonzero at finite $u$ and
are compatible with holding them fixed under variations of fields. The
endpoint coefficients remain free to vary, subject to the constraints and
matching conditions, and the time-independent electric and shift
transformations preserve these falloffs.

\subparagraph{Alternative radial boundary condition.}

The boundary conditions proposed by Shimizu and Sugishita~\cite{Shimizu:2025hfl} can be recovered by setting the leading asymptotic field $A_{(0)u}$ to zero, 
\begin{equation}
A_u\big|_{\mathbb S^1_\infty}=0
\qquad\Rightarrow\qquad
A_{(0)u}=0\,,
\label{Au}
\end{equation}
and additionally imposing $A_{(1)r}=0$. Indeed, the leading Hamilton equation
$\dot{A}_{(1)r}=\frac{1}{2}\,A_{(0)u}$ in \eqref{dot(0)} implies only that
$A_{(1)r}$ is independent of $u$. The stronger falloff
$A_r=\mathcal{O}(r^{-3/2})$ therefore requires setting this integration
function to zero.
Compared to \eqref{rad},  the only differences are the faster falloffs $A_u=\mathcal{O}(r^{-1/2})$, $A_r=\mathcal{O}(r^{-3/2})$, and $w_u=\mathcal{O}(r^{-1/2})$, which make these conditions more restrictive.

These stronger conditions lead to the same differentiability condition \eqref{diffH} and the same improved Hamiltonian $H_Q$ as in our case. The difference appears in the electric charge. Namely, preservation of the stronger falloff of $A_r$ excludes gauge transformations with a nonvanishing leading parameter, requiring $\varepsilon_{(0)}=0$. Consequently, the electric charge vanishes, in agreement with~\cite{Shimizu:2025hfl}.

Keeping the non-radiative mode $A_{(1)r}$ allows the weaker radial falloff
of $A_r$ and permits large gauge transformations with nontrivial electric
charges even when $A_{(0)u}=0$. Such a mode is also included in the general asymptotic expansion of Ref.~\cite{Shimizu:2025hfl}.

\section{Edge modes on a black hole background} \label{Ch:BTZ}

Black hole horizons provide a natural setting to study a null boundary at finite radius. They can support infinite-dimensional symmetries, as shown in four dimensions~\cite{Donnay:2016ejv}. We consider Maxwell theory in the probe approximation on a fixed Ba\~{n}ados--Teitelboim--Zanelli (BTZ) black hole background
\cite{Banados:1992wn,Carlip:1995qv}, whose horizon is located at $r=r_{+}$. The aim is to determine whether the edge structure found at null infinity in Minkowski space persists at the horizon, while also exploring a spacetime with different asymptotics.

\subparagraph{Static BTZ black hole.}

The BTZ black hole~\cite{Banados:1992wn} is a solution of the three-dimensional 
Einstein equations with negative cosmological constant $\Lambda =-1/\ell ^{2}$, where $\ell $ is the anti-de Sitter (AdS) radius. We restrict our analysis to the static, electrically neutral geometry, 
\begin{equation}
\diff s^{2}=-N^{2}(r)\,\diff t^{2}+\frac{\diff r^{2}}{N^{2}(r)}+r^{2}\diff
\varphi ^{2}\,,\qquad N^{2}(r)= \frac{r^{2}-r_{+}^{2}}{\ell ^{2}}\,,  \label{BTZ}
\end{equation}
where $t\in\mathbb R$, $r\geq0$, and $\varphi\simeq  \varphi+2\pi$. The horizon is located at $r=r_{+}$, with 
\begin{equation}
r_{+}\equiv \ell \sqrt{M}\,,\qquad M > 0\,,  \label{horizonte}
\end{equation}
where $M$ is the mass parameter.
The Schwarzschild-like coordinates are singular at $r=r_{+}$ and therefore
cover the exterior and interior regions as separate patches.

The limiting case $M=0$ corresponds to the massless BTZ geometry, while $M=-1
$ gives global AdS$_3$. Thus, the energy measured relative to the global AdS vacuum is proportional to $M+1$.

The vacuum BTZ geometry is locally AdS. The black hole interpretation comes from the global identifications of AdS rather than from a local
curvature singularity~\cite{Banados:1992gq}. This property is lost when matter is
included. For example,
the electrically charged BTZ solution in Einstein--Maxwell theory is not locally of constant curvature and develops a curvature singularity at 
$r=0$.

As in the Minkowski case, we first introduce coordinates adapted to the null boundary at $r=r_{+}$.

\subparagraph{Eddington--Finkelstein coordinates.}

In the BTZ geometry, null coordinates cannot be introduced using the flat
space definition \eqref{u}. Instead, we introduce the tortoise coordinate  defined in the exterior
region $r>r_{+}$,
\begin{equation}
\frac{\diff r_{* }}{\diff r}=\frac{1}{N^{2}}\,,\qquad r_{*
}\rightarrow -\infty \quad \text{as}\quad r\rightarrow r_{+}\,,  \label{N}
\end{equation}
such that the $(t,r_*)$ sector of the metric becomes conformally flat, 
\begin{equation}
\diff s^{2}=N^{2}\left( -\diff t^{2}+\diff r_{* }^{2}\right) +r^{2}\diff\varphi ^{2}\,.
\end{equation}
Then the retarded or advanced time can be defined as 
\begin{equation}
u=t-\epsilon r_{* }\,,\qquad \epsilon =\pm 1\,.
\end{equation}
The metric takes the Eddington--Finkelstein (EF) form~\cite{Tamburino:1966zz}
\begin{equation}
\diff s^{2}=g_{\mu\nu}(x)\, \diff x^\mu \diff x^\nu=-N^{2}\,\diff u^{2}-2\epsilon \,\diff u\,\diff r+r^{2}\diff
\varphi ^{2}\,,\quad N^{2}=\frac{r^{2}-r_{+}^{2}}{\ell ^{2}}\,,
\label{EF}
\end{equation}
adapted to null geodesics. These coordinates
are regular at the horizon and extend
over the range $r>0$, covering the exterior and interior regions
in a single patch. 
The metric, inverse metric, and volume element are summarized as
\begin{equation}
g_{\mu \nu }=\left( 
\begin{array}{ccc}
-N^{2} & -\epsilon  & 0 \\ 
-\epsilon  & 0 & 0 \\ 
0 & 0 & r^{2}
\end{array}
\right) ,\quad g^{\mu \nu }=\left( 
\begin{array}{ccc}
0 & -\epsilon  & 0 \\ 
-\epsilon  & N^{2} & 0 \\ 
0 & 0 & \frac{1}{r^{2}}
\end{array}
\right) ,\quad \sqrt{g}=r\,.  \label{gEF}
\end{equation}
The Bondi--Sachs metric \eqref{metric} is recovered by setting
$N=1$.

\subparagraph{Hamiltonian analysis.}

We proceed as in the previous section using the BTZ background \eqref{gEF}. The radial momentum \eqref{pi.r} and two primary constraints \eqref{primary} have the same form as in Minkowski space because they do not depend on $N$.
The total Hamiltonian is 
\begin{equation}
H=\int\limits_{\Xi} \diff^{2}x\,\left[\frac{1}{2r}(\pi ^{r})^{2}+\frac{r}{2}\, N^{2}\,(\pi
^{\varphi })^{2}-A_{u}\nabla _{i}\pi ^{i}+\nabla _{i}(A_{u}\pi
^{i})\right]+\Pi[w_u]+S[w]\,, \label{Hbtz}
\end{equation}
where the smeared constraints $\Pi[w_u]$ and $S[w]$ are given by \eqref{v.suav}. This expression reduces to the Minkowski result for $N=1$.

The Hamilton equations coincide with those of the Minkowski case
\eqref{dot} for all canonical variables except the azimuthal component of
the gauge field, whose evolution is
\begin{equation}
\dot{A}_{\varphi }=rN^{2}\pi ^{\varphi }+\partial _{\varphi}A_{u}+w\,. 
\label{dotAx}
\end{equation}
Since the primary constraints are unchanged, and the first one does not depend on $A_{\varphi}$, the secondary constraint remains the same as in \eqref{secondary}, as does the integral operator $\Omega$ defined in
\eqref{Omega}. The difference is that for $\Lambda <0$ there is no null infinity, so the asymptotic conditions are
different, as we will see below. 

The condition $\dot \psi \approx 0$ is  identically satisfied, whereas preservation of the constraint $\chi $ requires
\begin{equation}
\dot{\chi}=-\frac{2\epsilon }{\sqrt{r}}\,\partial _{r}\left( \frac{w}{\sqrt{r}}\right) -\frac{\epsilon }{r}\,\partial _{r}\left( rN^{2}\pi
^{\varphi }\right) +\frac{\epsilon }{r^{2}}\,\partial _{\varphi }\pi
^{r}=0\,.
\end{equation}%
Solving this equation determines the multiplier $w$ as in \eqref{W}. Its general solution contains an undetermined contribution $\sqrt{r}\, W(u,\varphi )$ and a determined part $\bar{w}$ given by
\begin{equation}
\bar{w}(x)=\frac{\sqrt{r}}{2}\int\limits^{r}\frac{\diff r'}{\sqrt{r'}}\left.\left[ -\partial _{r}\left( rN^{2}\pi ^{\varphi }\right) +\frac{1}{r}\,\partial _{\varphi }\pi ^{r}\right]\right|_{r=r'} \,.
\label{bar w'}
\end{equation}

\subparagraph{Asymptotic conditions.}

EF coordinates are smooth across the horizon, but the Hamiltonian phase space is defined on the region accessible to the exterior observer, such that the horizon acts as an inner boundary of the
Hamiltonian region. The associated boundary terms correspond to a causal
boundary relevant for the exterior Hamiltonian evolution.

We require regularity of the canonical fields and
multipliers near the horizon. In the exterior vicinity of $r_+$, they expand as 
\begin{equation}
f(x)=\sum_{n\geq 0}f_{+(n)}(u,\varphi )\,\rho ^{n}\,, \qquad \rho =r-r_{+}\,,
\label{rho}
\end{equation}
where $\rho$ is the radial coordinate measured from the horizon. 
Using\begin{equation}
\frac{1}{r} =  \frac{1}{r_+} -\frac{\rho}{r_{+}^2}+\mathcal{O}(\rho ^2)\,, \qquad N^{2} =\frac{2r_{+}\rho}{\ell ^{2}} + \frac{\rho ^{2}}{\ell ^{2}}\,,
\qquad \partial _{r}=\partial _{\rho }\,,  \label{r--rho}
\end{equation}
it can be shown that the expansion \eqref{rho} of the fields is consistent with Hamilton equations.

Near the AdS boundary, however, the growth $N^{2}=\mathcal{O}(r^{2})$
imposes a stronger restriction, through the term $rN^{2}\pi ^{\varphi }$ in
the evolution of $A_{\varphi }$ given by \eqref{dotAx}. The multiplier $w$
now follows from 
\begin{equation}
\partial _{r}\!\left( \frac{w}{\sqrt{r}}\right) =\frac{1}{2\sqrt{r}}\left[
-\partial _{r}\!\left( rN^{2}\pi ^{\varphi }\right) +\frac{1}{r}\,\partial
_{\varphi }\pi ^{r}\right] ,  \label{w diff}
\end{equation}
whose solution reduces to \eqref{W}--\eqref{bar w} for $N=1$. As in the
Minkowski radiative sector, we take $\pi ^{r}=\mathcal{O}(r^{-1/2})$ and
require $w=\sqrt{r}\,\tilde{w}+\mathcal{O}(r^{-1/2})$, so that $\tilde{w}$
plays the role of the news. Then $\partial _{r}(w/\sqrt{r})$ must be
integrable at infinity, and Eq.~\eqref{w diff} requires $\pi ^{\varphi }=
\mathcal{O}(r^{-7/2})$. Its $r^{-3/2}$ coefficient thus vanishes, and the
Hamilton equation for $\pi ^{\varphi }$ at this order gives $\tilde{w}=0$.
The Gauss constraint and $\chi \approx 0$ then remove the remaining leading
coefficients. Only subleading fields and stationary gauge data remain. 

In addition, unlike at null infinity, where the shifted coefficient is free boundary data, here the $r^{-3/2}$ coefficient of $\pi^\varphi$ must vanish. Since the zero mode transformation \eqref{trShift} shifts it by
$\frac{\epsilon}{2}\,\eta_{(0)}$, see \eqref{eta0}, it is allowed only for $\eta_{(0)}=0$. 

Moreover, the zero mode of the multiplier is now fully determined. Fixing the integration constant in $w$ by $\bar{w}(r_{+})=0$ and using
$N^{2}(r_{+})=0$, fields regular at the horizon give $F_{u\varphi}|_{r_{+}}=w(r_{+})=\sqrt{r_{+}}\,W$, so $W$ is the horizon
analogue of the news. The condition $\tilde{w}=0$ at infinity then
expresses it in terms of the canonical fields,
\begin{equation}
W=\int\limits_{r_{+}}^{\infty}\frac{\diff r}{2\sqrt{r}}\left[\partial_{r}\!
\left(rN^{2}\pi^{\varphi}\right)-\frac{1}{r}\,\partial_{\varphi}\pi^{r}
\right],
\end{equation}
so that $W$ is no longer a free function.

The leading radiative sector is therefore absent at the AdS boundary, and
the radiation crossing the horizon is determined by the fields on the null
slice. This is consistent with the fact that asymptotically AdS spacetimes
do not have a null boundary. Their conformal boundary is timelike, and the
reflecting boundary conditions impose vanishing net energy flux through it
\cite{Holzegel:2015jwa}, so nothing enters or leaves the spacetime there.
Radiation is not removed from the spacetime, because the subleading fields
still propagate in the bulk, and radiation can cross the event horizon
even when infinity is reflecting. With these boundary conditions, however,
the additional zero mode sector is absent.

This is why, in what follows, we focus on a near-horizon phase space $\Gamma
_{\mathrm{H}}$, with the canonical variables defined on $\Xi \setminus
\{\infty \}$. There $W$ is free horizon data, and the associated charge is a
quasilocal horizon charge. This is in the spirit of quasilocal constructions
of horizon symmetries, where the phase space is defined by boundary
conditions in a near-horizon region and the corresponding surface charges
are evaluated at the horizon~\cite{Donnay:2016ejv}. The additional charge
thus exists on the truncated near-horizon phase space $\Gamma _{\mathrm{H}}$, but not on the phase space with the standard AdS boundary conditions. This restriction concerns only the zero mode sector: configurations in the $C$-twisted sector, with half-integer angular modes, are compatible
with the standard AdS boundary conditions (see Section~\ref{Ch:Exact}), but in them $W$ is determined by the canonical fields.

\subparagraph{Differentiability of the Hamiltonian.}

The total Hamiltonian must be differentiable in order to define
the evolution through Poisson brackets. Varying \eqref{Hbtz} and using the notation of Sec.~\ref{Ch:Minkowski}, we obtain 
\begin{equation}
\delta H=\mathscr D+(\delta S[w])_{ \mathbb{S}^1_+} -\oint\limits_{\mathbb{S}
^{1}_+}\diff\varphi \, \delta A_{u}\,\pi ^{r} \,.
\end{equation}
The additional minus sign compared to \eqref{diffH} reflects the fact that $\mathbb{S}_+^1$ is an inner boundary. The last term vanishes upon fixing the leading value of $A_{u}$ at the horizon, 
\begin{equation}
\delta A_{+(0)u}=0\qquad \text{on } \; \mathbb{S}_{+}^{1}\,,  \label{Dir+}
\end{equation}
since its integrand reduces to $\delta A_{+(0)u}\,\pi _{+(0)}^{r}=0$. As for the constraint term $S[w]$, its variation contains the boundary contribution  
\begin{equation}
(\delta S[w])_{\mathbb{S}^{1}_+}=\epsilon\oint\limits_{\mathbb{S}^{1}_+}\diff\varphi \, \frac{w}{r}\,\delta A_{\varphi }\,.
\end{equation}
It can therefore be cancelled by adding a boundary charge whose variation is 
\begin{equation}
\delta Q_S[w]=-\epsilon\oint\limits_{\mathbb{S}_{+}^{1}}\diff\varphi \,\frac{w}{r}\,\delta A_{\varphi }\,.
\end{equation}
The determined part of the multiplier behaves as 
\begin{equation}
\bar{w}=\mathcal{O}(\rho )\,,\qquad \rho =r-r_{+}\rightarrow 0\,,
\end{equation}
while $A_{\varphi }$ is finite. Hence the horizon contribution of $\bar{w}$
vanishes, and we have $Q_{S}[\bar{w}]=0$.
The nontrivial contribution comes from the homogeneous zero mode of $w$. From \eqref{dotAx}, it behaves as
\begin{equation}
w_{+(0)}=\sqrt{r_{+}}\,W(u,\varphi) =F_{u\varphi }|_{r_{+}}\,,
\label{w+(0)}
\end{equation}
and the charge is supported at the horizon,
\begin{equation}
Q_{S}[w]=Q_{S,+}[W]=-\frac{\epsilon }{\sqrt{r_{+}}}\oint\limits_{\mathbb{S}_{+}^{1}}\diff\varphi \,WA_{+(0)\varphi }\,.
\end{equation}
Thus, the differentiable Hamiltonian on the near-horizon phase space $\Gamma_H$ is
\begin{equation}
H_{Q}=H+Q_{S,+}[W]\,.
\end{equation}

\subparagraph{Quasilocal charges.}

The presence of first class constraints does not by itself guarantee
nontrivial symmetries or charges, since the result depends on the boundary conditions.
We first construct the $\mathrm U(1)$ generator with local parameter $\varepsilon (x)$
using the Castellani method~\cite{Castellani:1981us}. It has the same form as in the previous section, Eqs.~\eqref{G.U(1)} and \eqref{G}, namely, 
\begin{equation}
G_{Q}[\varepsilon ]=\int\limits_\Xi \diff^{2}x\,\left( \varepsilon \,\psi -\dot{\varepsilon}
\,\pi ^{u}\right) +Q[\varepsilon ]\,,  \label{genU(1)}
\end{equation}
where $Q[\varepsilon ]\equiv Q_{r \to r_+}[\varepsilon ]$ is the boundary term required for differentiability.
As in Section~\ref{Ch:Minkowski}, the improved generators also carry a
surface term at the other end of the radial domain, but we study only the horizon contributions for the fields and parameters in $\Gamma_\mathrm{H}$.

Because the boundary is now located at $r_+$, we have different boundary conditions.
Regularity of the transformation $\delta _{\varepsilon }A_{\mu }=-\partial
_{\mu }\varepsilon $ at the
horizon gives the behaviour of the local parameter,
\begin{equation}
\varepsilon   =\varepsilon _{+(0)}(\varphi )+\mathcal{O}(\rho )\,,\qquad
\rho =r-r_{+}\rightarrow 0\,.
\end{equation}
Near the horizon, the subleading coefficients $\varepsilon _{+(n)}(u,\varphi )$, $n\geq 1$, may depend on $u$. In all
cases, $\delta \varepsilon =0$, because the parameter is field-independent. The variation of the generator then gives 
\begin{equation}
\label{varQ}
\delta Q[\varepsilon ]=  \oint\limits_{\mathbb{S}^{1}_{+}}\diff\varphi \,
\varepsilon \,\delta \pi ^{r} \,,
\end{equation}
and the electric charge is supported
only at the horizon, 
\begin{equation}
Q[\varepsilon ]= \oint\limits_{\mathbb{S}_{+}^{1}}\diff\varphi \,\varepsilon
\,\pi ^{r}=  \oint\limits_{\mathbb{S}_{+}^{1}}\diff\varphi \,\varepsilon
_{+(0)}\,\pi _{+(0)}^{r}\,.  \label{Q u(1)}
\end{equation}
Again, the additional minus sign compared to \eqref{Qel} is because $\mathbb{S}_+^1$ is an inner boundary.
 
On the other hand, the shift symmetry generator is 
\begin{equation}
G_{S}[\eta ]=\int\limits_\Xi \diff^{2}x\,\eta \,\chi +Q_{S}[\eta ]\,,\qquad \eta =\sqrt{r}\,\eta _{+(0)}(\varphi )\,,
\label{generador asym}
\end{equation}
where the parameter $\eta$ is restricted to have the radial dependence of the zero mode of $\Omega$, so that the bulk part of $G_S$ selects the first class sector of $\chi$. It is taken to be independent of $u$ for simplicity and in analogy with the null infinity analysis. The boundary term $Q_S$ cancels the boundary contribution from the variation of the bulk generator, provided that
\begin{equation}
    \delta Q_{S}[\eta]= -\epsilon
\oint\limits_{\mathbb S^1_+}\diff\varphi\,
\frac{1}{r}\,\eta \, \delta A_\varphi \,,
\end{equation}
where the minus sign follows from the orientation of the inner boundary. Near the horizon, 
\begin{equation}
\eta =\sqrt{r_{+}}\,\eta _{+(0)}(\varphi )+\mathcal{O}(\rho )\,,
\label{sqrt(r)}
\end{equation}
and one obtains 
\begin{equation}
Q_{S}[\eta ]=-\epsilon
\oint\limits_{\mathbb S^1_+}\diff\varphi\,
\frac{1}{r}\,\eta \, A_\varphi=-\frac{\epsilon }{\sqrt{r_{+}}}\oint\limits_{\mathbb{S}_{+}^{1}}\diff\varphi \,\eta _{+(0)}A_{+(0)\varphi }\,.  \label{qs BTZ}
\end{equation}
The resulting quasilocal charge $Q_S$ is the counterpart of the soft charge \eqref{Qshift} in Minkowski space, now supported at the BTZ horizon. Again, for $\eta_{(0)}=\mathrm{const}$, this charge is proportional to the holonomy integral of the Abelian gauge field evaluated at the horizon,  and therefore admits a magnetic interpretation analogous to the zero mode at null infinity, with the difference that the holonomy is finite in this case. It depends on the black hole mass through the horizon radius, 
\begin{equation}
\frac{1}{\sqrt{r_{+}}}=(\ell ^{2}M)^{-1/4}\,.
\label{ellM}
\end{equation}

With both boundary terms determined, the improved generators are differentiable and can now be used to obtain their action on the near-horizon phase space.

\subparagraph{Local transformations.}

The generators \eqref{genU(1)} and \eqref{generador asym} act on any
phase space function in the horizon phase space $\Gamma_{\mathrm H}$
according to \eqref{local}.  Their nontrivial action on the canonical fields is given by \eqref{trU(1)} and \eqref{trShift},
and on multipliers  by
\begin{equation}
\delta_{\varepsilon\eta} w_u=-\ddot \varepsilon\,,\qquad  \delta_{\varepsilon\eta} w=\epsilon N^2r\partial_r\left( \frac{\eta}{r} \right)\,,
\label{dwu,dw}
\end{equation}
reproducing the angle-dependent shift of $A_{\varphi }$ on the horizon, characteristic of null boundaries. 
These shift modes parametrize boundary configurations related by the
intrinsic shift transformations and may therefore be viewed as
Goldstone-like edge degrees of freedom.

Unlike at null infinity, here we do not include the compensating shift
\eqref{dAu} of $A_u$, since at the horizon it is of leading order and would
violate the Dirichlet condition \eqref{Dir+}. The multiplier law
\eqref{dwu,dw} is chosen so that the evolution equation of $A_\varphi$,
Eq.~\eqref{dotAx}, is preserved exactly.

The combined transformation $\delta _{\varepsilon \eta
}=\delta _{\varepsilon }+\delta _{\eta }$ is compatible with the boundary conditions. Indeed, using
\begin{eqnarray}
\delta _{\varepsilon \eta }A_{\mu } &=&\delta _{\varepsilon \eta }A_{+(0)\mu
}+\mathcal{O}(\rho )\,,  \notag \\
\delta _{\varepsilon \eta }\pi ^{\mu } &=&\delta _{\varepsilon \eta }\pi
_{+(0)}^{\mu }+\mathcal{O}(\rho )\,,
\end{eqnarray}
and 
\begin{equation}
\varepsilon =\varepsilon _{+(0)}+\mathcal{O}(\rho )\,,\qquad \eta =\sqrt{r_{+}}\,\eta _{+(0)}+\mathcal{O}(\rho )\,,
\end{equation}
one verifies from \eqref{trU(1)} and \eqref{trShift} that the transformations preserve the order of the near-horizon falloffs. 

The $\mathrm U(1)$ transformations are the usual gauge symmetries.
The shifts generated by $G_{S}[\eta]$ act canonically on the horizon phase
space $\Gamma_{\mathrm H}$, preserving the near-horizon falloffs and defining
finite charges, although they do not generally preserve the bulk Hamilton
equations. The obstruction comes from terms involving derivatives transverse
to the horizon. The shift charge is nevertheless a well-defined edge observable of the horizon phase space, in the sense of the paragraph ``Charges as edge observables'' in Section~\ref{Ch:Minkowski}, since both the charge and the
flow it generates involve only data intrinsic to the horizon. By contrast, in the asymptotic Minkowski case, the
corresponding transverse terms are subleading under the radiative falloffs,
so the leading Hamilton equations remain invariant.

\subparagraph{Algebra and central extension.}
On the constraint surface, the algebra of the generators \eqref{genU(1)} and 
\eqref{generador asym} reduces to  the
algebra of the corresponding horizon charges. Using \eqref{Q u(1)} and \eqref{qs BTZ}, we find two Abelian sectors, corresponding to the ordinary $\mathrm U(1)$ gauge symmetry and to the horizon shift symmetry, together
with the central term 
\begin{equation}
\{Q[\varepsilon ],Q_{S}[\eta ]\}^*=C_+[\varepsilon,\eta ]\,,
\end{equation}
where 
\begin{equation}
C_+[\varepsilon,\eta]=\frac{\epsilon }{r_{+}}\oint\limits_{\mathbb{S}
_{+}^{1}}\diff\varphi \, \varepsilon\,\partial _{\varphi }\eta =\frac{\epsilon }{\sqrt{r_{+}}}\oint\limits_{\mathbb{S}_{+}^{1}}\diff\varphi
\,\varepsilon _{+(0)}\,\partial _{\varphi }\eta _{+(0)}\,.
\end{equation}
Notice that the inner boundary orientation changes not only the signs of both surface charges, but also the sign of the induced boundary symplectic
form, and hence of the corresponding Dirac bracket. Therefore, the two minus signs in the charges do not cancel in their algebra, and the central term acquires a single relative minus sign with respect to the outer boundary case \eqref{C}.

The Fourier expansion is analogous to that in the Minkowski case, see Eqs.~\eqref{QFour}, except that the fields are now defined on the horizon. Denoting the Fourier modes of the horizon charges by $Q_{n}$ and $S_{n}$, we find
\begin{equation}
\{ S_{n},Q_{m}\}^*=-\frac{\mathrm{i}\epsilon}{2}\,m\kappa _{+}\,\delta _{n+m,0}\,,\qquad
\kappa _{+}=\frac{4\pi}{\sqrt{r_{+}}}=\frac{\kappa}{\sqrt{r_+}}\,, 
\end{equation}
where $n\in\mathbb Z$ for $\sigma=+1$, and $n\in\mathbb Z+\frac 12$ for $\sigma=-1$, as in Sec.~\ref{Ch:Minkowski}.

In this case, however, $Q_{n}$ and $S_{n}$ have different natural units
\footnote{With the normalization used in \eqref{lagrangiano}, the natural units are $[r_{+}]=[\ell ]=L$ and $[M]=1$, so the relevant Taylor coefficients have dimensions $[A_{+(0)\varphi }]=L^{1/2}$, $[\pi _{+(0)}^{r}]=L^{-1/2}$ and $[\varepsilon _{+(0)}] =L^{1/2}$. Therefore, the charge modes have dimensions $[Q_{n}]=L^{-1/2}$ and $[S_{n}]=1$, while the constant $[\kappa _{+}]=L^{-1/2}
$.}, so a dimensionless combination that diagonalizes the algebra is 
\begin{equation}
Q_{n}^{\pm }=\sqrt{r_{+}}Q_{n}\pm S_{n}\,.
\end{equation}
We obtain two independent Kac--Moody algebras, 
\begin{equation}
\{Q_{n}^{\pm },Q_{m}^{\pm }\}^{\ast } =\pm \mathrm{i}\epsilon n\kappa
\,\delta _{n+m,0}\,,  \qquad
\{Q_{n}^{+},Q_{m}^{-}\}^* =0\,.
\end{equation}
The level $\kappa $ does not depend on the BTZ background, 
\begin{equation}
\kappa =\sqrt{r_{+}}\kappa _{+}=4\pi \,.  \label{kappa}
\end{equation}
In this dimensionless normalization, the two Kac--Moody algebras have
universal levels of opposite sign, with $\kappa =4\pi $, as in the Minkowski
case.

It is also worth mentioning that, in the BTZ exterior, the angular circle surrounds the black hole and cannot be continuously shrunk to a point~\cite{Banados:1992gq}. The fields can thus return charge-conjugated after one circuit while
remaining locally smooth, so the twist is encoded in a nontrivial holonomy
around the black hole and does not require a localised defect in the exterior. In this sector the horizon charges have half-integer Fourier
modes, as in Section~\ref{Ch:Minkowski}, and the twisted algebra keeps the same
levels.

\subparagraph{Matching and balance laws.}

To discuss the balance laws, we make explicit in this paragraph the $u$-dependence of the charges $Q[\varepsilon ](u)$ and $Q_{S}[\eta ](u)$. In the null foliation \eqref{EF},  $r\to r_+$  corresponds to approaching future or past horizon ($\mathcal{H^+}$ or $\mathcal{H}^{-}$), depending on $\epsilon$. This resembles the null infinity structure in Minkowski spacetime (see also Fig.~\ref{Fig:three spacetimes}). Future and past horizons intersect at the bifurcation surface $\mathcal{B}$, which is topologically a circle $\mathbb{S}^1$. Following~\cite{Barnich:2001jy, Cheng:2022xyr, Mao:2023rca}, charges are naturally defined as integrals over the bifurcation surface, similarly to \eqref{bndy-integral}. 

Unlike at null infinity, no antipodal matching is needed at the
bifurcation surface, since the geometry is regular there and the fields can
be smoothly continued between $\mathcal H^{-}$ and $\mathcal H^{+}$
\cite{Mao:2023rca}. Nevertheless, the two signs of $\epsilon$ correspond to different EF patches. Since we treat these patches separately, their boundary data must be identified at $\mathcal B$, leading to the corresponding matching conditions for the horizon charges. Accordingly, as in \eqref{Qmatching}, the matching at $\mathcal B$ gives the horizon conservation laws:
\begin{equation}
Q[\varepsilon]\big|_{\mathcal H^{+}_{-}}
=
Q[\varepsilon]\big|_{\mathcal H^{-}_{+}}\,,\qquad Q_S[\eta]\big|_{\mathcal H^{+}_{-}}
=
Q_S[\eta]\big|_{\mathcal H^{-}_{+}}\,.
\end{equation}

As regards the evolution of the quasilocal charges, the analysis is analogous to the null infinity case, with the horizon replacing null infinity. On the reduced phase space, the boundary multiplier $W$ acts as a source through the central term, 
\begin{equation}
\dot{Q}[\varepsilon] = \{Q[\varepsilon],H_Q\}^* = C_{+} [\varepsilon,W]\,.
\end{equation}
Equivalently, in Fourier modes, 
\begin{equation}
\dot{Q}_n = \frac{\mathrm{i}\epsilon}{2}\,n\kappa_+\,W_n\,.
\end{equation}
Thus, in the untwisted sector, the zero mode $Q_0$ is conserved for arbitrary $W(u,\varphi)$, while the
higher electric modes are driven by the angular modes of the horizon multiplier.  Similarly, the evolution of the horizon observable $Q_{S}$ is also driven by the source $W$, as already discussed in the Minkowski case.

\subparagraph{Endpoint boundary conditions.}

In addition to convergence of the endpoint charges, we require the Maxwell
field to extend smoothly to the bifurcation circle $\mathcal{B}=\mathcal{H}%
^{+}\cap\mathcal{H}^{-}$ and choose a gauge in which the potential is also
regular there. For the black hole horizon, this circle is approached at $u_{\mathcal{B}}=\epsilon\infty$. An
affine parameter along the horizon behaves as $\lambda \propto \mathrm{e}^{-\epsilon \varkappa _{+}u}$, where $\varkappa _{+}=\frac{r_{+}}{\ell ^{2}}$
is the surface gravity, and vanishes at $\mathcal{B}$. Smoothness then
implies the endpoint behaviour 
\begin{equation}
\begin{array}[b]{llll}
A_{+(0)\varphi } & =A_{+(0)\varphi }^{\mathcal{B}}(\varphi )+\mathcal{O}(%
\mathrm{e}^{-\varkappa _{+}|u|})\,,\qquad \medskip  & A_{+(0)u}\, & =
\mathcal{O}(\mathrm{e}^{-\varkappa _{+}|u|})\,, \\[4pt]
\pi _{+(0)}^{r} & =\pi _{+(0)}^{r,\mathcal{B}}(\varphi )+\mathcal{O}(\mathrm{e}^{-\varkappa _{+}|u|})\,, & w_{+(0)} & =\mathcal{O}(\mathrm{e}^{-\varkappa
_{+}|u|})\,,
\end{array}
\label{endBTZ}
\end{equation}
with the same estimates for angular derivatives. Here $w_{+(0)}=\sqrt{r_{+}}
\,W$. These falloffs are compatible with the leading horizon Hamilton
equations and ensure convergence of the completion integrals, giving finite
electric and shift charges at $\mathcal{B}$.

\subparagraph{Alternative radial boundary condition.}

One may further restrict the horizon data by imposing 
\begin{equation}
A_{u}(r_{+})=A_{+(0)u}=0 \quad\Rightarrow\quad A_{u}=A_{+(1)u}\,\rho+%
\mathcal{O}(\rho^{2})\,.  \label{A+u}
\end{equation}
Consistency with the Hamilton equations then requires
\begin{equation}
w_{u}(r_{+})=w_{+(0)u}=0 \quad\Rightarrow\quad w_{u}=w_{+(1)u}\,\rho+%
\mathcal{O}(\rho^{2})\,,
\end{equation}
while all other canonical fields and multipliers may remain finite at the horizon.
The falloff \eqref{A+u} implies that the gauge parameter behaves near the horizon as
\begin{equation}
\varepsilon=\varepsilon_{+(0)}(\varphi) + \mathcal{O}(\rho)\,.   
\end{equation}
Hence, the boundary conditions are preserved by large $\mathrm U(1)$ gauge transformations. The horizon shift symmetry also preserves them, and the corresponding charges remain finite.

Thus, unlike at null infinity, the quasilocal horizon
construction does not require a nonvanishing  $A_{+(0)u}$.
It can consistently be set to zero without affecting the horizon symmetry algebra or the corresponding quasilocal charges.

\section{Edge modes in de Sitter space}

Another interesting background in which a null boundary appears at finite distance is de Sitter (dS) space. Global dS has no boundary, so one does not expect edge observables in the full spacetime. The situation changes, however, for an observer restricted to the static patch, whose spatial region is bounded by the cosmological horizon. Although this description does not make the cosmological expansion manifest, it provides a natural setting for analyzing Maxwell theory on a finite null boundary. In particular, one may expect an enhancement of the symmetry structure near the horizon, similarly to the Minkowski and BTZ cases.

\subparagraph{Static coordinates in dS space.}

The static patch of dS space is adapted to a static observer. It covers only the causal region accessible to a single static
observer, bounded by the cosmological horizon at $r=r_c$. For this reason, it is a natural coordinate system for an analysis focused on null
hypersurfaces and horizon physics. 
In $2+1$ dimensions, its metric has the same form as the BTZ metric \eqref{BTZ}, with the lapse function
\begin{equation}
N^{2}=1-\frac{r^{2}}{\ell^{2}}\,.
\label{dS}
\end{equation}
The radial coordinate range accessible to the observer is $0\leq r<\ell$, where $\ell$ is the dS radius. The cosmological horizon is located at $r=r_{c}=\ell $.

The horizon is a null surface, so it is convenient to introduce the EF
coordinates \eqref{EF}, which are regular at $r=r_{c}$. The important difference from the BTZ case is that the horizon is now the outer boundary of the static patch. The metric, inverse metric, and volume element are again given
by \eqref{gEF}, with $N^{2}$ replaced by  \eqref{dS}.

This also reverses the role of the advanced and retarded EF coordinates.
Indeed, while in the BTZ case the definition of the tortoise coordinate \eqref{N} gives $r_{*}\rightarrow-\infty$ as $r\rightarrow r_{+}$, for the dS static patch $r_{*}\rightarrow+\infty$ as $r\rightarrow r_{c}$. The sign difference
is already visible from the near-horizon behaviour,
\begin{equation}
N_{\rm BTZ}^{2}\sim \frac{2r_{+}}{\ell^{2}}\,(r-r_{+})\,,
\qquad
N_{\rm dS}^{2}\sim \frac{2}{\ell}\,(r_{c}-r)\,.
\end{equation}
Consequently, the future BTZ horizon is regular in advanced EF coordinates,
whereas the future dS horizon is regular in retarded EF coordinates. The opposite holds for the past horizons.

\subparagraph{Hamiltonian analysis.}

The local Hamiltonian analysis has the same form as in the BTZ case of Sec.~\ref{Ch:BTZ}, with $N^{2}$ replaced by its dS expression \eqref{dS}. In particular, the constraints, the integral operator $\Omega$ and its zero mode, the Hamilton equations, and the solution for the multiplier $w$ are unchanged in form. We therefore focus only on the differences associated with the boundary.

The main difference is the radial range. While the BTZ exterior covers $r_{+}\leq r<\infty$, the dS static patch extends over $0\leq r\leq r_{c}$, with $r_{c}=\ell$. The cosmological horizon is therefore an outer boundary,
whereas $r=0$ is geometrically a regular origin rather than a boundary. As in the Minkowski case, however, allowing the zero mode sector with $W\neq0$
produces a singularity at $r=0$,%
\footnote{For example $\dot{\pi}^{\varphi }\sim \frac{\epsilon }{2}\,r^{-3/2}W$ and $\dot{\pi}^{r}\sim \epsilon \,r^{-1/2}\partial _{\varphi }W$ near $r=0$.} where the source is located.
The corresponding horizon
phase space $\Gamma_c$ is therefore defined on a patch near the horizon, without requiring an extension of this sector to the origin, with the canonical variables defined on $\Xi\backslash \{0\}$.

At the cosmological horizon, we impose regularity of the canonical fields and multipliers. Introducing the inward radial distance, 
\begin{equation}
\rho =r_{c}-r \,,
\end{equation}
which differs by a sign from the BTZ definition \eqref{rho}, we expand the fields, conjugate momenta, and multipliers as
\begin{equation}
    f(x)=\sum_{n\geq0} f_{c(n)}(u,\varphi)\,\rho^n\,.
\end{equation}
The leading symmetry parameters are taken to be independent of $u$, as in
the previous horizon analysis.
Near $r=r_{c}=\ell $, we use
\begin{equation}
\frac{1}{r}=\frac{1}{\ell }+\mathcal{O}(\rho )\,,\qquad N^{2}=\frac{2\rho }{%
\ell }+\mathcal{O}(\rho ^{2})\,,\qquad \partial _{r}=-\partial _{\rho }\,,
\end{equation}
to show that the Hamilton equations are consistent with the regular near-horizon conditions.

As $r\rightarrow 0$, the canonical fields admit the same leading near-origin
behaviour \eqref{r=0Mink} and cancellation conditions \eqref{cancell} as in
Minkowski space. Since $N^{2}=1+\mathcal O(r^2)$ in this limit, curvature affects only the
subleading terms in this expansion. The same angular consistency condition
allows nontrivial half-angle modes $\mathrm{e}^{\pm \frac{\mathrm{i}}{2}\varphi }$. With the standard angular identification $\varphi \simeq \varphi+2\pi $, these modes can be accommodated by a charge-conjugation twist, $\Phi^s (\varphi +2\pi )=-\Phi^s (\varphi )$. With these antiperiodic boundary conditions, the background metric remains dS, even at the origin, with scalar curvature $R=6/\ell ^{2}$,
while the defect at $r=0$ arises from the field gluing conditions. The twisted identification is a global property of this sector and determines the allowed angular modes of the fields and
symmetry parameters entering the second charge. In global dS space the twist requires a second defect at the
antipodal point, which lies beyond the cosmological horizon and therefore
does not affect the static patch.\medskip

A differentiable Hamiltonian is obtained by imposing the Dirichlet boundary condition 
\begin{equation}
\delta A_{c(0)u}=0\qquad \text{on }\mathbb{S}_{c}^{1}\,,
\label{Dir-c}
\end{equation}
and adding the boundary term whose variation is
\begin{equation}
\delta Q_{S}[w]=\frac{\epsilon }{\ell}\oint\limits_{\mathbb{S}_{c}^{1}}
\diff\varphi \,w_{c(0)}\,\delta A_{c(0)\varphi }\,.
\end{equation}
Again we restrict to the quasilocal charges, such that  $Q_S[\eta ]\equiv Q_{S,r \to r_c}[\eta]$,  because we study only the horizon contributions for the fields and parameters in $\Gamma_c$.
The total charge also carries a
surface term near the origin.

The field-dependent particular solution $\bar{w}$ in \eqref{bar w'}
satisfies  
\begin{equation}
\bar{w}=\mathcal{O}(\rho )\,,\qquad \delta \bar{w}=\mathcal{O}(\rho
)\,,\qquad \rho \rightarrow 0\,,
\end{equation}
and therefore does not contribute at the cosmological horizon. Only the
homogeneous zero mode remains, with $\delta w_{c(0)}=0$, giving a nontrivial contribution
\begin{equation}
Q_{S}[w] =\frac{\epsilon }{\ell}\oint\limits_{\mathbb{S}_{c}^{1}}\diff\varphi \,w_{c(0)}A_{c(0)\varphi }=\frac{\epsilon }{\sqrt{\ell }}\oint\limits_{\mathbb{S}_{c}^{1}}\diff\varphi \,W\,A_{c(0)\varphi } \equiv Q_{S,c}[W]\,.
\end{equation}

\subparagraph{Charges and algebra.}

The construction of the improved generators is the same as in the BTZ
case, so we only discuss the final expressions. The $\mathrm U(1)$ generator \eqref{genU(1)} has the boundary term determined by \eqref{varQ}, with an overall minus sign because the cosmological horizon is the outer boundary. For a regular, field independent gauge parameter 
\begin{equation}
\varepsilon =\varepsilon _{c(0)}(\varphi )+\mathcal{O}(\rho )\,,
\end{equation}
the electric charge is 
\begin{equation}
Q[\varepsilon ]=-\oint\limits_{\mathbb{S}_{c}^{1}}\diff\varphi \,\varepsilon
\,\pi ^{r}=-\oint\limits_{\mathbb{S}_{c}^{1}}\diff\varphi \,\varepsilon
_{c(0)}\,\pi _{c(0)}^{r}\,.  
\end{equation}
The shift generator has the form \eqref{generador asym}, with 
\begin{equation}
\eta =\sqrt{r}\,\eta _{c(0)}(\varphi )\,,
\end{equation}
giving the analogue of \eqref{qs BTZ},
\begin{equation}
Q_{S}[\eta ]=\frac{\epsilon }{\ell }\oint\limits_{\mathbb{S}_{c}^{1}}
\diff\varphi \,\eta \,A_{\varphi }=\frac{\epsilon }{\sqrt{\ell }}
\oint\limits_{\mathbb{S}_{c}^{1}}\diff\varphi \,\eta _{c(0)}A_{c(0)\varphi
}\,. 
\end{equation}
Thus, the form of the charges is the same as in the BTZ case, with a different normalization and boundary orientation.

The corresponding transformations are the same as in the BTZ case, given by
\eqref{trU(1)}, \eqref{trShift} and \eqref{dwu,dw} with $N^2$ from \eqref{dS}, and they preserve the
cosmological horizon boundary conditions. As there, the compensating shift
\eqref{dAu} is not included, since it would violate the Dirichlet condition
\eqref{Dir-c}.

The two charges are Abelian, while their mixed bracket gives the central term 
\begin{equation}
\{Q[\varepsilon ],Q_{S}[\eta ]\}^* = -\frac{\epsilon }{\sqrt{\ell }}\oint\limits_{\mathbb{S}_{c}^{1}}\diff\varphi \,\varepsilon _{c(0)}\,\partial
_{\varphi }\eta _{c(0)} = C_{c}[\varepsilon,\eta ]\,.
\end{equation}
Expanding in Fourier modes as before gives two independent Abelian Kac--Moody algebras with levels $\pm\kappa$. 
Thus, the cosmological horizon supports the same centrally extended edge algebra as the BTZ horizon, with a universal level $\kappa=4\pi$. Geometrically, the difference is that the BTZ horizon is an inner boundary of the exterior region, whereas the cosmological horizon is the outer boundary of the static patch.

As in the previous cases, the multiplier $W(u,\varphi)$ acts as a boundary source: $Q_{0} =0$ in the untwisted sector (it is absent in the twisted one), and the higher electric modes obey balance equations determined by the corresponding angular modes of $W$. The shift charge $Q_{S}$ is an intrinsic horizon observable. 

The multiplier also has a radiative interpretation at the horizons.
For $h\in\{+,c\}$, with $r_c=\ell$, the leading Hamilton equation gives
$F_{u\varphi}|_{r=r_h}=w_{h(0)}=\sqrt{r_h}\,W$.
Thus, the normalized electromagnetic news is again
$\mathcal{N}_h=W$. A corresponding shear amplitude can be introduced
by integrating the news along the horizon generators,
$\dot{\mathcal{C}}_h=\mathcal{N}_h$, with its time-independent
part specified by the endpoint data. The residual shift preserves
the news while changing the time-independent part of the horizon
potential. 

No antipodal identification is required at the bifurcation surface, since the geometry is regular there and the fields can be smoothly continued between the two horizon branches. This is also transparent in the stretched horizon picture~\cite{Ball:2024hqe}.
This smoothness requirement, in a gauge regular at the bifurcation circle,
gives the same exponential endpoint falloffs as in Eq.~\eqref{endBTZ}, with $+\rightarrow c$, $r_{c}=\ell $ and the surface gravity $\varkappa _{c}=\frac{1}{\ell }$. In this case, the bifurcation circle is approached at $u_{\mathcal{B}} =-\epsilon \infty $, and the affine horizon parameter behaves as 
$\lambda \propto \mathrm{e}^{\epsilon \varkappa _{c}u}$.

\section{Exact solution in the twisted sector}
\label{Ch:Exact}

The near-origin behaviour \eqref{r=0Mink} was not used to define the phase space of the fields on the Minkowski background. We show that it is  nevertheless realized by an exact solution,  which provides an explicit configuration in $\Gamma_\infty$ with a $C$-twist defect at the origin. We use the dual scalar \eqref{scalar}. The Bianchi identity $\diff F=0$ is equivalent to the wave
equation $\Box \phi =0$. Writing a single mode as $\phi =\frac{1}{\sqrt{r}}\,\hat{\phi}(u,r)\,\mathrm{e}^{\mathrm{i}k\varphi }$, the wave equation becomes 
\begin{equation}
2\epsilon \,\partial _{u}\partial _{r}\hat{\phi}=\partial _{r}^{2}\hat{\phi}-
\frac{k^{2}-\frac{1}{4}}{r^{2}}\,\hat{\phi}\,.
\label{waveeq}
\end{equation}
In the sector \eqref{r=0Mink}, $k=\pm \frac{1}{2}$ and the centrifugal term
vanishes. The radial equation then reduces to the free $1+1$ wave equation,
with general solution $\hat{\phi}=f(u)+g(u+2\epsilon r)$, i.e., a
superposition of $f(t-\epsilon r)$ and $g(t+\epsilon r)$. The purely
retarded ($\epsilon =+1$) or purely advanced ($\epsilon =-1$) solution has $g=0$. Combining the two modes $k=\pm \frac{1}{2}$, it reads 
\begin{eqnarray}
\phi  &=&-\frac{1}{\sqrt{r}}\,\left[ \mathcal{F}(u)\cos \left( \dfrac{%
\varphi }{2}\right) +\mathcal{G}(u)\sin \left( \dfrac{\varphi }{2}\right) %
\right] \,,  \notag \\
W &=&\left. \epsilon \,\partial _{u}\left( \sqrt{r}\phi \right) \right\vert
_{r\rightarrow \infty }=-\epsilon \left[ \dot{\mathcal{F}}\cos \left( \dfrac{%
\varphi }{2}\right) +\dot{\mathcal{G}}\sin \left( \dfrac{\varphi }{2}\right) \right] \,,  \label{Wexact}
\end{eqnarray}
with two arbitrary functions $\mathcal{F}(u)$ and $\mathcal{G}(u)$, which are
the two modes of \eqref{excess1} given by $W_{\pm 1/2}=-\frac{\epsilon }{2}%
\left( \dot{\mathcal{F}}\mp \mathrm{i\,}\dot{\mathcal{G}}\right) $. They
determine the time-dependent strength $\sqrt{\mathcal{F}^{2}+\mathcal{G}^{2}}
$ of the twist defect. As we deal with a linear theory, we can use the
superposition principle to construct further solutions. The superposition
holds only within the twisted sector.

\begin{figure}[h!]
    \centering
\includegraphics[width=\textwidth]{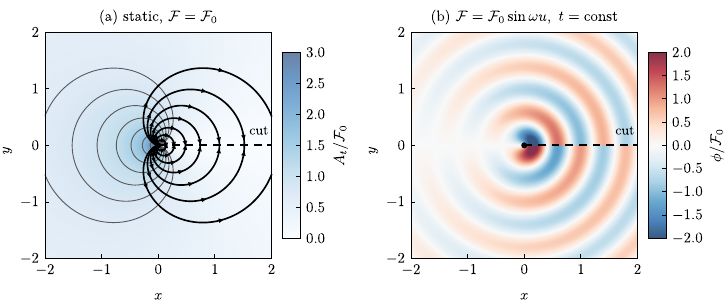}
    \caption{\it Exact solution in the $C$-twisted sector with $\mathcal{G}=0$,
$\epsilon=+1$ and the cut on the half-line $\varphi=0$.
(a) Static case $\mathcal{F}=\mathcal{F}_0$: the potential
$A_t=\frac{1}{\sqrt{r}}\,\mathcal{F}_0\sin(\varphi/2)$ (colour), the equipotentials
$r=a(1-\cos\varphi)$ (thin) and the field lines $r=a(1+\cos\varphi)$
(thick). Both families are cardioids with their cusp at the defect. The
electric field reverses across the cut, $\vec E\to-\vec E$.
(b) Outgoing wave $\mathcal{F}=\mathcal{F}_0\sin\omega u$: the dual scalar
$\phi$ on a $t=\mathrm{const}$ slice, which changes sign across the cut.
On a $u=\mathrm{const}$ slice the configuration is the static one rescaled
by $\mathcal{F}(u)$. The colour scales saturate near the defect.}
    \label{Fig:twist defect}
\end{figure}

The non-vanishing components of the canonical fields are%
\begin{eqnarray}
A_{u} &=&\left( -2\epsilon \sqrt{r}\,\dot{\mathcal{F}}+\frac{\mathcal{F}}{%
\sqrt{r}}\right) \sin \left( \dfrac{\varphi }{2}\right) +\left( 2\epsilon 
\sqrt{r}\,\dot{\mathcal{G}}-\frac{\mathcal{G}}{\sqrt{r}}\right) \cos \left( 
\dfrac{\varphi }{2}\right) \,,  \notag \\
A_{r} &=&-\frac{\epsilon }{\sqrt{r}}\,\left[ \mathcal{F}\sin \left( \dfrac{%
\varphi }{2}\right) -\mathcal{G}\cos \left( \dfrac{\varphi }{2}\right) %
\right] \,,  \notag \\
A_{\varphi } &=&-2\epsilon \sqrt{r}\,\left[ \mathcal{F}\cos \left( \dfrac{%
\varphi }{2}\right) +\mathcal{G}\sin \left( \dfrac{\varphi }{2}\right) %
\right] \,,
\end{eqnarray}%
and%
\begin{eqnarray}
\pi ^{r} &=&\frac{1}{2\sqrt{r}}\,\left[ \rule{0pt}{13pt}\mathcal{F}\sin
\left( \dfrac{\varphi }{2}\right) -\mathcal{G}\cos \left( \dfrac{\varphi }{2}%
\right) \right] \,,  \notag \\
\pi ^{\varphi } &=&-\frac{1}{2r^{3/2}}\,\left[ \rule{0pt}{13pt}\mathcal{F}%
\cos \left( \dfrac{\varphi }{2}\right) +\mathcal{G}\sin \left( \dfrac{%
\varphi }{2}\right) \right] \,.
\end{eqnarray}%
The Hamilton multiplier is $w=\sqrt{r}W$ with $\bar{w}=0$. This solution
satisfies the asymptotic boundary conditions \eqref{rad}. From \eqref{shear},
the shear is $\mathcal{C}=-\epsilon \left[ \mathcal{F}\cos \left( \frac{%
\varphi }{2}\right) +\mathcal{G}\sin \left( \frac{\varphi }{2}\right) \right]
$ and the news $\mathcal{N}=\dot{\mathcal{C}}=W$. The news has the manifestly gauge-invariant form, $W=\lim_{r\to\infty}r^{-1/2}F_{u\varphi}$.

The origin is a true electromagnetic singularity because the Maxwell invariant has the form 
\begin{equation}
F_{\mu \nu }F^{\mu \nu }=-\frac{\mathcal{F}^{2}+\mathcal{G}^{2}}{2r^{3}}-%
\frac{2\epsilon }{r^{2}}\left[ \mathcal{F}\cos \left( \frac{\varphi }{2}%
\right) +\mathcal{G}\sin \left( \frac{\varphi }{2}\right) \right] \left[ 
\dot{\mathcal{F}}\cos \left( \frac{\varphi }{2}\right) +\dot{\mathcal{G}}%
\sin \left( \frac{\varphi }{2}\right) \right] .
\end{equation}%
It diverges as $r^{-3}$ unless $\mathcal{F}=\mathcal{G}=0$, so the
singularity is physical. Note that the invariant $F_{\mu \nu }F^{\mu \nu }$
is $2\pi $-periodic, although $A_{\mu }$ and $F_{\mu \nu }$ are
antiperiodic. 

Rotating the configuration by $\beta $ rotates $(\mathcal{F},\mathcal{G})$
by $\beta /2$, up to moving the cut. Thus $\sqrt{\mathcal{F}^{2}+\mathcal{G}^{2}}$ is the strength of the defect, and $\mathcal{G}/\mathcal{F}$ is its
orientation. If the ratio depends on $u$, the defect rotates in time. 

For further analysis, we take $\mathcal{G}=0$ for simplicity, which fixes the orientation. The Figure~\ref{Fig:twist defect} shows the solution in the plane in the static case $\mathcal{F}=\mathcal{F}_0$ and the wave case $\mathcal{F}=\mathcal{F}_0\sin(\omega u)$. The cut $\varphi=\varphi_0$ is
the half-line where the fields are glued using $C$, that is, where they
change sign. We choose $\varphi_0 =0$. Its position is unobservable:
moving it flips the sign of the fields in a wedge, while $C$-invariant
quantities such as $F_{\mu \nu }F^{\mu \nu }$ and the energy density are
unchanged. Only the holonomy around the defect is physical. 

If an actual boundary is placed on the half-line, the same fields describe a
perfect electric conductor along it, carrying induced surface charges, and
currents when $\mathcal{F}$ depends on $u$. The $\mathcal{G}$-part alone
would satisfy the magnetic conductor condition instead.

Regarding the near-origin falloffs \eqref{r=0Mink}, all momenta and the gauge
field have exactly the required powers, except the radial component of the
gauge field that behaves as $A_{r}\sim r^{-1/2}$, and not as proposed $%
A_{r}\sim r^{3/2}$. However, there is a gauge-equivalent field that has this behavior without changing the asymptotic fields. Let us take $A'=A+
\diff \Lambda $, with 
\begin{equation}
\Lambda =2\epsilon \,\Phi (r)\sqrt{r}\,\mathcal{F}(u)\sin \left( \frac{\varphi }{2}\right) + \frac25 \, r^2\sqrt{r} \tilde A_r (x)\,,
\label{A'}
\end{equation}
where $\Phi =1$ near the origin and $\Phi =0$ near infinity, such that 
\begin{eqnarray}
A'_{u} &=&\dfrac{\mathcal{F}(u)}{\sqrt{r}}\,\sin \left( \frac{\varphi }{2}\right) +\dfrac{2}{5}\,r^{2}\sqrt{r}\,\dot{\tilde{A}}_{r}\,, \qquad
A'_{r} =r\sqrt{r}\,\tilde{A}_{r}\,,  \notag \\
A'_{\varphi } &=&-\epsilon \sqrt{r}\,\mathcal{F}(u)\cos \left( 
\frac{\varphi }{2}\right) +\dfrac{2}{5}\,r^{2}\sqrt{r}\,\partial _{\varphi }
\tilde{A}_{r}\,,
\end{eqnarray}%
reproducing the right near-origin behaviours.

The charges \eqref{Qel} and \eqref{Qshift} can be computed by identifying
the leading coefficients  
\begin{equation}
\pi _{(1)}^{r}=\frac{1}{2}\,\mathcal{F}(u)\,\sin \left( \frac{\varphi }{2}%
\right) \,,\qquad A_{(0)\varphi }=-2\epsilon \,\mathcal{F}(u)\cos \left( 
\frac{\varphi }{2}\right) \,.
\end{equation}%
In the twisted sector the parameters $\varepsilon _{(0)}$ and $\eta _{(0)}$
are antiperiodic, so the integrands of the charges are periodic. The evaluated charges at a generic cut $u$ (as in the paragraph ``Matching and balance laws'') are 
\begin{equation}
Q[\varepsilon ](u)=-\frac{\mathcal{F}(u)}{2}\int\limits_{0}^{2\pi }\mathrm{d}%
\varphi \,\varepsilon _{(0)}\sin \left( \frac{\varphi }{2}\right) \,,\qquad
Q_{S}[\eta ](u)=-2\mathcal{F}(u)\int\limits_{0}^{2\pi }\mathrm{d}\varphi
\,\eta _{(0)}\cos \left( \frac{\varphi }{2}\right) \,.
\end{equation}
Only the modes $n=\pm \frac{1}{2}$ are nonvanishing,\footnote{With $\mathcal G\neq0$, we obtain $Q_{\pm1/2}=\frac{\pi}{2}(\mathcal G\pm \mathrm{i} \mathcal F)$ and $S_{\pm1/2}=-2\pi(\mathcal F\mp \mathrm{i}\mathcal G)$. The two modes then describe the defect directly -- their moduli give its strength, $\sqrt{\mathcal F^2+\mathcal G^2}$, and their phases give its orientation.} 
\begin{equation}
Q_{\pm 1/2}=\pm \frac{\mathrm{i}\pi }{2}\,\mathcal{F}(u)\,,\qquad S_{\pm
1/2}=-2\pi \,\mathcal{F}(u)\,.
\end{equation}%
There are no global charges, since the twisted sector has no Fourier zero mode.  The
angle-dependent charges are nevertheless nontrivial. For example, 
\begin{equation}
Q\left[ \varepsilon _{(0)}=\sin \left( \frac{\varphi }{2}\right) \right] =-%
\frac{\pi }{2}\,\mathcal{F}(u)\,,\qquad Q_{S}\left[ \eta _{(0)}=\cos \left( 
\frac{\varphi }{2}\right) \right] =-2\pi \,\mathcal{F}(u)\,.
\end{equation}
Their $u$-dependence is driven by $\dot{\mathcal{F}}$, that is, by the news $W$, in agreement with the balance laws \eqref{Qbalance} and \eqref{dotQs}.

The inner surface terms of Section~\ref{Ch:Minkowski} can also be evaluated.
For this solution, $\sqrt{r}\,\pi^r$ and $A_\varphi/\sqrt{r}$ are
independent of $r$. Hence, for parameters with the profiles $\varepsilon=\sqrt{r}\,\varepsilon_{(0)}$ and $\eta=\sqrt{r}\,\eta_{(0)}$ at all $r$, the
inner and outer terms coincide, $Q_{S,r\to0}[\eta]=Q_{S,r\to\infty}[\eta]$,
and similarly for $Q[\varepsilon]$, so the improved generators vanish on
shell. The charges at infinity thus measure the data carried by the defect,
just as the flux of a Coulomb field at infinity measures the charge of its
source. Since $Q_S$ is not gauge invariant, its inner value changes under
gauge transformations that do not vanish at the origin, such as \eqref{A'}.
Its gauge-invariant part, determined by the dual scalar $\sqrt{r}\,\phi=-
\mathcal{F}\cos\left(\frac{\varphi}{2}\right)$, is the
same at both ends.

The energy can be computed using the energy-momentum tensor 
\begin{equation}
T_{\mu \nu }=F_{\mu \alpha }F_{\nu }{}^{\alpha }-\tfrac{1}{4}\,g_{\mu \nu
}F^{2}=\nabla _{\mu }\phi \nabla _{\nu }\phi -\tfrac{1}{2}\,g_{\mu \nu
}\,(\nabla \phi )^{2}\,,  \label{T}
\end{equation}%
with $F_{\mu \nu }=\sqrt{g}\varepsilon _{\mu \nu \rho }\nabla ^{\rho }\phi $
and $\varepsilon _{ur\varphi }=1$. In the static case$\,\mathcal{F}=\mathcal{%
F}_{0}=\mathrm{const}$, we have in the inertial coordinates%
\begin{equation}
A=\frac{\mathcal{F}_{0}}{\sqrt{r}}\,\sin \left( \dfrac{\varphi }{2}\right) \,%
\diff t+\diff \left[ -4\epsilon \mathcal{F}_{0}\sqrt{r}\,\sin \left( 
\dfrac{\varphi }{2}\right) \right] \,,  
\end{equation}%
and a static electric field satisfies $|\vec{E}|^{2}=\frac{\mathcal{F}%
_{0}^{2}}{4r^{3}}$. The field energy for $r>r_{0}$ is 
\begin{equation}
E(r>r_{0})=\int\limits_{r_{0}}^{\infty }\diff r\,r\int\limits_{0}^{2\pi }%
\diff \varphi \,\tfrac{1}{2}\,|\vec{E}|^{2}=\frac{\pi \mathcal{F}_{0}^{2}%
}{4r_{0}}\,.  
\end{equation}
Thus, it diverges like $1/r_{0}$, exactly as the Coulomb self-energy of a
point charge in $3+1$ dimensions. The magnetic field is $
B(t,r,\varphi )=\frac{\mathcal{\dot{F}}(u)}{\sqrt{r}}\,\cos \left( \frac{
\varphi }{2}\right) $ and it vanishes in the static limit. 

The energy flux, or the power at infinity, is obtained from $T_{uu}$.
Using \eqref{T}, it is satisfied $rT_{uu}\rightarrow W^{2}$ as $r\rightarrow
\infty $, thus%
\begin{equation}
P(u)=\oint\limits_{\mathbb{S}_{\infty }^{1}}r\,T_{uu}\,\diff \varphi
=\int\limits_{0}^{2\pi }\diff \varphi \,W^{2}=\pi \,\mathcal{\dot{F}}^{2},  
\end{equation}
or $\pi (\mathcal{\dot{F}}^{2}+\dot{\mathcal{G}}^{2})$ for the more general
solution. For $\epsilon =+1$, $P$ is radiated through $\mathscr{I}^{+}$
while nothing comes in from $\mathscr{I}^{-}$, so the energy is supplied by
the worldline $r=0$. For $\epsilon =-1$ the same power enters from $%
\mathscr{I}^{-}$ and is absorbed at $r=0$. 

Keeping both branches, for instance $\hat{\phi}=f(t-r)-f(t+r)$, gives a
standing wave with finite energy near the origin, in which radiation coming
from $\mathscr I^{-}$ is re-emitted to $\mathscr I^{+}$. The purely retarded
solution instead requires the defect to supply the radiated power.

The obtained twisted sector possesses electromagnetic memory. From \eqref{Wexact}, the shear amplitude $\mathcal{F}$ is fixed by the news $W$ only up to a constant: $\mathcal{F}(u)=\mathcal{F}(-\infty )-\epsilon \int_{-\infty }^{u}\hat{W}(u')\,\diff u'$, where $\hat{W}$ is the amplitude in $W$ multiplying $\cos\left(\frac \varphi 2\right)$. Therefore, $W=0$ does not imply a vanishing field. Namely, it leaves the static field $\mathcal{F}=\mathcal{F}_{0}$,
unless $\mathcal{F}(-\infty )=0$. After a burst of news with $\int \hat{W}\,
\diff u\neq 0$, a static twisted field with $\Delta \mathcal{F}
=-\epsilon \int \hat{W}\,\diff u$ remains, describing an electromagnetic
memory effect in the twisted sector. This is exactly the kick memory $\Delta 
\mathcal{C}=\int \mathcal{N}\,\diff u$ previously discussed. The shift 
\eqref{trShift} with $\eta _{(0)}=v\cos \left( \frac{\varphi }{2}\right) $ maps this solution to the one with $\mathcal{F} \rightarrow \mathcal{F}-\epsilon v$, up to a gauge transformation. 
 For $u$-independent $v$, it leaves the news invariant and shifts the static twisted field, that is, the memory.

The obtained solution is the three-dimensional counterpart of the retarded field of a time-dependent scalar point source in four dimensions, $\psi =\frac{1}{r}\,q(t-r)$. Indeed, for $\phi =\frac{1}{\sqrt{r}}\,\hat{\phi}\,\mathrm{e}^{%
\mathrm{i}k\varphi }$ in three dimensions and $\psi =\frac{1}{r}\,\hat{\psi}%
\,Y_{\ell m}$ in four, the radial equations coincide when $|k|=\ell +\frac{1%
}{2}$, so the half-angle modes $k=\pm \frac{1}{2}$ correspond to $\ell =0$.
Accordingly, both solutions propagate without tails, radiate a power
quadratic in $\dot{\mathcal{F}}$ (respectively $\dot{q}$), have a
self-energy diverging as $1/r_{0}$, and exhibit a memory effect. In
four-dimensional Maxwell theory, monopole radiation is forbidden by charge
conservation. In the twisted sector there is no zero mode and no conserved $\mathrm U(1)$ charge, so the lowest mode radiates freely. There is also a geometric counterpart. Uplifted to a $z$-independent field
in four dimensions, the twist defect becomes a straight string around
which the fields return charge-conjugated, that is, an Alice
string~\cite{Schwarz:1982ec,Alford:1990mk}. The solution then describes
the field radiated by such a string.  In the half-angle sector this
radiation is free of tails, unlike cylindrical waves from an ordinary line
source. The static limit
reproduces the classical edge behaviour of fields near a conducting half-plane~\cite{Sommerfeld:1896,Meixner:1972}. To our knowledge, the time-dependent solution
has not been discussed as a twisted sector of three-dimensional Maxwell
theory or in the context of null-boundary charges. Topological $\mathbb{Z}_{2}$-defects in free field theories in four dimensions were discussed recently in \cite{Bianchi:2021snj}.

\subparagraph{Comments on the BTZ and de Sitter backgrounds.}

Twisted sectors are not specific to flat space, but their realization
depends on the background. For the metrics (3.6) with lapse $N^{2}$, the
ansatz $\phi =\frac{1}{\sqrt{r}}\,\hat{\phi}(u,r)\,\mathrm{e}^{\mathrm{i}k\varphi }$
gives instead of \eqref{waveeq}
\begin{equation}
2\epsilon \,\partial _{u}\partial _{r}\hat{\phi}=\partial _{r}(N^{2}\partial
_{r}\hat{\phi})-U\hat{\phi}\,,\qquad
U=\frac{(N^{2})'}{2r}+\frac{k^{2}-\frac{1}{4}\,N^{2}}{r^{2}}\,.
\end{equation}
For $k=\pm \frac{1}{2}$, the effective radial potential $N^2U$ vanishes in Minkowski space only, so the closed solution with arbitrary $\mathcal{F}(u)$ is special to flat space. In the dS static patch, $U=-\frac{3}{4\ell ^{2}}$, and the half-angle modes are Legendre functions of degree $\frac{1}{2}$ in $\frac{r}{\ell}$. For example, the
static twisted field regular at the cosmological horizon is 
\begin{equation}
    \phi =-\mathcal{F}_{0} \,\frac{P_{1/2}(\frac r\ell )}{\sqrt{r}\,P_{1/2}(0)} \, \cos \left( \frac{\varphi}{2}\right)\,.
\end{equation}
It reduces to the Minkowski solution for $r\ll \ell $ and is
singular at the observer's worldline. In global dS space, a loop around one pole of the spatial sphere also encircles the antipodal one, so the twist requires a pair of defects, on the worldlines of two antipodal
static observers. In the BTZ black hole, instead, the angular circle is not
contractible \cite{Banados:1992gq}. The $C$-twisted sector is then a smooth
holonomy sector around the black hole, which requires no defect, in analogy
with the choice of periodic or antiperiodic spinors on BTZ. The radial
equation is hypergeometric, with $k\in \mathbb{Z}+\frac{1}{2}$. Near the horizon the effective potential $N^{2}U$ vanishes, so the modes propagate
freely and $W$ plays the role of the news, while near the AdS boundary the
potential grows and the standard reflecting conditions remove the leading radiative coefficient, as in the untwisted case.

\section{Conclusions }

We have analyzed Maxwell theory in three spacetime dimensions in the radiative sector using a Hamiltonian formulation adapted to null foliations~\cite{Gonzalez:2023yrz,Gonzalez:2024rho,Dordevic:2026dx}. The main result is that null boundaries support two independent quasilocal charges, or edge observables. The first is the electric charge $Q[\varepsilon]$, associated with large $\mathrm U(1)$ gauge transformations. The second, $Q_{S}[\eta ]$, arises from the zero mode sector of a constraint characteristic of the null foliation. Its parameter has a
fixed radial profile, with $\eta/\sqrt{r}$ constant along the null
generators, and the transformation it generates shifts the angular component
of the gauge field. We refer to it as the magnetic charge since, for
constant $\eta_{(0)}$ in the untwisted sector, it measures the holonomy of
the leading boundary connection.

In Minkowski spacetime, we found radiative boundary conditions for which the Hamiltonian is differentiable and both charges are finite and nonvanishing. These conditions
are weaker than the falloffs of~\cite{Shimizu:2025hfl}, which set the
leading component of $A_u$ to zero  and require the faster falloff $A_r=\mathcal{O}(r^{-3/2})$. The
latter excludes large gauge transformations and eliminates the electric
charge, whereas keeping the non-radiative coefficient $A_{(1)r}$ allows
them. With the relaxed conditions, $Q$ and $Q_S$ are both nontrivial, and
their mixed Dirac bracket contains a central extension. In Fourier modes, the boundary observables form two independent Abelian Kac--Moody algebras with opposite levels $\pm\kappa$. Antipodal matching relates the
endpoint charges at $\mathscr I_-^+$ and $\mathscr I_+^-$, while their evolution along each null boundary is governed by a balance law.

The same Hamiltonian mechanism occurs at finite distance null boundaries. On a fixed BTZ background, the black hole horizon supports the same pair of
charges and the same centrally extended algebra, and so does the
cosmological horizon of the dS static patch. The two Kac--Moody algebras
have the same universal levels $\pm \kappa $, with $\kappa =4\pi $, in all
three cases. The geometry enters only through the relative normalization of
the electric and magnetic modes, via the horizon radius. Since the geometry
is treated in the probe approximation, $r_{+}$ and $M$ are fixed parameters
rather than dynamical variables. The difference between the two finite distance cases is geometric: the BTZ horizon is an inner boundary of the exterior region, whereas the cosmological horizon is the outer boundary of the static patch. Edge modes on stretched dS horizons have also been studied from a different viewpoint in~\cite{Ball:2024hqe}.

These examples show that the additional soft sector is not tied to null infinity, but originates from the zero mode structure of the canonical
formulation on null hypersurfaces~\cite{Dordevic:2026dx}. On a null surface, the
characteristic constraint fixes the radial dependence of the multiplier $w$
up to an integration function $W(u,\varphi)$, which is constant along each
null generator. This undetermined part is the zero mode behind the second
charge, and on shell it coincides with the news. Since it is constant along
the generators, it cannot be confined near one end of the radial domain,
and it can be prescribed at one end only, namely the one where the charges
are measured. This is a familiar feature of field theories on light
fronts~\cite{Maskawa:1975hx,Heinzl:1998kz}, and it makes the canonical description naturally
quasilocal~\cite{Chandrasekaran:2018aop}. In our examples, the charges are
measured at null infinity, at the black hole horizon and at the cosmological horizon. The other end is the origin in Minkowski and de
Sitter space, and the AdS boundary in BTZ. Regularity at the origin, or
reflecting conditions at the AdS boundary, would restrict the zero mode,
so the phase spaces $\Gamma_\infty$, $\Gamma_{\mathrm{H}}$ and $\Gamma_c$
leave that end open. At finite distance,
these edge degrees of freedom are associated with the chosen subregion
rather than interpreted as additional local degrees of freedom of the global spacetime. Gluing complementary regions relates their boundary data through
matching conditions and Gauss's law. 

The analysis also reveals a role for global structure. Since charge
conjugation is a symmetry of pure Maxwell theory, the phase space contains,
besides the ordinary sector, a $C$-twisted sector in which all fields,
multipliers and symmetry parameters change sign around the origin or the black hole, depending on the background metric. The
topological defect lies in the space of field configurations, not in the
geometry, in
analogy with twist defects for electromagnetic duality~\cite{Shao:2025qvf} and with Alice strings~\cite{Schwarz:1982ec,Alford:1990mk}. In this sector the angular modes are half-integer, and the edge algebra becomes the twisted version of the same
Kac--Moody algebra, with the same levels and no zero modes. How the twist is realized depends on the global geometry. In Minkowski space a single defect sits at the origin. In global de Sitter space a pair of defects is required, on the worldlines of two antipodal static observers, each enclosed by its own cosmological horizon. In the BTZ black hole the angular circle is not contractible, so the twist is a smooth holonomy around the black hole and
needs no defect.

The twisted sector is not empty. In Minkowski space we found the general
purely retarded or advanced solution with half-angle dependence, which
contains two arbitrary functions $\mathcal{F}(u)$ and $\mathcal{G}(u)$.
The solution is the three-dimensional
counterpart of the retarded field of a time-dependent scalar point source in four dimensions, since the half-angle modes obey the same radial equation as
the four-dimensional monopole. For half-angle parameters, the magnetic shift
maps such solutions into each other exactly.

We have also clarified the physical status of the charges. Both are finite,
differentiable functionals, invariant under proper gauge transformations,
and hence well-defined edge observables on the reduced phase space. Their
Hamiltonian flows preserve the symplectic form, the constraints and the
boundary conditions. They do not necessarily commute with the Hamiltonian. Like
angular momentum, which generates rotations and is a useful observable even
when the Hamiltonian is not rotationally invariant, they evolve according to
balance laws driven by the boundary flux. Their centrally extended algebra
is a kinematical property of the Poisson structure, which makes $Q$ and $Q_S$
canonically conjugate. The charges are tied to the radiative data at the boundary: the electromagnetic shear $\mathcal C$, the news $\mathcal N=\dot{\mathcal C}$, and the electric charge aspect $\mathcal{Q}
=\epsilon\,\partial_\varphi \mathcal C$. The magnetic shift changes the
time-independent part of the shear while preserving the news, as
supertranslations act on the gravitational shear, and the change of the
shear between two cuts is the electromagnetic kick memory.

Another common feature is the role of the zero mode multiplier $W$. On shell it coincides with the news, so the undetermined part of the multiplier carries the radiative data, and it acts as a boundary source in the balance
laws. Through the central extension, it generates the balance
equation for the electric charge. In Fourier modes, the global electric charge vanishes
in the untwisted sector and is absent in the twisted one, whereas the higher electric modes are sourced by the
corresponding angular modes of $W$. This is structurally similar to familiar flux-balance laws at null infinity, such as the Bondi mass-loss equation~\cite{Sachs:1962wk,Strominger:2017zoo,Flanagan:2015pxa}, with $W$ playing at the same time the role of a boundary multiplier and of the news.  A related finite region example can be found in Maxwell--Chern--Simons theory on a disk, where edge observables form a $\mathrm U(1)$ Kac--Moody algebra and their conservation depends on the spatial boundary sources~\cite{GBarbero:2022gnx}. When the exterior medium is a superconductor, the parameter appearing in the boundary terms in~\cite{GBarbero:2022gnx} can be related to the penetration depth of the fields into the medium. In our case, the non-conservation of the angle-dependent electric edge observables has an analogous flux interpretation, being driven by the boundary source across the horizon.

For constant $\eta_{(0)}$, the magnetic charge is related to the magnetic
flux through the holonomy of the leading boundary connection $A_{(0)}$. It is not conserved under the full bulk Hamiltonian evolution, which also involves transverse derivatives and flux terms that are not intrinsic to the null surface. It would therefore be interesting to determine whether an effective
intrinsic dynamics can be constructed in which the magnetic shift is a
symmetry and leads to a conserved charge, in particular at horizons. Such a construction would require separating the
tangential and transverse components of the bulk equations and specifying
the transverse derivatives through appropriate boundary data. 
A related decomposition has been carried out for asymptotically flat three-dimensional Einstein--Maxwell theory in~\cite{Bosma:2023sxn}, with both radiative and Coulombic contributions. Another approach uses a double-null $2+2$ decomposition, as in four-dimensional
Einstein--Maxwell--scalar theory~\cite{Madler:2025ibn}. In four-dimensional gravity, on the other hand,  a $3+1$ decomposition of the bulk spacetime combined with the
induced $2+1$ geometry of the horizon was developed in~\cite{Gourgoulhon:2005ng}.

The zero mode mechanism should also hold beyond the backgrounds considered here whenever the null constraint structure admits zero modes and boundary
conditions can be chosen such that the corresponding charges are finite and integrable. A natural example is the rotating BTZ black hole. In co-rotating
EF coordinates, one expects the same mechanism at the outer horizon, with the normalization of the edge observables depending on the angular momentum only through the horizon radius $r_+(M,J)$.  The extremal case requires an independent analysis, since $N^2$ develops a double zero at the horizon and the near-horizon expansion and regularity conditions change qualitatively.

The boundary charge algebras found here may also be related
to centrally extended bulk current algebras associated with
generalized global symmetries~\cite{Gaiotto:2014kfa}.
Such algebras have been studied in four-dimensional Maxwell
theory~\cite{Hofman:2018lfz,Hofman:2024oze}, in scalar theories
with mixed anomalies~\cite{Vitouladitis:2025zoy}, and more generally in $p$-form gauge theories~\cite{Arbalestrier:2026lna}.
Connections between higher-form and asymptotic symmetries
have also been discussed in~\cite{Lake:2018dqm}. Maxwell--scalar duality in three dimensions could help determine
whether the electric and magnetic charges found here provide
an edge realization of these bulk current algebras at
asymptotic and finite-distance null boundaries.

A further question concerns memory effects and soft theorems. The relation of the magnetic shift to memory is already visible here: it changes the time-independent part of the shear, whose change between two cuts is the
electromagnetic kick memory, and the exact solution exhibits a memory effect in the twisted sector. At null infinity, the relation between large gauge symmetries, soft photon theorems and electromagnetic
memory is well established~\cite{Strominger:2017zoo,He:2014cra,Kapec:2014zla}. At finite distance, Maxwell edge modes have been related to an operational quasilocal memory effect in~\cite{Araujo-Regado:2024dpr}. More recently, a finite distance generalization of gravitational displacement memory has been formulated on null Carrollian screens~\cite{Diaz:2026igh}. These
results suggest that the relation of the magnetic charge to
soft theorems, and its quasilocal counterpart at horizons, deserve further investigation.

Recent work in gravity has also established a direct relation between the
infrared phase space at null infinity and the phase space of finite causal
diamonds or finite cuts of null hypersurfaces~\cite{Ciambelli:2025fbo,Ciambelli:2026pwi}. In these constructions, the soft and Goldstone data are
encoded in fluctuations of the finite surface and their canonical partners.
Our probe field analysis keeps the background geometry fixed and does not introduce analogous geometric edge variables. One could also investigate whether a similar phase space map exists for the electromagnetic zero mode sector.

Finally, the present analysis can be extended by including backreaction and by considering non-Abelian gauge theories or gravity. The three examples
studied here indicate that zero modes characteristic of null foliations provide a robust mechanism for generating quasilocal edge observables and
centrally extended current algebras at null boundaries, and that
charge-conjugation twisted sectors, with half-integer modes, enrich this
structure.

\section*{Acknowledgments}

We thank Felipe Diaz, Oscar Fuentealba, Ayan Mukhopadhyay and Radouane Gannouji for their valuable comments and suggestions, which helped improve this manuscript. We also thank the anonymous referee for useful questions that helped us substantially improve this manuscript.
We are grateful to the seminar audiences at the Yukawa Institute for Theoretical Physics, the Yau Mathematical Sciences Center, and the Center for Gravitation and Cosmology at Yangzhou University, where preliminary versions of this work were presented, for their helpful questions and observations. This work was partially funded by FONDECYT Regular Grants No.~1230492 and No.~1231779. The work of D.D.~was supported by the funding provided by the Faculty of Physics, University of Belgrade, through grant number 451-03-137/2025-03/200162 by the Ministry of Science, Technological Development and Innovations of the Republic of Serbia. T.V.~acknowledges
financial support from CIC, Universidad Michoacana de San Nicol\'{a}s de Hidalgo, Mexico. A.M.~acknowledges financial support from Pontificia Universidad Cat\'{o}lica de Valpara\'{\i}so through the PAIM scholarship, and thanks Universidad Michoacana de San Nicol\'{a}s de Hidalgo for its hospitality during the part of this work.

\subparagraph{Use of AI tools.} During the revision of this work in response to the referee report, Claude Opus~5.5 (Anthropic) and ChatGPT Astra 6 (OpenAI) were used to refine the physical interpretation, cross-check numerical factors, and prepare the code for Fig.~\ref{Fig:twist defect}. Claude Opus~5.5 also helped obtain and generalize the exact solution in Section~\ref{Ch:Exact}. All results were verified independently by the authors, who take full responsibility for the content of this paper.

\appendix
\section{Symplectic derivation of the charge algebra}
\label{Appendix}

We derive the quasilocal electromagnetic charges and their algebra on the Minkowski background using the symplectic formulation. 
The covariant phase space consists of solutions of the field equations, so the constraints are satisfied by construction. The algebra of
Hamiltonian charges can then be computed directly by contracting the presymplectic form with their associated Hamiltonian vector fields. This method does not require construction of the reduced phase space and Dirac brackets.

Varying the Maxwell Lagrangian gives the symplectic potential current density $\theta ^{\mu } (A,\delta A) $, which is a one-form in field space. Its exterior derivative in field space defines the symplectic current density two-form, $\omega ^{\mu }$. Explicitly, they are given by
\begin{eqnarray}
\theta ^{\mu } &=& -rF^{\mu \nu }\delta A_{\nu }\,,\notag \\
\omega ^{\mu } &=& \delta \theta ^{\mu }=-\delta (rF^{\mu \nu })\wedge \delta A_{\nu }\,.
\end{eqnarray}

On a null hypersurface $\Xi $ of constant $u$, the presymplectic potential one-form and the canonical symplectic two-form are, respectively,
\begin{eqnarray}
\Theta  &=&\int\diff ^{2}x\,\theta ^{u}=\int \diff ^{2}x\, \pi ^{\mu }\delta A_{\mu } \,,  \notag \\
\mathit{\Omega } &=&\delta \Theta =\int \diff ^{2}x\,\delta \pi ^{\mu
}\wedge \delta A_{\mu }\,.
\end{eqnarray}
For the boundary conditions \eqref{rad}, the presymplectic potential has a logarithmic divergence. For large $r$ and a fixed reference length $r_0$, the divergent part of $\Theta $ is 
\begin{equation}
\Theta ^{\mathrm{div}}=\oint\limits_{\mathbb{S}_{\infty }^{1}}\diff \varphi \,\log \left( \frac{r}{r_0}\right) \left( \pi _{(2)}^{u}\delta 
A_{(0)u}+\pi _{(1)}^{r}\delta A_{(1)r}+\pi _{(2)}^{\varphi }\delta A_{(0)\varphi }\right) .  
\end{equation}
The first term vanishes because of the Dirichlet boundary conditions on $A_{(0)u}$, as given by \eqref{Au fijo}. The remaining terms can be transformed using the leading order of the
characteristic and Gauss constraints, 
\begin{equation}
\chi _{(2)}=\pi _{(2)}^{\varphi }-\frac{\epsilon }{2}\,A_{(0)\varphi
}+\epsilon \partial _{\varphi }A_{(1)r}\approx 0\,,\qquad \psi _{(2)}=-\frac 12\,\pi _{(1)}^{r}+\partial _{\varphi }\pi _{(2)}^{\varphi }\approx 0\,.
\label{(0)constr}
\end{equation}
Substituting the Gauss constraint and integrating by parts in the angle, we
obtain%
\begin{equation}
\oint \diff \varphi \,\left( \pi _{(1)}^{r}\delta A_{(1)r}+\pi
_{(2)}^{\varphi }\delta A_{(0)\varphi }\right) =\oint \diff \varphi
\,\pi _{(2)}^{\varphi }\,\delta \left( A_{(0)\varphi }-2\partial _{\varphi
}A_{(1)r}\right) \,.
\end{equation}%
Then we substitute the characteristic constraint, and we get%
\begin{eqnarray}
\oint \diff \varphi \,\pi _{(2)}^{\varphi }\,\delta \left( A_{(0)\varphi
}-2\partial _{\varphi }A_{(1)r}\right)  &=&\frac{\epsilon }{2}\oint \mathrm{d%
}\varphi \,\left( A_{(0)\varphi }-2\partial _{\varphi }A_{(1)r}\right)
\delta \left( A_{(0)\varphi }-2\partial _{\varphi }A_{(1)r}\right)   \notag
\\
&=&\frac{\epsilon }{4}\oint \diff \varphi \,\delta \left( A_{(0)\varphi
}-2\partial _{\varphi }A_{(1)r}\right) ^{2}\,.
\end{eqnarray}
Consequently, on the constraint surface, 
\begin{equation}
\Theta ^{\mathrm{div}}=\delta  \oint \diff \varphi \,\frac{\epsilon }{4}\log \left( \frac{r}{r_0}\right) \,\left( A_{(0)\varphi
}-2\partial _{\varphi }A_{(1)r}\right) ^{2} \quad \Rightarrow \quad \delta \Theta ^{\mathrm{div}}=0\,.  
\end{equation}
The logarithmic term therefore drops out of $\mathit{\Omega}$. The remaining terms are $\mathcal{O}(r^{-2})$, so the radial integral converges and leads to a finite symplectic form $\mathit{\Omega}$.

Near the origin, with the falloff of the fields \eqref{r=0Mink} that satisfy \eqref{cancell}, analogous term $r^{-1}\tilde{\pi}^{\varphi
}\delta \tilde{A}_{\varphi }$ in $\Theta$ equals $r^{-1}\frac{\epsilon }{4}\,\delta (\tilde{A}_{\varphi }^{2})$ on the constraint surface, by $\tilde{\pi}%
^{\varphi }=\frac{\epsilon }{2}\tilde{A}_{\varphi }$, and is therefore also
exact. Thus, the symplectic form $\mathit{\Omega}$ is finite also in the limit $r\to 0$.\medskip

The Hamiltonian vector fields $X$ that preserve the symplectic structure are given by transformations \eqref{trU(1)}, \eqref{trShift} and \eqref{dAu},
\begin{eqnarray}
X_{\varepsilon } &=&-\int \diff ^{2}x\,\partial _{\mu }\varepsilon \,
\frac{\delta }{\delta A_{\mu }}\,,  \notag \\
X_{\eta } &=&\int \diff ^{2}x\,\left[ \frac{2\epsilon}{r}\, \partial_\varphi\eta \,\frac{\delta }{\delta
A_{u}}+\eta \,\frac{\delta }{\delta
A_{\varphi }}+\frac{\epsilon }{r}\,\partial _{\varphi }\eta \,\frac{\delta }{\delta \pi ^{r}}-\epsilon \partial _{r}\left( \frac{\eta }{r}\right) \,\frac{\delta }{\delta \pi ^{\varphi }}\right] \,.
\end{eqnarray}
Indeed, using Cartan's identity $\mathscr L _{X}=i_{X}\delta +\delta i_{X}$, with the contractions of $\mathit \Omega$ given by
\begin{eqnarray}
i_{X_{\varepsilon }}\mathit{\Omega}  &=&\int \diff ^{2}x\,\left( \partial
_{i}\varepsilon \,\delta \pi ^{i}+\dot{\varepsilon}\,\delta \pi ^{u}\right)
\,,  \notag \\
i_{X_{\eta }}\mathit{\Omega}   &=&\int \diff ^{2}x\,\left[ \frac{\epsilon }{r}
\,\partial _{\varphi }\eta \,\delta A_{r}-\epsilon \partial _{r}\left( \frac{\eta }{r}\right) \,\delta A_{\varphi } -\frac{2\epsilon }{r}\,\partial _{\varphi }\eta \,\delta \pi ^{u}
-\eta \,\delta \pi ^{\varphi }\right]
\,,
\end{eqnarray}
it follows that these transformations preserve the symplectic form, 
\begin{equation}
\mathscr L _{X_{\varepsilon }}\mathit{\Omega }=0\,,\qquad \mathscr L _{X_{\eta }}
\mathit{\Omega }=0\,.
\end{equation}
The charge $Q$ can, therefore, be obtained as
\begin{equation}
\mathscr L _{X}\mathit{\Omega }=\delta i_{X}\mathit{\Omega }=0\quad \Rightarrow
\quad i_{X}\mathit{\Omega }=-\delta Q\,,
\end{equation}
because $i_{X}\mathit{\Omega }$ is locally exact. On the covariant phase space, the bulk parts of the improved generators vanish, and the obtained surface contributions reproduce the charge variations. Explicitly, we find
\begin{eqnarray}
i_{X_{\varepsilon }}\mathit{\Omega}   &=&\oint \diff \varphi \,\varepsilon \,\delta \pi ^{r}=-\delta Q[\varepsilon ]\,,  \notag \\
i_{X_{\eta }}\mathit{\Omega}   &=&-\epsilon \oint \diff \varphi \,\frac{\eta }{r}\,\delta A_{\varphi }=-\delta Q_{S}[\eta ]\,,
\end{eqnarray}
and, for the field-independent parameters, we recover the charges \eqref{Qel} and \eqref{Qshift'}.

The bracket of these charges is computed through a double contraction 
\begin{equation}
\{Q_{1},Q_{2}\} =i_{X_{1}}i_{X_{2}} \mathit{\Omega}  =\int\mathrm d^2x\,
\left[
(i_{X_2}\delta\pi^\mu)(i_{X_1}\delta A_\mu)
-(i_{X_1}\delta\pi^\mu)(i_{X_2}\delta A_\mu)
\right] \,,
\end{equation}
where $
i_X\delta A_\mu=\delta_X A_\mu$. 
It is straightforward to show that the electric-electric contraction vanishes because $X_{\varepsilon }(\pi ^{\mu })=0$ and, for two shifts, it vanishes because $\eta _{[1}\partial _{r}\left( \eta
_{2]}/r\right) =0$. The mixed contraction is
\begin{equation}
i_{X_{\varepsilon }}i_{X_{\eta }}\mathit{\Omega}  =-\epsilon \int\diff ^{2}x\,\partial _{r}\left( \frac{\varepsilon }{r}\,\partial _{\varphi
}\eta \right) \,,
\end{equation}
leading to the known central charge \eqref{C}.
A similar construction, without the $A_u$ component of $X_\eta$, can be performed on the BTZ and dS backgrounds.

\printbibliography

@article{GBarbero:2022gnx,
    author = "G. Barbero, J. F. and D\'\i{}az, B. and Margalef-Bentabol, J. and Villase\~nor, E. J. S.",
    title = "{Edge observables of the Maxwell-Chern-Simons theory}",
    eprint = "2204.06073",
    archivePrefix = "arXiv",
    primaryClass = "hep-th",
    journal = "Phys. Rev. D",
    volume = "106",
    number = "2",
    pages = "025011",
    year = "2022"
}

@article{Shimizu:2025hfl,
    author = "Shimizu, K. and Sugishita, S.",
    title = "{Asymptotic symmetry and confinement in three-dimensional QED}",
    eprint = "2503.20173",
    archivePrefix = "arXiv",
    primaryClass = "hep-th",
    reportNumber = "KUNS-3041",
    doi = "10.1103/x92g-9w2h",
    journal = "Phys. Rev. D",
    volume = "112",
    number = "10",
    pages = "105001",
    year = "2025"
}

@article{Bosma:2023sxn,
    author = "Bosma, J. and Geiller, M. and Majumdar, S. and Oblak, B.",
    title = "{Radiative asymptotic symmetries of 3D Einstein-Maxwell theory}",
    eprint = "2311.09156",
    archivePrefix = "arXiv",
    primaryClass = "hep-th",
    journal = "SciPost Phys.",
    volume = "16",
    number = "4",
    pages = "092",
    year = "2024"
}

@article{Ball:2024hqe,
    author = "Ball, A. and Law, Y. T. A. and Wong, G.",
    title = "{Dynamical edge modes and entanglement in Maxwell theory}",
    eprint = "2403.14542",
    archivePrefix = "arXiv",
    primaryClass = "hep-th",
    journal = "JHEP",
    volume = "09",
    pages = "032",
    year = "2024"
}

@article{Regge:1974zd,
    author = "Regge, T. and Teitelboim, C.",
    title = "{Role of Surface Integrals in the Hamiltonian Formulation of General Relativity}",
    reportNumber = "Print-74-0988 (IAS,PRINCETON)",
    journal = "Annals Phys.",
    volume = "88",
    pages = "286",
    year = "1974"
}

@article{Geiller:2017xad,
    author = "Geiller, M.",
    title = "{Edge modes and corner ambiguities in 3d Chern\textendash{}Simons theory and gravity}",
    eprint = "1703.04748",
    archivePrefix = "arXiv",
    primaryClass = "gr-qc",
    journal = "Nucl. Phys. B",
    volume = "924",
    pages = "312--365",
    year = "2017"
}

@article{Donnelly:2016auv,
    author = "Donnelly, W. and Freidel, L.",
    title = "{Local subsystems in gauge theory and gravity}",
    eprint = "1601.04744",
    archivePrefix = "arXiv",
    primaryClass = "hep-th",
    journal = "JHEP",
    volume = "09",
    pages = "102",
    year = "2016"
}

@book{Strominger:2017zoo,
    author = "Strominger, A.",
    title = "{Lectures on the Infrared Structure of Gravity and Gauge Theory}",
    eprint = "1703.05448",
    archivePrefix = "arXiv",
    primaryClass = "hep-th",
    isbn = "978-0-691-17973-5",
    month = "3",
    year = "2017"
}

@article{Gonzalez:2023yrz,
    author = "Gonz\'alez, H. A. and Labrin, O. and Miskovic, O.",
    title = "{Kac-Moody symmetry in the light front of gauge theories}",
    eprint = "2304.03211",
    archivePrefix = "arXiv",
    primaryClass = "hep-th",
    reportNumber = "PUCV-UAI-23/03",    
    journal = "JHEP",
    volume = "06",
    pages = "165",
    year = "2023"
}

@article{Gonzalez:2024rho,
    author = "Gonz\'alez, H. A. and Labrin, O. and Miskovic, O.",
    title = "{Asymptotic structure of scalar-Maxwell theory at the null boundary}",
    eprint = "2407.13866",
    archivePrefix = "arXiv",
    primaryClass = "hep-th",
    journal = "Phys. Rev. D",
    volume = "111",
    number = "2",
    pages = "025011",
    year = "2025"
}

@article{Castellani:1981us,
    author = "Castellani, L.",
    title = "{Symmetries in Constrained Hamiltonian Systems}",
    reportNumber = "ITP-SB-81-5",
    journal = "Annals Phys.",
    volume = "143",
    pages = "357",
    year = "1982"
}

@article{Brown:1986nw,
    author = "Brown, J. D. and Henneaux, M.",
    title = "{Central Charges in the Canonical Realization of Asymptotic Symmetries: An Example from Three-Dimensional Gravity}",
    journal = "Commun. Math. Phys.",
    volume = "104",
    pages = "207--226",
    year = "1986"
}

@article{Geiller:2017whh,
    author = "Geiller, M.",
    title = "{Lorentz-diffeomorphism edge modes in 3d gravity}",
    eprint = "1712.05269",
    archivePrefix = "arXiv",
    primaryClass = "gr-qc",
    journal = "JHEP",
    volume = "02",
    pages = "029",
    year = "2018"
}

@article{Hosseinzadeh:2018dkh,
    author = "Hosseinzadeh, V. and Seraj, A. and Sheikh-Jabbari, M. M.",
    title = "{Soft Charges and Electric-Magnetic Duality}",
    eprint = "1806.01901",
    archivePrefix = "arXiv",
    primaryClass = "hep-th",
    reportNumber = "IPM-P-2018-030",
    journal = "JHEP",
    volume = "08",
    pages = "102",
    year = "2018"
}

@article{Speranza:2017gxd,
    author = "Speranza, A. J.",
    title = "{Local phase space and edge modes for diffeomorphism-invariant theories}",
    eprint = "1706.05061",
    archivePrefix = "arXiv",
    primaryClass = "hep-th",
    journal = "JHEP",
    volume = "02",
    pages = "021",
    year = "2018"
}

@article{Banados:1992wn,
    author = "Ba\~nados, M. and Teitelboim, C. and Zanelli, J.",
    title = "{The Black hole in three-dimensional space-time}",
    eprint = "hep-th/9204099",
    archivePrefix = "arXiv",
    reportNumber = "PRINT-92-0151 (CHILE), IASSNS-HEP-92-29",
    journal = "Phys. Rev. Lett.",
    volume = "69",
    pages = "1849--1851",
    year = "1992"
}

@article{Henneaux:2018gfi,
    author = "Henneaux, M. and Troessaert, C.",
    title = "{Asymptotic symmetries of electromagnetism at spatial infinity}",
    eprint = "1803.10194",
    archivePrefix = "arXiv",
    primaryClass = "hep-th",
    journal = "JHEP",
    volume = "05",
    pages = "137",
    year = "2018"
}

@article{Donnay:2016ejv,
    author = "Donnay, L. and Giribet, G. and Gonz{\'a}lez, H. A. and Pino, M.",
    title = "{Extended Symmetries at the Black Hole Horizon}",
    eprint = "1607.05703",
    archivePrefix = "arXiv",
    primaryClass = "hep-th",
    journal = "JHEP",
    volume = "09",
    pages = "100",
    year = "2016"
}

@article{Prohazka:2017equ,
    author = {Prohazka, S. and Salzer, J. and Sch{\"o}ller, F.},
    title = "{Linking Past and Future Null Infinity in Three Dimensions}",
    eprint = "1701.06573",
    archivePrefix = "arXiv",
    primaryClass = "hep-th",
    journal = "Phys. Rev. D",
    volume = "95",
    number = "8",
    pages = "086011",
    year = "2017"
}

@article{Esmaeili:2019hom,
    author = "Esmaeili, E.",
    title = "{Asymptotic Symmetries of Maxwell Theory in Arbitrary Dimensions at Spatial Infinity}",
    eprint = "1902.02769",
    archivePrefix = "arXiv",
    primaryClass = "hep-th",
    journal = "JHEP",
    volume = "10",
    pages = "224",
    year = "2019"
}

@article{Carlip:1995qv,
    author = "Carlip, S.",
    title = "{The (2+1)-Dimensional black hole}",
    eprint = "gr-qc/9506079",
    archivePrefix = "arXiv",
    reportNumber = "UCD-95-15",
    journal = "Class. Quant. Grav.",
    volume = "12",
    pages = "2853--2880",
    year = "1995"
}

@article{Banados:1992gq,
    author = "Banados, M. and Henneaux, M. and Teitelboim, C. and Zanelli, J.",
    title = "{Geometry of the (2+1) black hole}",
    eprint = "gr-qc/9302012",
    archivePrefix = "arXiv",
    reportNumber = "IASSNS-HEP-92-81",
    journal = "Phys. Rev. D",
    volume = "48",
    pages = "1506--1525",
    year = "1993",
    note = "[Erratum: Phys.Rev.D 88, 069902 (2013)]"
}

@article{Holzegel:2015jwa,
  author        = {Holzegel, G. and Luk, J. and Smulevici, J. and Warnick, C.},
  title         = {{Asymptotic properties of linear field equations in anti-de Sitter space}},
  journal       = {Commun. Math. Phys.},
  volume        = {374},
  number        = {2},
  pages         = {1125--1178},
  year          = {2020},
  eprint        = {1502.04965},
  archivePrefix = {arXiv},
  primaryClass  = {gr-qc}
}

@article{deBoer:2013gz,
  author        = {de Boer, J. and Jottar, J. I.},
  title         = {Thermodynamics of Higher Spin Black Holes in AdS$_3$},
  journal       = {JHEP},
  volume        = {01},
  pages         = {023},
  year          = {2014},
  eprint        = {1302.0816},
  archivePrefix = {arXiv},
  primaryClass  = {hep-th}
}

@article{Ferlaino:2013vga,
  author        = {Ferlaino, M. and Hollowood, T. J. and Kumar, S. P.},
  title         = {Asymptotic symmetries and thermodynamics of higher spin black holes in AdS$_3$},
  journal       = {Phys. Rev. D},
  volume        = {88},
  number        = {6},
  pages         = {066010},
  year          = {2013},
  eprint        = {1305.2011},
  archivePrefix = {arXiv},
  primaryClass  = {hep-th}
}

@article{Alessio:2020ioh,
    author = "Alessio, F. and Barnich, G. and Ciambelli, L. and Mao, P. and Ruzziconi, R.",
    title = "{Weyl charges in asymptotically locally AdS$_3$ spacetimes}",
    eprint = "2010.15452",
    archivePrefix = "arXiv",
    primaryClass = "hep-th",
    reportNumber = "CJQS-2021-002",
    journal = "Phys. Rev. D",
    volume = "103",
    number = "4",
    pages = "046003",
    year = "2021"
}

@article{Ciambelli:2023bmn,
    author = "Ciambelli, L. and D'Alise, A. and D'Esposito, V. and {\DJ}or{\dj}evi\'c, D. and Fern{\'a}ndez-Silvestre, D. and Varrin, L.",
    title = "{Cornering quantum gravity}",
    eprint = "2307.08460",
    archivePrefix = "arXiv",
    primaryClass = "hep-th",
    journal = "PoS",
    volume = "QG-MMSchools",
    pages = "010",
    year = "2024"
}

@article{Flanagan:2015pxa,
    author = "Flanagan, {\'E}. {\'E}. and Nichols, D. A.",
    title = "{Conserved charges of the extended Bondi-Metzner-Sachs algebra}",
    eprint = "1510.03386",
    archivePrefix = "arXiv",
    primaryClass = "hep-th",
    journal = "Phys. Rev. D",
    volume = "95",
    number = "4",
    pages= "044002",
    year = "2017",
    note = "[Erratum: Phys.Rev.D 108, 069902 (2023)]"
}

@article{Sachs:1962wk,
    author = "Sachs, R. K.",
    title = "{Gravitational waves in general relativity. 8. Waves in asymptotically flat space-times}",
    journal = "Proc. Roy. Soc. Lond. A",
    volume = "270",
    pages = "103--126",
    year = "1962"
}

@article{Ashtekar:1999,
    author = "Ashtekar, A. and Corichi, A. and Krasnov, K.",
    title = "{Isolated Horizons: the Classical Phase Space}",
    journal = "Adv. Theor. Math. Phys.",
    volume = "3",
    pages = "419--478",
    year = "1999"
}

@article{Ashtekar:2000,
    author = "Ashtekar, A. and Fairhurst, S. and Krishnan, B.",
    title = "{Isolated horizons: Hamiltonian evolution and the first law}",
    journal = "Phys. Rev. D",
    volume = "62",
    pages = "104025",
    year = "2000"
}

@article{Corichi:2025,
    author = "Corichi, A. and Reyes, J. D. and Vuka\v{s}inac, T.",
    title = "{On covariant and canonical Hamiltonian formalisms: weakly isolated horizons}",
    journal = "Class. Quantum Grav.",
    volume = "42",
    pages = "175022",
    year = "2025"
}

@article{Tamburino:1966zz,
    author = "Tamburino, L. A. and Winicour, J. H.",
    title = "{Gravitational Fields in Finite and Conformal Bondi Frames}",
    journal = "Phys. Rev.",
    volume = "150",
    pages = "1039--1053",
    year = "1966"
}

@article{Araujo-Regado:2024dpr,
    author = "Araujo-Regado, G. and Hoehn, P. A. and Sartini, F. and Tomova, B.",
    title = "{Soft edges: the many links between soft and edge modes}",
    eprint = "2412.14548",
    archivePrefix = "arXiv",
    primaryClass = "hep-th",
    journal = "JHEP",
    volume = "07",
    pages = "180",
    year = "2025"
}

@article{Simic:2023exz,
    author = "Simi{\'c}, D.",
    title = "{Note on asymptotic symmetry of massless scalar field at null infinity}",
    eprint = "2309.06148",
    archivePrefix = "arXiv",
    primaryClass = "hep-th",
    journal = "Phys. Rev. D",
    volume = "108",
    number = "8",
    pages = "085017",
    year = "2023"
}

@article{Bondi:1962px,
    author = "Bondi, H. and van der Burg, M. G. J. and Metzner, A. W. K.",
    title = "{Gravitational waves in general relativity. VII. Waves from
    axisymmetric isolated systems}",
    journal = "Proc. Roy. Soc. Lond. A",
    volume = "269",
    pages = "21--52",
    year = "1962",
}

@article{He:2014cra,
    author = "He, T. and Mitra, P. and Porfyriadis, A. P. and Strominger, A.",
    title = "{New Symmetries of Massless QED}",
    eprint = "1407.3789",
    archivePrefix = "arXiv",
    primaryClass = "hep-th",
    journal = "JHEP",
    volume = "10",
    pages = "112",
    year = "2014",
}

@article{Kapec:2014zla,
    author = "Kapec, D. and Lysov, V. and Strominger, A.",
    title = "{Asymptotic Symmetries of Massless QED in Even Dimensions}",
    eprint = "1412.2763",
    archivePrefix = "arXiv",
    primaryClass = "hep-th",
    journal = "Adv. Theor. Math. Phys.",
    volume = "21",
    pages = "1747--1767",
    year = "2017",
}

@article{Kapec:2015ena,
    author = "Kapec, D. and Pate, M. and Strominger, A.",
    title = "{New Symmetries of QED}",
    eprint = "1506.02906",
    archivePrefix = "arXiv",
    primaryClass = "hep-th",
    journal = "Adv. Theor. Math. Phys.",
    volume = "21",
    pages = "1769--1785",
    year = "2017",
}

@article{Strominger:2013lka,
    author = "Strominger, A.",
    title = "{Asymptotic Symmetries of Yang-Mills Theory}",
    eprint = "1308.0589",
    archivePrefix = "arXiv",
    primaryClass = "hep-th",
    journal = "JHEP",
    volume = "07",
    pages = "151",
    year = "2014",
}

@article{Mao:2017wvx,
    author = "Mao, P. and Wu, J.",
    title = "{Note on asymptotic symmetries and soft gluon theorems}",
    eprint = "1704.05740",
    archivePrefix = "arXiv",
    primaryClass = "hep-th",
    journal = "Phys. Rev. D",
    volume = "96",
    number = "6",
    pages = "065023",
    year = "2017",
}

@article{Maskawa:1975hx,
    author = "Maskawa, T. and Yamawaki, K.",
    title = "{The Problem of $P^{+}=0$ Mode in the Null-Plane Field Theory
    and Dirac's Method of Quantization}",
    journal = "Prog. Theor. Phys.",
    volume = "56",
    pages = "270--283",
    year = "1976",
}

@article{Heinzl:1998kz,
    author = "Heinzl, T.",
    title = "{Light-Cone Dynamics of Particles and Fields}",
    eprint = "hep-th/9812190",
    archivePrefix = "arXiv",
    year = "1998"
}

@article{Alexandrov:2014rba,
    author = "Alexandrov, S. and Speziale, S.",
    title = "{First Order Gravity on the Light Front}",
    eprint = "1412.6057",
    archivePrefix = "arXiv",
    primaryClass = "gr-qc",
    journal = "Phys. Rev. D",
    volume = "91",
    number = "6",
    pages = "064043",
    year = "2015",
}

@article{Nagarajan:1985xn,
    author = "Nagarajan, R. and Goldberg, J. N.",
    title = "{Canonical Formalism on a Null Surface: The Scalar and the Electromagnetic Fields}",
    journal = "Phys. Rev. D",
    volume = "31",
    pages = "1354--1362",
    year = "1985"
}

@article{Majumdar:2022fut,
    author = "Majumdar, S.",
    title = "{Residual gauge symmetry in light-cone electromagnetism}",
    eprint = "2212.10637",
    archivePrefix = "arXiv",
    primaryClass = "hep-th",
    journal = "JHEP",
    volume = "02",
    pages = "215",
    year = "2023"
}

@article{Dirac:1949cp,
    author = "Dirac, P. A. M.",
    title = "{Forms of Relativistic Dynamics}",
    journal = "Rev. Mod. Phys.",
    volume = "21",
    pages = "392--399",
    year = "1949"
}

@article{Campiglia:2015qka,
    author = "Campiglia, M. and Laddha, A.",
    title = "{Asymptotic symmetries of QED and Weinberg{\textquoteright}s soft photon theorem}",
    eprint = "1505.05346",
    archivePrefix = "arXiv",
    primaryClass = "hep-th",
    journal = "JHEP",
    volume = "07",
    pages = "115",
    year = "2015"
}

@article{Campiglia:2017dpg,
    author = "Campiglia, M. and Coito, L. and Mizera, S.",
    title = "{Can scalars have asymptotic symmetries?}",
    eprint = "1703.07885",
    archivePrefix = "arXiv",
    primaryClass = "hep-th",
    journal = "Phys. Rev. D",
    volume = "97",
    number = "4",
    pages = "046002",
    year = "2018"
}

@article{Campiglia:2018see,
    author = "Campiglia, M. and Freidel, L. and Hopfmueller, F. and Soni, R. M.",
    title = "{Scalar Asymptotic Charges and Dual Large Gauge Transformations}",
    eprint = "1810.04213",
    archivePrefix = "arXiv",
    primaryClass = "hep-th",
    reportNumber = "TIFR/TH/18-22",
    journal = "JHEP",
    volume = "04",
    pages = "003",
    year = "2019"
}

@article{Benguria:1976in,
    author = "Benguria, R. and Cordero, P. and Teitelboim, C.",
    title = "{Aspects of the Hamiltonian Dynamics of Interacting Gravitational Gauge and Higgs Fields with Applications to Spherical Symmetry}",
    reportNumber = "Print-77-0408 (PRINCETON)",
    journal = "Nucl. Phys. B",
    volume = "122",
    pages = "61--99",
    year = "1977"
}

@article{Steinhardt:1979it,
    author = "Steinhardt, P. J.",
    title = "{Problems of Quantization in the Infinite Momentum Frame}",
    reportNumber = "HUTP-79-A033",
    journal = "Annals Phys.",
    volume = "128",
    pages = "425",
    year = "1980"
}

@article{Blagojevic:1993fp,
    author = "Blagojevi{\'c}, M. and Vuka\v{s}inac, T.",
    title = "{Hamiltonian analysis of SL(2,R) symmetry in Liouville theory}",
    eprint = "hep-th/9311032",
    archivePrefix = "arXiv",
    journal = "Class. Quant. Grav.",
    volume = "11",
    pages = "1155--1175",
    year = "1994"
}

@article{Goldberg:1991pb,
    author = "Goldberg, J. N.",
    title = "{Selfdual Maxwell field on a null cone}",
    journal = "Gen. Rel. Grav.",
    volume = "23",
    pages = "1403--1413",
    year = "1991"
}

@article{Mao:2023rca,
    author = "Mao, P. and Zhang, K.",
    title = "{Soft theorems in de Sitter spacetime}",
    eprint = "2308.08861",
    archivePrefix = "arXiv",
    primaryClass = "hep-th",
    journal = "JHEP",
    volume = "01",
    pages = "044",
    year = "2024"
}

@article{Cheng:2022xyr,
    author = "Cheng, P. and Mao, P.",
    title = "{Soft theorems in curved spacetime}",
    eprint = "2206.11564",
    archivePrefix = "arXiv",
    primaryClass = "hep-th",
    journal = "Phys. Rev. D",
    volume = "106",
    number = "8",
    pages = "L081702",
    year = "2022"
}

@article{Barnich:2001jy,
    author = "Barnich, G. and Brandt, F.",
    title = "{Covariant theory of asymptotic symmetries, conservation laws and central charges}",
    eprint = "hep-th/0111246",
    archivePrefix = "arXiv",
    reportNumber = "ULB-TH-01-19, MPI-MIS-94-2001",
    journal = "Nucl. Phys. B",
    volume = "633",
    pages = "3--82",
    year = "2002"
}

@article{Lee:1990nz,
    author = "Lee, J. and Wald, R. M.",
    title = "{Local symmetries and constraints}",
    journal = "J. Math. Phys.",
    volume = "31",
    pages = "725--743",
    year = "1990"
}

@article{Dordevic:2026dx, 
  author = "Đorđevic, D.  and  Mišković, O.  and  Montecinos, A.  and  Vukašinac, T.",
  title = "{Soft charges and zero modes at null boundaries}",
eprint = "2607.28543",
 archivePrefix = "arXiv",
  journal = "Proceedings of Science for the Conference School ``Foundations of General-Relativistic Gauge Field Theory''",
  pages = "005",
  year =  "held at Politecnico di Torino, Italy, in March 2026, FGRGFT2026"
}

@article{Diaz:2026igh,
    author = {Diaz, F. and H{\"u}sn{\"u}gil, S. and Labrin, O. and Sanhueza, L.},
    title = "{Gravitational Memory Beyond Null Infinity through Finite-Distance Carrollian Screens}",
    eprint = "2607.18675",
    archivePrefix = "arXiv",
    primaryClass = "hep-th",
    month = "7",
    year = "2026"
}

@article{Ciambelli:2025fbo,
    author = "Ciambelli, L. and He, T. and Zurek, K. M.",
    title = "{From Asymptotically Flat Gravity to Finite Causal Diamonds}",
    eprint = "2512.09018",
    archivePrefix = "arXiv",
    primaryClass = "hep-th",
    reportNumber = "CALT-TH 2025-039",
    doi = "10.1103/lbm1-vkks",
    journal = "Phys. Rev. Lett.",
    volume = "136",
    number = "19",
    pages = "191501",
    year = "2026"
}

@article{Ciambelli:2026pwi,
    author = "Ciambelli, L. and He, T. and Klinger, M. S. and Zurek, K. M.",
    title = "{Mapping the Infrared Phase Space of Gravity to Finite Subregions}",
    eprint = "2606.12515",
    archivePrefix = "arXiv",
    primaryClass = "hep-th",
    reportNumber = "CALT-TH 2026-023",
    month = "6",
    year = "2026"
}

@article{tong_gaugetheory,
    author    = {Tong, D.},
    title     = {Lectures on Gauge Theory},
    year      = {2018},
    institution = {University of Cambridge},
    note      = {Lecture notes available at the University of Cambridge}
}

@article{Turner:2019wnh,
    author = "Turner, C.",
    title = "{Dualities in 2+1 Dimensions}",
    eprint = "1905.12656",
    archivePrefix = "arXiv",
    primaryClass = "hep-th",
    doi = "10.22323/1.349.0001",
    journal = "PoS",
    volume = "Modave2018",
    pages = "001",
    year = "2019"
}

@article{Maggiore:2019wie,
    author = "Maggiore, N.",
    title = "{Conserved chiral currents on the boundary of 3D Maxwell theory}",
    eprint = "1902.01901",
    archivePrefix = "arXiv",
    primaryClass = "hep-th",
    doi = "10.1088/1751-8121/ab045a",
    journal = "J. Phys. A",
    volume = "52",
    number = "11",
    pages = "115401",
    year = "2019"
}

@article{Gourgoulhon:2005ng,
    author = "Gourgoulhon, E. and Jaramillo, J. L.",
    title = "{A 3+1 perspective on null hypersurfaces and isolated horizons}",
    eprint = "gr-qc/0503113",
    archivePrefix = "arXiv",
    doi = "10.1016/j.physrep.2005.10.005",
    journal = "Phys. Rept.",
    volume = "423",
    pages = "159--294",
    year = "2006"
}

@article{Madler:2025ibn,
    author = {M{\"a}dler, T. and Gannouji, R. and Gallo, E.},
    title = "{Characteristic initial value problems for the Einstein-Maxwell-scalar field equations in spherical symmetry}",
    eprint = "2503.24162",
    archivePrefix = "arXiv",
    primaryClass = "gr-qc",
    doi = "10.1103/f5fc-zpld",
    journal = "Phys. Rev. D",
    volume = "111",
    number = "12",
    pages = "124015",
    year = "2025"
}

@article{Iyer:1994ys,
    author = "Iyer, V. and Wald, R. M.",
    title = "{Some properties of Noether charge and a proposal for dynamical black hole entropy}",
    eprint = "gr-qc/9403028",
    archivePrefix = "arXiv",
    doi = "10.1103/PhysRevD.50.846",
    journal = "Phys. Rev. D",
    volume = "50",
    pages = "846--864",
    year = "1994"
}

@article{Bondi:1960jsa,
    author = "Bondi, H.",
    title = "{Gravitational Waves in General Relativity}",
    doi = "10.1038/186535a0",
    journal = "Nature",
    volume = "186",
    number = "4724",
    pages = "535--535",
    year = "1960"
}

@article{Mukherjee:2023ihb,
    author = "Mukherjee, Jyotirmoy",
    title = "{Entanglement entropy and the boundary action of edge modes}",
    eprint = "2310.14690",
    archivePrefix = "arXiv",
    primaryClass = "hep-th",
    doi = "10.1007/JHEP06(2024)113",
    journal = "JHEP",
    volume = "06",
    pages = "113",
    year = "2024"
}

@article{Hofman:2018lfz,
    author = "Hofman, D.~M. and Iqbal, N.",
    title = "{Goldstone modes and photonization for higher form symmetries}",
    eprint = "1802.09512",
    archivePrefix = "arXiv",
    primaryClass = "hep-th",
    doi = "10.21468/SciPostPhys.6.1.006",
    journal = "SciPost Phys.",
    volume = "6",
    number = "1",
    pages = "006",
    year = "2019"
}

@article{Hofman:2024oze,
    author = "Hofman, D.~M. and Vitouladitis, S.",
    title = "{Generalised symmetries and state-operator correspondence for nonlocal operators}",
    eprint = "2406.02662",
    archivePrefix = "arXiv",
    primaryClass = "hep-th",
    doi = "10.1007/JHEP02(2025)061",
    journal = "JHEP",
    volume = "02",
    pages = "061",
    year = "2025"
}

@article{Vitouladitis:2025zoy,
    author = "Vitouladitis, S.",
    title = "{Higher-form anomalies and state-operator correspondence beyond conformal invariance}",
    eprint = "2507.01104",
    archivePrefix = "arXiv",
    primaryClass = "hep-th",
    doi = "10.21468/SciPostPhys.20.1.011",
    journal = "SciPost Phys.",
    volume = "20",
    number = "1",
    pages = "011",
    year = "2026"
}

@article{Arbalestrier:2026lna,
    author = "Arbalestrier, A. and Paznokas, E. and Vitouladitis, S.",
    title = "{The state/defect correspondence}",
    eprint = "2606.12560",
    archivePrefix = "arXiv",
    primaryClass = "hep-th",
    month = "6",
    year = "2026"
}

@article{Gaiotto:2014kfa,
    author = "Gaiotto, D. and Kapustin, A. and Seiberg, N. and W., Brian",
    title = "{Generalized Global Symmetries}",
    eprint = "1412.5148",
    archivePrefix = "arXiv",
    primaryClass = "hep-th",
    doi = "10.1007/JHEP02(2015)172",
    journal = "JHEP",
    volume = "02",
    pages = "172",
    year = "2015"
}

@article{Lake:2018dqm,
    author = "Lake, E.",
    title = "{Higher-form symmetries and spontaneous symmetry breaking}",
    eprint = "1802.07747",
    archivePrefix = "arXiv",
    primaryClass = "hep-th",
    month = "2",
    year = "2018"
}

@article{Freidel:2018fsk,
    author = "Freidel, L. and Pranzetti, D.",
    title = "{Electromagnetic duality and central charge}",
    eprint = "1806.03161",
    archivePrefix = "arXiv",
    primaryClass = "hep-th",
    doi = "10.1103/PhysRevD.98.116008",
    journal = "Phys. Rev. D",
    volume = "98",
    number = "11",
    pages = "116008",
    year = "2018"
}

@article{Pasterski:2015zua,
    author = "Pasterski, S.",
    title = "{Asymptotic Symmetries and Electromagnetic Memory}",
    eprint = "1505.00716",
    archivePrefix = "arXiv",
    primaryClass = "hep-th",
    doi = "10.1007/JHEP09(2017)154",
    journal = "JHEP",
    volume = "09",
    pages = "154",
    year = "2017"
}

@article{Shao:2025qvf,
    author = "Shao, S.-H. and Zhong, S.",
    title = "{Where non-invertible symmetries end: twist defects for electromagnetic duality}",
    eprint = "2509.21279",
    archivePrefix = "arXiv",
    primaryClass = "hep-th",
    reportNumber = "MIT-CTP/5930, YITP-SB-2025-17",
    doi = "10.1007/JHEP01(2026)118",
    journal = "JHEP",
    volume = "01",
    pages = "118",
    year = "2026"
}

@article{Chandrasekaran:2018aop,
    author = "Chandrasekaran, V. and Flanagan, {\'E}.~{\'E}. and Prabhu, K.",
    title = "{Symmetries and charges of general relativity at null boundaries}",
    eprint = "1807.11499",
    archivePrefix = "arXiv",
    primaryClass = "hep-th",
    doi = "10.1007/JHEP11(2018)125",
    journal = "JHEP",
    volume = "11",
    pages = "125",
    year = "2018",
    note = "[Erratum: JHEP 07, 224 (2023)]"
}

@article{Polyakov:1977,
    author = "Polyakov, A.~M.",
    title = "{Quark confinement and topology of gauge theories}",
    doi = "10.1016/0550-3213(77)90086-4",
    journal = "Nuclear Physics B",
    volume = "120",
    number = "3",
    pages = "429--458",
    year = "1977"
}

@article{Sommerfeld:1896,
    author  = "Sommerfeld, A.",
    title   = "{Mathematische Theorie der Diffraction}",
    journal = "Mathematische Annalen",
    volume  = "47",
    pages   = "317--374",
    year    = "1896",
    doi     = "10.1007/BF01447273"
}

@article{Meixner:1972,
    author  = "Meixner, J.",
    title   = "{The behavior of electromagnetic fields at edges}",
    journal = "IEEE Transactions on Antennas and Propagation",
    volume  = "20",
    number  = "4",
    pages   = "442--446",
    year    = "1972",
    doi     = "10.1109/TAP.1972.1140243"
}

@article{Schwarz:1982ec,
    author  = "Schwarz, A.~S.",
    title   = "{Field theories with no local conservation of the electric charge}",
    journal = "Nucl. Phys. B",
    volume  = "208",
    pages   = "141--158",
    year    = "1982",
    doi     = "10.1016/0550-3213(82)90190-0"
}

@article{Alford:1990mk,
    author  = "Alford, M.~G. and Benson, K. and
               Coleman, S.~R. and March-Russell, J. and
               Wilczek, F.",
    title   = "{Interactions and excitations of non-Abelian vortices}",
    journal = "Phys. Rev. Lett.",
    volume  = "64",
    pages   = "1632--1635",
    year    = "1990",
    doi     = "10.1103/PhysRevLett.64.1632",
    note    = "[Erratum: Phys. Rev. Lett. 65, 668 (1990)]"
}

@article{Bianchi:2021snj,
    author = "Bianchi, L. and Chalabi, A. and Proch{\'a}zka, V. and Robinson, Brandon and Sisti, Jacopo",
    title = "{Monodromy defects in free field theories}",
    eprint = "2104.01220",
    archivePrefix = "arXiv",
    primaryClass = "hep-th",
    reportNumber = "UUITP- 16/21",
    doi = "10.1007/JHEP08(2021)013",
    journal = "JHEP",
    volume = "08",
    pages = "013",
    year = "2021"
}

@article{Dixon:1986qv,
    author  = "Dixon, L.~J. and Friedan, D. and
               Martinec, E.~J. and Shenker, Stephen H.",
    title   = "{The Conformal Field Theory of Orbifolds}",
    journal = "Nucl. Phys. B",
    volume  = "282",
    pages   = "13--73",
    year    = "1987",
    doi     = "10.1016/0550-3213(87)90676-6"
}

@article{Wald:1999wa,
    author = "Wald, R.~M. and Zoupas, A.",
    title = "{A General definition of 'conserved quantities' in general relativity and other theories of gravity}",
    eprint = "gr-qc/9911095",
    archivePrefix = "arXiv",
    doi = "10.1103/PhysRevD.61.084027",
    journal = "Phys. Rev. D",
    volume = "61",
    pages = "084027",
    year = "2000"
}

\end{document}